\documentclass[12pt]{article}
\usepackage{amsmath,amssymb,array,calc,rotating,epsfig,psfrag,amscd, cite}
\usepackage{epsfig}
\usepackage{amsmath}
\usepackage{hyperref}
\hypersetup{
    colorlinks,
    citecolor=black,
    filecolor=black,
    linkcolor=black,
    urlcolor=black
}
\makeatletter
\renewcommand\section{\@startsection {section}{1}{\z@}%
                                   {-3.5ex \@plus -1ex \@minus -.2ex}
                                   {2.3ex \@plus.2ex}%
                                   {\normalfont\large\bfseries}}
\renewcommand\subsection{\@startsection{subsection}{2}{\z@}%
                                     {-3.25ex\@plus -1ex \@minus -.2ex}%
                                     {1.5ex \@plus .2ex}%
                                     {\normalfont\bfseries}}
\makeatother

\newenvironment{acknowledgments}{%
\vskip 3.25ex
\noindent {\bf Acknowledgments}
}

\newcommand{\be}{\begin{equation}}
\newcommand{\ee}{\end{equation}}
\newcommand{\bea}{\begin{eqnarray}}
\newcommand{\eea}{\end{eqnarray}}

\newcommand{\al}{\alpha}
\renewcommand{\d}{\delta}
\newcommand{\e}{\epsilon}
\newcommand{\G}{\Gamma}
\newcommand{\g}{\gamma}

\newcommand{\la}{\lambda}
\newcommand{\m}{\mu}

\newcommand{\vp}{\varphi}

\newcommand{\hlf}{\frac{1}{2}}

\newcommand{\non}{\nonumber}

\newcommand{\p}{\partial}

\newcommand{\R}{\mathbb{R}}
\newcommand{\rr}{\rightarrow}

\newcommand{\SO}{\operatorname{SO}}

\newcommand{\lp}{\left(}
\newcommand{\rp}{\right)}
\newcommand{\ls}{\left[}
\newcommand{\rs}{\right]}

\newcommand{\hph}[1]{{\hphantom{#1}}}

\newcommand{\mcA}{\mathcal{A}}
\newcommand{\mcD}{\mathcal{D}}

\newcommand{\mcN}{\mathcal{N}}
\newcommand{\mcO}{\mathcal{O}}
\newcommand{\whf}{\widehat{f}}
\newcommand{\whg}{\widehat{g}}
\newcommand{\wtmcA}{{\widetilde{\mathcal{A}}}}
\newcommand{\wtmcO}{{\widetilde{\mathcal{O}}}}
\newcommand{\wtPi}{{\widetilde{\Pi}}}

\begin{document}
\begin{titlepage}

\begin{center}

\hfill         MI-TH-1530

\vskip 2 cm
{\Large \bf Scalar-Vector Bootstrap}\\
\vskip 1.25 cm {Fernando Rejon-Barrera\footnote{email address: F.G.RejonBarrera@uva.nl}$^{a}$ and Daniel Robbins\footnote{email address:
drobbins@physics.tamu.edu}}$^{b}$\\
{\vskip 0.5cm $^{a}$ \it Institute for Theoretical Physics, University of Amsterdam, \\ Science Park 904, Postbus 94485, 1090 GL, Amsterdam, The Netherlands \\}

{\vskip 0.5cm $^{b}$ \it Department of Physics, Texas A{\&}M University, \\ College Station, TX 77843, USA \\}

\end{center}
\vskip 2 cm

\begin{abstract}
\baselineskip=18pt
We work out all of the details required for implementation of the conformal bootstrap program applied to the four-point function of two scalars and two vectors in an abstract conformal field theory in arbitrary dimension.  This includes a review of which tensor structures make appearances, a construction of the projectors onto the required mixed symmetry representations, and a computation of the conformal blocks for all possible operators which can be exchanged.  These blocks are presented as differential operators acting upon the previously known scalar conformal blocks.  Finally, we set up the bootstrap equations which implement crossing symmetry.  Special attention is given to the case of conserved vectors, where several simplifications occur.

\end{abstract}

\end{titlepage}

\tableofcontents

\pagestyle{plain}
\baselineskip=19pt
\section{Introduction}

\vspace{0.5cm}

The data of an abstract unitary conformal field theory (CFT) in $D$ dimensions is encoded by the spectrum of primary operators and their OPEs, which are in turn specified by a finite number of real constants for each triplet of primary operators.  From this information, we can in principle compute any correlation function by iteratively performing OPEs to reduce the correlator to a two-point function.  This procedure should not depend on which order we perform OPEs, and the equivalence of different procedures puts constraints on which sets of data can correspond to consistent CFTs.  In particular, for a four-point function we can divide the four operators into pairs in three different ways, or channels.  Equivalence between these channels is called crossing symmetry, and the general endeavor of exploring the constraints on CFT data which are imposed by crossing symmetry is known as the conformal bootstrap program.

Since the revival of the conformal bootstrap program in recent years~\cite{Rattazzi}, several research groups have obtained both numerical (bounds on operator dimensions, OPE coefficients and central charges) \cite{Rychkov2009,Caracciolo2010,Poland,Rattazzi2011,Rattazzi2011a,Rychkov2011,Polanda,El-Showk2012a,El-Showk2014,El-Showk2014a,Caracciolo2014,Kos2014} and analytical results (determination of anomalous dimensions and OPE coefficients) \cite{Fitzpatrick2012,Komargodski2013,Fitzpatrick2014a,Kaviraj2015,Alday2015,Kaviraj2015a}, as well as studies in theories with global symmetries \cite{Vichi,Kos2014a,Bae2014,Nakayama2014,Chester2014,Kos2015} or supersymmetries \cite{Rattazzi2011a,Beem2013,Alday2014,Bashkirov2013,Fitzpatrick:2014oza,Khandker:2014mpa,Berkooz2014,Alday2015a,Nakayama2014a,Chester2014a,Beem2014,Bobev2015a,Bobev2015}. So far these results have arisen from bootstrapping 4-point functions of scalar (in relation to the Lorentz subgroup of the conformal group) operators\footnote{There is a notable recent exception of~\cite{Iliesiu:2015qra}, which considers bootstrapping fermionic operators in 3D.}, whose conformal blocks were computed in \cite{Dolan2001,Dolan2004,Dolan2011} for $D=2,4,6$ (in any even dimension they can be computed recursively) with numerical approximations  given in \cite{Hogervorst,Hogervorst2013,Kos2014a} for any dimension $D$.

However, the consistency conditions from scalar correlators are only a small part of the (infinitely) many conditions that the bootstrap program imposes.  One expects more interesting and universal bounds to arise from bootstrapping 4-point functions of operators with spin, such as the stress-tensor or conserved currents. The main obstacle in tackling these problems is that the full set of conformal blocks for spinning correlators is not readily available yet. Partial progress has been made in this direction. In \cite{Costa2011} it was observed that there is a class of conformal blocks of tensor 4-point functions that can be related (via differential operators) to the well known scalar blocks of \cite{Dolan2001,Dolan2004,Dolan2011}. However, the class of conformal blocks derived in this way is associated to the exchange of traceless symmetric operators $\mcO$, whereas tensor correlator bootstrap requires, in addition, the exchange of mixed-symmetric operators $\mcA$. Later, in \cite{Simmons-Duffin2014,Costa2014} it was shown that conformal blocks associated to $\mcA$ can be calculated as a (finite) sum of scalar blocks evaluated at zero spin which, in principle, can be done by a computer. However, the numerical evaluation of these blocks is quite resource intensive due to the fact that the number of terms in the sum increases rapidly with the spin of $\mcA$. In numerical computations one might get away with if the maximum spin of $\mcA$ is not too large, but this approach is hopeless in the analytic bootstrap, where one needs to have control over the conformal blocks at very high spin \cite{Fitzpatrick2012,Komargodski2013}. Therefore the objective of this paper is to start building explicit closed form expressions of spinning conformal blocks that can be used in the analytic bootstrap and for efficient numerical evaluation\footnote{During the preparation of this draft, \cite{Echeverri2015} appeared which generalizes \cite{Costa2011} and proposes a relation between spinning (not necessarily bosonic) blocks associated to mixed-symmetric exchange, to more basic ``seed'' conformal blocks in 4D. However, the ``seed'' blocks were not presented yet. In the language of that paper, our work provides the ``seed'' blocks for the $[k+1,1]$ representation.}.

To start with, in section \ref{sec:TensorStructures} we classify all of the tensor structures which can appear in the three- and four-point functions which concern us in this paper (namely three-point functions with either two scalars or a scalar and a vector, along with a third operator, and four-point functions with four, three, or two scalars with zero, one, or two vectors).  We pay special attention to the information obtained from exchanging two operators, especially when the operators are identical.  In section \ref{subsec:ConservedVectors} we work out the extra information available when the vector operator is conserved.

Section \ref{sec:ShadowFormalism} reviews the shadow formalism, and computes the three-point coefficients for shadow operators in terms of the three-point functions of the original operators, the results of which are needed for the computation of the conformal blocks.

Section \ref{sec:ConformalBlocks} is the heart of the paper, in which we compute the conformal blocks which are needed to implement the bootstrap program with two scalars and two vectors.  Using the shadow formalism and the results from appendix \ref{app:BuildingBlocks} (various identities obeyed by the building blocks of our correlators), appendix \ref{app:Projectors} (where we compute the projection operators corresponding to the Lorentz representations of exchanged operators; in particular the results of appendix \ref{subapp:PiA} for the mixed symmetry exchange are some of the novel ingredients which really allows us to compute the required blocks), and appendix \ref{app:Integrals} (where we evaluate all of the required basic integrals), we compute the required integrals and perform the monodromy projection to finally obtain the conformal blocks. The new result in this section is the computation of the mixed-symmetric blocks, which relies on the contraction formula (\ref{eq:mixedcontraction}). Further details on how this formula is derived, are given in appendix \ref{app:contractiondetails} where we also present one of the two new contractions that appear in conformal blocks of four vectors, as evidence that our methods can be applied in more general situations. Our results are all written in terms of differential operators acting on scalar conformal blocks.

Then in section \ref{sec:Bootstrap} we set up the bootstrap program for four-point functions of two scalars and two vectors.  In particular, we examine the case of conserved vectors, in which case several simplifications occur.  Finally, we summarize our results and look forward to future directions in section \ref{sec:Conclusions}.

The next step in this program is to use the results of this paper to obtain bounds (numerical or analytical) on the data of a general class of CFTs, and in particular for a CFT with a conserved primary vector operator and an associated continuous global symmetry.  More formally, one would like to use the techniques developed in the present work to set up the bootstrap for even more complicated four-point functions.  In particular, the cases of four vectors (in particular conserved currents), or correlators involving conserved stress tensors, would be of great interest.  The real prize would be to implement the bootstrap with four conserved stress-energy tensors, thus gleaning extremely general information about the space of consistent unitary CFTs.

\section{Tensor structures}
\label{sec:TensorStructures}

Conformal invariance places strong constraints on the form of correlation functions.  We will focus on correlations of primary operators.  Correlation functions of descendants can of course be obtained from those of primaries.  For two-point functions of primary operators, conformal invariance fixes the result up to an overall constant, and it is conventional to normalize the primary operators themselves to remove that remaining ambiguity.  Each three-point function is determined up to a finite number of constants, each one multiplying a different tensor structure.  These same constants appear in the operator product expansion (OPE).  For four-point functions (and higher, though we won't go beyond four-point in this paper), there are a finite number of tensor structures.  These tensor structures are multiplied not by constants in general, but by functions of the conformally invariant cross-ratios.  Since the four-point function can in principle be evaluated by splitting into two pairs of operators and then using operator product expansions to reduce the problem to a sum of two-point functions, it follow that the functions multiplying the tensor structures are determined by the spectrum of primary operators and 
the constants which appear in their three-point functions.

In this section we will determine the tensor structures which can appear in the four-point function of scalars and up to two vectors, and in the three-point functions which act as intermediate stages in the evaluation.  The techniques are well established~\cite{Dolan2001,Dolan2004,Dolan2011}, and especially in~\cite{Costa:2011mg}, but we give a self-contained presentation in order to establish our conventions and to put emphasis on the properties that will be most relevant for our purposes.  In subsequent sections we will compute the functions which multiply these tensor structures in terms of the underlying data of the CFT.

\subsection{Embedding space}

When considering the consequences of conformal invariance, it is often useful to make use of embedding space.  This is a $(D+2)$-dimensional space, with coordinates $P^A$ and metric
\be
ds^2=\eta_{AB}dP^AdP^B=-dP^+dP^-+\d_{ab}dP^adP^b,
\ee
on which the conformal group $\SO(D+1,1)$ acts linearly (we will be working with Euclidean signature in physical space throughout this paper).  The $D$-dimensional physical space is identified with a null projective surface.  The map to physical coordinates is given by
\be
x^a=P^a/P^+,
\ee
while we can do the inverse map by sending a point in physical space to a particular point on the projective null line,
\be
\label{eq:EmbeddingMap}
P^A(x)=(P^+,P^-,P^a)=(1,x^2,x^a),\qquad \eta_{AB}P^A(x)P^B(x)=0.
\ee

Now consider a tensor function of three coordinates (to serve as an example) on embedding space,
\be
F_{A_1\cdots A_k,B_1\cdots B_\ell,C_1\cdots C_m}(P_1,P_2,P_3),
\ee
which is homogeneous in each variable (so that it is well defined on projective hypersurfaces),
\be
F_{A_1\cdots A_k,B_1\cdots B_\ell,C_1\cdots C_m}(\la_1P_1,\la_2P_2,\la_3P_3)=\la_1^{-\Delta_1}\la_2^{-\Delta_2}\la_3^{-\Delta_3}F_{A_1\cdots A_k,B_1\cdots B_\ell,C_1\cdots C_m}(P_1,P_2,P_3),
\ee
and is transverse in the sense that
\begin{multline}
P_1^{A_1}F_{A_1\cdots A_k,B_1\cdots B_\ell,C_1\cdots C_m}(P_1,P_2,P_3)=\cdots=P_1^{A_k}F_{A_1\cdots A_k,B_1\cdots B_\ell,C_1\cdots C_m}(P_1,P_2,P_3)\\
=P_2^{B_1}F_{A_1\cdots A_k,B_1\cdots B_\ell,C_1\cdots C_m}(P_1,P_2,P_3)=\cdots=P_2^{B_\ell}F_{A_1\cdots A_k,B_1\cdots B_\ell,C_1\cdots C_m}(P_1,P_2,P_3)\\
=P_3^{C_1}F_{A_1\cdots A_k,B_1\cdots B_\ell,C_1\cdots C_m}(P_1,P_2,P_3)=\cdots=P_3^{C_m}F_{A_1\cdots A_k,B_1\cdots B_\ell,C_1\cdots C_m}(P_1,P_2,P_3)\\
=0.
\end{multline}

We can map this function to a tensor function on physical space by
\begin{multline}
f_{a_1\cdots a_k,b_1\cdots b_\ell,c_1\cdots c_m}(x_1,x_2,x_3)\\
=\frac{\p P_1^{A_1}}{\p x_1^{a_1}}\cdots\frac{\p P_1^{A_k}}{\p x_1^{a_k}}\frac{\p P_2^{B_1}}{\p x_2^{b_1}}\cdots\frac{\p P_2^{B_\ell}}{\p x_2^{b_\ell}}\frac{\p P_3^{C_1}}{\p x_3^{c_1}}\cdots\frac{\p P_3^{C_m}}{\p x_3^{c_m}}F_{A_1\cdots A_k,B_1\cdots B_\ell,C_1\cdots C_m}(P_1,P_2,P_3),
\end{multline}
where we use the map (\ref{eq:EmbeddingMap}).  Because we are mapping from a null hypersurface, different embedding space tensors can map to the same physical tensor if they are related by
\be
\label{eq:EmbeddingGaugeShifts}
F'_{A_1\cdots A_k,B_1\cdots B_\ell,C_1\cdots C_m}=F_{A_1\cdots A_k,B_1\cdots B_\ell,C_1\cdots C_m}+P_{1\,A_1}\Lambda_{A_2\cdots A_k,B_1\cdots B_\ell,C_1\cdots C_m},
\ee
for any choice $\Lambda_{A_2\cdots A_k,B_1\cdots B_\ell,C_1\cdots C_m}$, and similarly for each of the other indices.  We will sometimes refer to this redundancy (somewhat sloppily) as gauge freedom.

The resulting function $f_{a_1\cdots a_k,b_1\cdots b_\ell,c_1\cdots c_m}(x_1,x_2,x_3)$ transforms as a conformal tensor of weights $\Delta_1$, $\Delta_2$, and $\Delta_3$ under conformal transformations of $x_1$, $x_2$, or $x_3$ respectively.  It turns out that a converse is also true; any function which transforms as a tensor of weights $\Delta_i$ can be obtained from a homogenous (of weights $\Delta_i$) transverse tensor in embedding space, unique up to equivalences of the form (\ref{eq:EmbeddingGaugeShifts}).

Thus, in order to determine the possible form of correlation functions of given operators, we need only determine the homogenous transverse tensors in embedding space up to the equivalences.  In embedding space there are not many different objects we can build.  Any scalar must be built out of scalar products of distinct $P_i$'s, and it will be useful to define
\be
P_{ij}=-2\eta_{AB}P_i^AP_j^B.
\ee
In physical space, this simply projects down to $x_{ij}^2$, where $x_{ij}^a=x_i^a-x_j^a$.  To ensure that free indices are transverse, it will also be useful to define (for distinct $i$, $j$, and $k$)
\be
\label{eq:KDef}
K^{(ijk)}_A=\frac{P_{ik}P_{j\,A}-P_{ij}P_{k\,A}}{\lp P_{ij}P_{ik}P_{jk}\rp^{1/2}},
\ee
which is transverse with respect to $P_i^A$, and antisymmetric in $j$ and $k$, and projects down to
\be
k^{(ijk)}_a=\frac{x_{ij}^2\lp x_{ik}\rp_a-x_{ik}^2\lp x_{ij}\rp_a}{\lp x_{ij}^2x_{ik}^2x_{jk}^2\rp^{1/2}},
\ee
and for distinct $i$ and $j$, both
\be
N^{(ij)}_{A_1A_2}=\eta_{A_1A_2}+\frac{2}{P_{ij}}\lp P_{i\,A_1}P_{j\,A_2}+P_{j\,A_1}P_{i\,A_2}\rp,
\ee
which is transverse in both indices with respect to $P_i$ (or $P_j$) and projects to $\d_{ab}$, and
\be
M^{(ij)}_{AB}=\eta_{AB}+\frac{2}{P_{ij}}P_{j\,A}P_{i\,B},
\ee
which is transverse to $P_i^A$ in the first index and $P_j^B$ in the second index, and projects down to
\be
m^{(ij)}_{ab}=\d_{ab}-\frac{2}{x_{ij}^2}\lp x_{ij}\rp_a\lp x_{ij}\rp_b.
\ee
Note that these building blocks, particularly (\ref{eq:KDef}) are defined to be scale invariant.

Finally, note that if $F_{A_1\cdots A_k}$ (we suppress other indices for now) transforms in a given way under permutations, then its projection $f_{a_1\cdots a_k}$ will inherit the same transformation and will thus transform as the corresponding representation of the rotation group $\SO(D)$.  For example, if $F_{A_1\cdots A_k}$ is invariant under permutations of its indices, then $f_{a_1\cdots a_k}$ will be a symmetric tensor.  If $F_{A_1\cdots A_k}$ is also traceless, then so will be $f_{a_1\cdots a_k}$.

Appendix \ref{app:BuildingBlocks} contains several useful formulae and identities for these structures in physical space.

\subsection{Two-point functions}

As we will see in the next subsection, the primary operators we will need in this paper fall into two classes of irreducible representations of the rotation group $\SO(D)$.  We either have totally symmetric traceless tensors of spin $\ell$, $\mcO_{a_1\cdots a_\ell}(x)$, which includes scalars and vectors as special cases, or we have mixed symmetry tensors $\mcA_{a_1a_2b_1\cdots b_k}(x)$ which are completely traceless, are antisymmetric in $a_1$ and $a_2$, are totally symmetric in the $b_i$, and which vanish when antisymmetrized over any three indices.  In terms of Young tableaux, the $\mcO_{a_1\cdots a_\ell}$ are represented by a horizontal row of $\ell$ boxes, while $\mcA_{a_1a_2b_1\cdots b_k}$ are represented by one row of $k+1$ boxes and a second row with only one box (equivalently one column with two boxes and $k$ columns of one box each).  For each of these cases we construct projectors onto the given representation in Appendix \ref{app:Projectors}.  For $\mcO_{a_1\cdots a_\ell}$ and $\mcA_{a_1a_2b_1\cdots b_k}$ we use projectors
\be
\Pi^{(\ell)\,b_1\cdots b_\ell}_{a_1\cdots a_\ell},\qquad\mathrm{and}\qquad\wtPi^{(k)\,c_1c_2d_1\cdots d_k}_{a_1a_2b_1\cdots b_k},
\ee
given in (\ref{eq:STPi}) and (\ref{eq:GeneralMSPi}) respectively.  We can also write the projectors in embedding space by simply taking the expressions in Appendix \ref{app:Projectors} and replacing each $\d_{ab}$ with $N^{(ij)}_{AB}$, with $i$ labeling the operator being projected, and $j$ being an arbitrarily chosen other variable (the choice is not physically relevant and can be changed by a gauge transformation (\ref{eq:EmbeddingGaugeShifts}).

It is well known that we can diagonalize the space of primary operators with respect to the two-point correlation functions, so we will only need to compute the two-point function of either a pair or $\mcO$ operators or a pair of $\mcA$ operators.  Indeed, if we have
\be
h_{a_1\cdots a_\ell;b_1\cdots b_\ell}(x_1,x_2)=\left\langle\mcO_{a_1\cdots a_\ell}(x_1)\mcO_{b_1\cdots b_\ell}(x_2)\right\rangle,
\ee
then this must descend from a tensor $H_{A_1\cdots A_\ell,B_1\cdots B_\ell}$ in embedding space.  In order to get the symmetric traceless representation, we must be able to put the indices on projectors $\Pi^{(\ell)}$.  By transversality, each $A$ index must be carried by either $P_{1\,A}$, $N^{(12)}_{AA'}$, or $M^{(12)}_{AB}$.  The first possibility is pure gauge and can be discarded.  The second possibility, which projects down to $\d_{aa'}$ will be eliminated when multiplied by the projector $\Pi^{(\ell)}$, and so can also be discarded (though it will appear in the projectors themselves).  This leaves only the third possibilty.  In order to get the correct homogeneity property we must include the appropriate power of $P_{12}$.  Finally, then, we are left with the form
\be
H_{A_1\cdots A_\ell,B_1\cdots B_\ell}(P_1,P_2)=P_{12}^{-\Delta_\mcO}\Pi^{(\ell)\,C_1\cdots C_\ell}_{A_1\cdots A_\ell}\Pi^{(\ell)\,D_1\cdots D_\ell}_{B_1\cdots B_\ell}M^{(12)}_{C_1D_1}\cdots M^{(12)}_{C_\ell D_\ell},
\ee
which projects down to
\be
h_{a_1\cdots a_\ell,b_1\cdots b_\ell}(x_1,x_2)=\lp x_{12}^2\rp^{-\Delta_\mcO}\Pi^{(\ell)\,c_1\cdots c_\ell}_{a_1\cdots a_\ell}\Pi^{(\ell)\,d_1\cdots d_\ell}_{b_1\cdots b_\ell}m^{(12)}_{c_1d_1}\cdots m^{(12)}_{c_\ell d_\ell}.
\ee

The same reasoning gives\footnote{Making use of the symmetries of $\wtPi^{(k)}$, one can show that any other arrangement of indices on the $m^{(12)}$'s, e.g.\ replacing $m^{(12)}_{e_2g_2}m^{(12)}_{f_1h_1}$ by $m^{(12)}_{e_2h_1}m^{(12)}_{f_1g_2}$, is equivalent to the one given.}
\begin{multline}
\left\langle\mcA_{a_1a_2b_1\cdots b_k}(x_1)\mcA_{c_1c_2d_1\cdots d_k}(x_2)\right\rangle\\
=\lp x_{12}^2\rp^{-\Delta_\mcA}\wtPi^{(k)\,e_1e_2f_1\cdots f_k}_{a_1a_2b_1\cdots b_k}\wtPi^{(k)\,g_1g_2h_1\cdots h_k}_{c_1c_2d_1\cdots d_k}m^{(12)}_{e_1g_1}m^{(12)}_{e_2g_2}m^{(12)}_{f_1h_1}\cdots m^{(12)}_{f_kh_k}.
\end{multline}

\subsection{Three-point functions}

Similarly, three-point correlation functions can be lifted to embedding space.  If all operators are in irreducible representations, then $N^{(ij)}_{A_1A_2}$ should again only appear in projectors, so all indices will be carried by either $K^{(ijk)}_A$ or by $M^{(ij)}_{AB}$.

\subsubsection{$\langle SS\mcO\rangle$}

If the first two operators are scalars, then all the indices of the remaining operator (after projection) must be carried by $K^{(312)}_A$.  This will be vanishing for any irreducible representation except for the symmetric traceless representation.  Then the three-point correlator $\langle\phi_1(x_1)\phi_2(x_2)\mcO_{a_1\cdots a_\ell}(x_3)\rangle$ lifts to an embedding space tensor
\begin{multline}
F_{A_1\cdots A_\ell}(P_1,P_2,P_3)\\
=\la_{12\mcO}P_{12}^{\hlf\lp -\Delta_1-\Delta_2+\Delta_\mcO\rp}P_{13}^{\hlf\lp -\Delta_1+\Delta_2-\Delta_\mcO\rp}P_{23}^{\hlf\lp\Delta_1-\Delta_2-\Delta_\mcO\rp}\Pi^{(\ell)\,B_1\cdots B_\ell}_{A_1\cdots A_\ell}K^{(312)}_{B_1}\cdots K^{(312)}_{B_\ell},
\end{multline}
which projects down to
\begin{multline}
\left\langle\phi_1(x_1)\phi_2(x_2)\mcO_{a_1\cdots a_\ell}(x_3)\right\rangle=\la_{12\mcO}\lp x_{12}^2\rp^{\hlf\lp -\Delta_1-\Delta_2+\Delta_\mcO\rp}\lp x_{13}^2\rp^{\hlf\lp -\Delta_1+\Delta_2-\Delta_\mcO\rp}\\
\times\lp x_{23}^2\rp^{\hlf\lp\Delta_1-\Delta_2-\Delta_\mcO\rp}\Pi^{(\ell)\,b_1\cdots b_\ell}_{a_1\cdots a_\ell}k^{(312)}_{b_1}\cdots k^{(312)}_{b_\ell}.
\end{multline}
Here $\la_{12\mcO}$ is a constant real number (in a unitary CFT), which is otherwise arbitrary.

If the two scalars are identical, then the the result has the form
\be
\left\langle\phi(x_1)\phi(x_2)\mcO_{a_1\cdots a_\ell}(x_3)\right\rangle=\la_{\phi\phi\mcO}\lp x_{12}^2\rp^{\hlf\lp -2\Delta_\phi+\Delta_\mcO\rp}\lp x_{13}^2x_{23}^2\rp^{-\hlf\Delta_\mcO}\Pi^{(\ell)\,b_1\cdots b_\ell}_{a_1\cdots a_\ell}k^{(312)}_{b_1}\cdots k^{(312)}_{b_\ell},
\ee
and this result should be invariant under the exchange of $x_1$ and $x_2$, which in turn forces $\ell$ to be even (otherwise the result changes sign under this exchange, since we get one factor of $-1$ from each $k^{(312)}$).

\subsubsection{$\langle SV\mcO\rangle$}

Next we consider the three-point function with one scalar $\phi$, one vector $v_a$, and one other operator.  This correlator will lift to an embedding tensor $F_{AB_1\cdots B_m}(P_1,P_2,P_3)$ where $A$ is transverse to $P_2$ and the $B$ indices are transverse to $P_3$.  The $A$ index can only be carried by either $K^{(213)}_A$ or $M^{(23)}_{AB}$, and then the remaining $B$ indices (after being projected by the appropriate rotation group projector) must be carried by $K^{(312)}_B$.  In the latter case, no two of the indices carried by $K^{(312)}_B$ can be antisymmetric.  So the third operator can only be either totally symmetric $\mcO$, or it can be an $\mcA$ in the mixed symmetry representation described above, where one of the first two $B$ indices is carried by $M^{(23)}_{AB}$, and the rest by $K^{(312)}_B$'s.

In the case where the third operator is totally symmetric, then there are two structures which can arise, with embedding space form
\begin{multline}
F_{AB_1\cdots B_\ell}(P_1,P_2,P_3)=P_{12}^{\hlf\lp -\Delta_\phi-\Delta_v+\Delta_\mcO\rp}P_{13}^{\hlf\lp -\Delta_\phi+\Delta_v-\Delta_\mcO\rp}P_{23}^{\hlf\lp\Delta_\phi-\Delta_v-\Delta_\mcO\rp}\\
\times\Pi^{(\ell)\,C_1\cdots C_\ell}_{B_1\cdots B_\ell}\ls\al_{\phi v\mcO}K^{(213)}_AK^{(312)}_{B_1}\cdots K^{(312)}_{B_\ell}+\beta_{\phi v\mcO}M^{(23)}_{AB_1}K^{(312)}_{B_2}\cdots K^{(312)}_{B_\ell}\rs,
\end{multline}
which projects down to
\begin{multline}
\left\langle\phi(x_1)v_a(x_2)\mcO_{b_1\cdots b_\ell}(x_3)\right\rangle=\lp x_{12}^2\rp^{\hlf\lp -\Delta_\phi-\Delta_v+\Delta_\mcO\rp}\lp x_{13}^2\rp^{\hlf\lp -\Delta_\phi+\Delta_v-\Delta_\mcO\rp}\lp x_{23}^2\rp^{\hlf\lp\Delta_\phi-\Delta_v-\Delta_\mcO\rp}\\
\times\Pi^{(\ell)\,c_1\cdots c_\ell}_{b_1\cdots b_\ell}\ls\al_\mcO k^{(213)}_ak^{(312)}_{c_1}\cdots k^{(312)}_{c_\ell}+\beta_\mcO m^{(23)}_{ac_1}k^{(312)}_{c_2}\cdots k^{(312)}_{c_\ell}\rs.
\end{multline}

If $\ell=0$, then we only have the first term labeled by a constant $\al_{\phi v\mcO}$.  If $\ell>0$, then we have two distinct possible tensor structures labeled by two real constant numbers $\al_{\phi v\mcO}$ and $\beta_{\phi v\mcO}$.

\subsubsection{$\langle SV\mcA\rangle$}

Similar considerations for the case where the third operator has mixed symmetry show that the three-point correlation function will have the form
\begin{multline}
\left\langle\phi(x_1)v_a(x_2)\mcA_{b_1b_2c_1\cdots c_k}(x_3)\right\rangle=\g_{\phi v\mcA}\lp x_{12}^2\rp^{\hlf\lp -\Delta_\phi-\Delta_v+\Delta_\mcA\rp}\lp x_{13}^2\rp^{\hlf\lp -\Delta_\phi+\Delta_v-\Delta_\mcA\rp}\\
\times\lp x_{23}^2\rp^{\hlf\lp\Delta_\phi-\Delta_v-\Delta_\mcA\rp}\wtPi^{(k)\,d_1d_2e_1\cdots e_k}_{b_1b_2c_1\cdots c_k}m^{(23)}_{ad_1}k^{(312)}_{d_2}k^{(312)}_{e_1}\cdots k^{(312)}_{e_k},
\end{multline}
with $\g_{\phi v\mcA}$ as a real constant.

\subsection{Four-point functions}

The case of four-point functions proceeds similarly, with the main difference being that there are cross-ratios
\be
U=\frac{P_{12}P_{34}}{P_{13}P_{24}},\qquad V=\frac{P_{14}P_{23}}{P_{13}P_{24}},
\ee
in embedding space, or
\be
\label{eq:CrossRatios}
u=\frac{x_{12}^2x_{34}^2}{x_{13}^2x_{24}^2},\qquad v=\frac{x_{14}^2x_{23}^2}{x_{13}^2x_{24}^2},
\ee
in physical space.  Then each tensor structure is accompanied by a function of the cross-ratios rather than by just a constant.  

\subsubsection{$\langle SSSS\rangle$}

For the case of four scalars, we have
\begin{multline}
\label{eq:SSSSstructs}
\left\langle\phi_1(x_1)\phi_2(x_2)\phi_3(x_3)\phi_4(x_4)\right\rangle\\
=\lp\frac{x_{14}^2}{x_{13}^2}\rp^{\hlf\lp\Delta_3-\Delta_4\rp}\lp\frac{x_{24}^2}{x_{14}^2}\rp^{\hlf\lp\Delta_1-\Delta_2\rp}\lp x_{12}^2\rp^{-\hlf\lp\Delta_1+\Delta_2\rp}\lp x_{34}^2\rp^{-\hlf\lp\Delta_3+\Delta_4\rp}q(u,v),
\end{multline}
where $q(u,v)$ is an (a priori) arbitrary function of the cross-ratios $u$ and $v$.  The factor multiplying $q(u,v)$, which does the work in ensuring that the correlator scales correctly, will appear often, and so it is convenient to abbreviate it.  Thus, we define
\be
X_{\Delta_1,\Delta_2,\Delta_3,\Delta_4}=\lp\frac{x_{14}^2}{x_{13}^2}\rp^{\hlf\lp\Delta_3-\Delta_4\rp}\lp\frac{x_{24}^2}{x_{14}^2}\rp^{\hlf\lp\Delta_1-\Delta_2\rp}\lp x_{12}^2\rp^{-\hlf\lp\Delta_1+\Delta_2\rp}\lp x_{34}^2\rp^{-\hlf\lp\Delta_3+\Delta_4\rp},
\ee
and sometimes we will simply write $X_{\Delta_i}$ for short.

In the case that all four scalars are identical, we have
\be
\left\langle\phi(x_1)\phi(x_2)\phi(x_3)\phi(x_4)\right\rangle=\lp x_{12}^2x_{34}^2\rp^{-\Delta_\phi}q(u,v),
\ee
and invariance under exchange of $x_1$ with $x_2$ implies that
\be
q(u,v)=q(u/v,1/v),
\ee
while under exchange of $x_1$ and $x_3$ we have
\be
\label{eq:SSSSIdenticalCS}
q(u,v)=\lp\frac{u}{v}\rp^{\Delta_\phi}q(v,u).
\ee
Other permutations of the $x_i$ give no new information about the function $q(u,v)$.

\subsubsection{$\langle SVSS\rangle$ or $\langle SSSV\rangle$}

Let us now consider the four-point function of three scalars and one vector (in the second position to start).  In principle the free index could be carried, in embedding space, by any of the three possibilities $K^{(213)}_A$, $K^{(214)}_A$, or $K^{(234)}_A$, but it turns out that there is a linear relation (\ref{eq:kLinComb})
\be
K^{(213)}_A=V^{1/2}K^{(214)}_A-U^{1/2}K^{(234)}_A,
\ee
so only two of the combinations are independent, and we can write (after projecting to physical space)
\be
\label{eq:SVSSstructs}
\left\langle\phi_1(x_1)v_a(x_2)\phi_3(x_3)\phi_4(x_4)\right\rangle=X_{\Delta_1,\Delta_v,\Delta_3,\Delta_4}
\ls q_1(u,v)k^{(214)}_a+q_2(u,v)k^{(234)}_a\rs.
\ee

If $\phi_3$ and $\phi_4$ are identical, then symmetry under exchange of $x_3$ and $x_4$ implies that
\be
q_1(u/v,1/v)=v^{\hlf\lp\Delta_v-\Delta_1-1\rp}q_1(u,v),\ q_2(u/v,1/v)=-v^{\hlf\lp\Delta_v-\Delta_1\rp}\lp q_2(u,v)+\lp\frac{u}{v}\rp^\hlf q_1(u,v)\rp,
\ee
while if $\phi_1$ and $\phi_3$ are identical, then we have
\be
q_2(u,v)=\lp\frac{u}{v}\rp^{\hlf\lp\Delta_\phi+\Delta_v\rp}q_1(v,u).
\ee
If all three scalars are identical, then both sets of constraints hold.

The situation when the vector is in fourth position is completely analogous (we put primes on the $q_i'$ to distinguish them from the SVSS functions),
\be
\label{eq:SSSVstructs}
\left\langle\phi_1(x_1)\phi_2(x_2)\phi_3(x_3)v_a(x_4)\right\rangle=X_{\Delta_1,\Delta_2,\Delta_3,\Delta_v}
\ls q_1'(u,v)k^{(412)}_a+q_2'(u,v)k^{(432)}_a\rs.
\ee

\subsubsection{$\langle SVSV\rangle$}

Finally, consider a four-point function of two scalars and two vectors,
\be
f_{ab}(x_1,x_2,x_3,x_4)=\left\langle\phi_1(x_1)v_{2\,a}(x_2)\phi_3(x_3)v_{4\,b}(x_4)\right\rangle.
\ee
In embedding space, the indices of the corresponding tensor can be either carried by $M^{(24)}_{AB}$ or else both indices are carried by $K$'s.  There are two independent choices of $K$ possible for each index, and so there are five possible tensor structures altogether,
\begin{multline}
\label{eq:SVSVstructs}
f_{ab}=
X_{\Delta_i}\ls q_0(u,v)m^{(24)}_{ab}+q_{11}(u,v)k^{(214)}_ak^{(412)}_b+q_{12}(u,v)k^{(214)}_ak^{(432)}_b\right.\\
\left. +q_{21}(u,v)k^{(234)}_ak^{(412)}_b+q_{22}(u,v)k^{(234)}_ak^{(432)}_b\rs.
\end{multline}

If the two scalars are identical, then $x_1$-$x_3$ exchange gives constraints
\bea
\label{eq:IdenticalScalarsq_0}
q_0(v,u) &=& \lp\frac{v}{u}\rp^{\hlf\lp\Delta_\phi+\Delta_2\rp}q_0(u,v),\\
q_{21}(u,v) &=& \lp\frac{u}{v}\rp^{\hlf\lp\Delta_\phi+\Delta_2\rp}q_{12}(v,u),\\
q_{22}(u,v) &=& \lp\frac{u}{v}\rp^{\hlf\lp\Delta_\phi+\Delta_2\rp}q_{11}(v,u).
\eea
If the two vectors are identical, then exchanging $x_2$ and $x_4$ while also exchanging the indices $a$ and $b$, gives
\bea
\label{eq:IdenticalVectorsq_0}
q_0(v,u) &=& \lp\frac{v}{u}\rp^{\hlf\lp\Delta_3+\Delta_v\rp}q_0(u,v),\\
q_{11}(v,u) &=& \lp\frac{v}{u}\rp^{\hlf\lp\Delta_3+\Delta_v\rp}q_{11}(u,v),\\
q_{21}(u,v) &=& \lp\frac{u}{v}\rp^{\hlf\lp\Delta_3+\Delta_v\rp}q_{12}(v,u),\\
q_{22}(v,u) &=& \lp\frac{v}{u}\rp^{\hlf\lp\Delta_3+\Delta_v\rp}q_{22}(u,v).
\eea
Finally if we have two identical scalars and two identical vectors, then we can combine the constraints and determine
\bea
\label{eq:IdenticalOpsq_0}
q_0(v,u) &=& \lp\frac{v}{u}\rp^{\hlf\lp\Delta_\phi+\Delta_v\rp}q_0(u,v),\\
q_{11}(v,u) &=& \lp\frac{v}{u}\rp^{\hlf\lp\Delta_\phi+\Delta_v\rp}q_{11}(u,v),\\
q_{21}(u,v) &=& \lp\frac{u}{v}\rp^{\hlf\lp\Delta_\phi+\Delta_v\rp}q_{12}(v,u),\\
q_{22}(u,v) &=& q_{11}(u,v).
\eea
Thus in this case we have one unconstrained function $q_{12}(u,v)$, and two constrained functions $q_0(u,v)$ and $q_{11}(u,v)$, with $q_{21}(u,v)$ and $q_{22}(u,v)$ determined in terms of the others.

\subsection{Conserved vectors}
\label{subsec:ConservedVectors}

Many of the structures discussed above simplify somewhat if we are dealing with conserved vectors, which obey $\p^av_a(x)=0$ inside correlation functions.  From the vector-vector two-point function, we have
\bea
0 &=& \p_1^b\left\langle v_b(x_1)v_a(x_2)\right\rangle=\p_1^b\ls\lp x_{12}^2\rp^{-\Delta_v}m^{(12)}_{ba}\rs\non\\
&=& \lp x_{12}^2\rp^{-\Delta_v}\ls -\frac{2\Delta_vx_{12}^b}{x_{12}^2}m^{(12)}_{ba}-\frac{2\lp D-1\rp x_{12\,a}}{x_{12}^2}\rs\non\\
&=& 2\lp x_{12}^2\rp^{-\Delta_v-1}\lp\Delta_v-D+1\rp x_{12\,a}.
\eea
Thus we conclude that $\Delta_v=D-1$ for a conserved vector in $D$-dimensions, i.e.\ it saturates the unitarity bound.

Turning next to three-point functions, we have (for $\ell>0$)
\bea
0 &=& \p_2^b\left\langle\phi(x_1)v_b(x_2)\mcO_{a_1\cdots a_\ell}(x_3)\right\rangle\non\\
&=& \p_2^b\left\{\lp x_{12}^2\rp^{\hlf\lp -\Delta_\phi+\Delta_\mcO-D+1\rp}\lp x_{13}^2\rp^{\hlf\lp -\Delta_\phi-\Delta_\mcO+D-1\rp}\lp x_{23}^2\rp^{\hlf\lp\Delta_\phi-\Delta_\mcO-D+1\rp}\right.\non\\
&& \qquad\left.\times\Pi^{(\ell)\,c_1\cdots c_\ell}_{b_1\cdots b_\ell}\ls\al_{\phi v\mcO}k^{(213)}_ak^{(312)}_{c_1}\cdots k^{(312)}_{c_\ell}+\beta_{\phi v\mcO}m^{(23)}_{ac_1}k^{(312)}_{c_2}\cdots k^{(312)}_{c_\ell}\rs\vphantom{\lp x_{12}^2\rp^{\hlf\lp -\Delta_\phi+\Delta_\mcO-D+1\rp}}\right\}\non\\
&=& \lp x_{12}^2\rp^{\hlf\lp -\Delta_\phi+\Delta_\mcO-D\rp}\lp x_{13}^2\rp^{\hlf\lp -\Delta_\phi-\Delta_\mcO+D\rp}\lp x_{23}^2\rp^{\hlf\lp\Delta_\phi-\Delta_\mcO-D\rp}\\
&& \quad\times\ls\al_{\phi v\mcO}\lp\Delta_\phi-\Delta_\mcO\rp+\beta_{\phi v\mcO}\lp\Delta_\phi-\Delta_\mcO+D+\ell-2\rp\rs\Pi^{(\ell)\,b_1\cdots b_\ell}_{a_1\cdots a_\ell}k^{(312)}_{b_1}\cdots k^{(312)}_{b_\ell}.\non
\eea
where we have made use of the fact that $\Delta_v=D-1$.  For this expression to vanish, we require
\be
\label{eq:ConservedVec3pt}
\lp\Delta_\phi-\Delta_\mcO\rp\al_{\phi v\mcO}+\lp\Delta_\phi-\Delta_\mcO+D+\ell-2\rp\beta_{\phi v\mcO}=0.
\ee

For $\ell=0$, we simply set $\beta_{\phi v\mcO}=0$ in the above equation, and require either $\al_{\phi v\mcO}=0$ or $\Delta_\phi=\Delta_\mcO$.  Actually, we can assign some more physical significance to this case by first recalling that we expect each conserved primary vector operator to be associated to a one parameter continuous global symmetry of our CFT.  Now pick a particular conformal weight $\Delta$ and consider all scalar operators $\phi_i$ that have that weight.  Form a matrix $\al_{ij}$ by taking three point functions with the conserved vector $v_a$,
\be
\left\langle\phi_i(x_1)v_a(x_2)\phi_j(x_3)\right\rangle=\lp x_{12}^2x_{23}^2\rp^{-\hlf\lp D-1\rp}\lp x_{13}^2\rp^{\hlf\lp D-1\rp-\Delta}\al_{ij}k^{(213)}_a.
\ee
Then $\al_{ij}$ is antisymmetric in its indices.  Since we are free to make orthonormal (with respect to the normalized two-point functions) rotations on the space of $\phi_i$, we can always take a basis in which $\al_{ij}$ is block diagonal,
\be
\al_{ij}=\lp\begin{matrix} 0 & -Q_1 & & & & & & \\ Q_1 & 0 & \cdots & & & & & \\ & \vdots & \ddots & & & & & \\ & & & 0 & -Q_n & & & \\ & & & Q_n & 0 & & \cdots & \\ & & & & & 0 & & \\ & & & & \vdots & & \ddots & \\ & & & & & & & 0 \end{matrix}\rp.
\ee
Here $n$ is just the number of charged scalars with weight $\Delta$.  In this basis, we say that for $i>2n$, $\phi_i$ is neutral under the global symmetry.  We can combine the others into complex combinations $\varphi_i=\phi_{2i-1}+i\phi_{2i}$, and we can say that $\varphi_i$ has charge\footnote{We have chosen a normalization for the charge that is convenient from the point of view of an abstract CFT, since it is given simply by the three-point function of primary fields (which have themselves been normalized by their two-point functions).  However, it may well differ from other well-motivated normalizations.  For example, in the case of a free complex scalar in $D>2$ dimensions, and the usual global $\operatorname{U}(1)$ symmetry, our definition gives the scalar a charge of $\sqrt{\frac{D-2}{2}}$.} $Q_i$.

For the other three-point functions, $\langle\phi v\mcA\rangle$, a similar calculation shows that conservation is automatic once we impose that $\Delta_v=D-1$.  Conservation gives no other constraints in this case.

For four-point functions with conserved vectors, the coefficient functions must obey linear differential equations.  For example, in the $\langle SVSS\rangle$ amplitude, if the vector is conserved then the functions $q_1$ and $q_2$ must obey
\begin{multline}
0=\ls\Delta_1-2u\p_u+\lp -uv^{-1}-1+v^{-1}\rp v\p_v\rs q_1+\lp\frac{u}{v}\rp^\hlf\ls\frac{\Delta_1}{2}\lp 1+u^{-1}v-u^{-1}\rp\right.\\
\left. +\frac{D-1}{2}\lp -1+u^{-1}v-u^{-1}\rp+\lp -1-u^{-1}v+u^{-1}\rp u\p_u-2v\p_v\rs q_2.
\end{multline}

And in the case of $\langle SVSV\rangle$, if the vector at $x_4$ is conserved (so in particular $\Delta_4=D-1$), then we have
\bea
\label{eq:SVSVConserved1}
0 &=& \lp\Delta_1-\Delta_2-\Delta_3+D-1\rp q_0\non\\
&& \quad +\ls\lp\Delta_1-\Delta_2\rp+\hlf\lp\Delta_3-1\rp\lp -1-u^{-1}+u^{-1}v\rp+\frac{D}{2}\lp 1-u^{-1}+u^{-1}v\rp\rs q_{11}\non\\
&& \quad +\ls\hlf\lp\Delta_1-\Delta_2+D\rp\lp u^{1/2}v^{-1/2}+u^{-1/2}v^{1/2}-u^{-1/2}v^{-1/2}\rp\right.\non\\
&& \qquad\left. +\hlf\Delta_3\lp -u^{1/2}v^{-1/2}+u^{-1/2}v^{1/2}+u^{-1/2}v^{-1/2}\rp\rs q_{12}-u^{-1/2}v^{1/2}q_{21}-2v\p_vq_0\non\\
&& \quad +\lp 1-u-v\rp\p_uq_{11}-2v\p_vq_{11}-2u^{1/2}v^{1/2}\p_uq_{12}\non\\
&& \quad +\lp -u^{1/2}v^{1/2}-u^{-1/2}v^{3/2}+u^{-1/2}v^{1/2}\rp\p_vq_{12},
\eea
and
\bea
\label{eq:SVSVConserved2}
0 &=& \lp\Delta_3+D-1\rp q_0-u^{1/2}v^{-1/2}q_{12}\non\\
&& \quad +\ls\vphantom{\frac{D}{2}}\lp\Delta_1-\Delta_2\rp u^{1/2}v^{-1/2}+\hlf\Delta_3\lp -u^{1/2}v^{-1/2}+u^{-1/2}v^{1/2}-u^{-1/2}v^{-1/2}\rp\right.\non\\
&& \qquad\left. +\frac{D}{2}\lp u^{1/2}v^{-1/2}+u^{-1/2}v^{1/2}-u^{-1/2}v^{-1/2}\rp\rs q_{21}\non\\
&& \quad +\ls\hlf\lp\Delta_1-\Delta_2+D-1\rp\lp 1+uv^{-1}-v^{-1}\rp+\hlf\Delta_3\lp 1-uv^{-1}+v^{-1}\rp\rs q_{22}\non\\
&& \quad -2u\p_uq_0+\lp -u^{3/2}v^{-1/2}-u^{1/2}v^{1/2}+u^{1/2}v^{-1/2}\rp\p_uq_{21}-2u^{1/2}v^{1/2}\p_vq_{21}\non\\
&& \quad -2u\p_uq_{22}+\lp 1-u-v\rp\p_vq_{22}.
\eea

\section{Shadow formalism}
\label{sec:ShadowFormalism}

As an intermediate step in the calculation of conformal blocks, we will need to define {\it{shadow operators}}.  Given any local primary operator $\mcO_{a_1\cdots a_n}(x)$ of conformal weight $\Delta$, we can define its shadow operator
\be\label{eq:shadowopDef}
\wtmcO_{a_1\cdots a_\ell}(x_1)=\Pi^{(\ell)\,b_1\cdots b_\ell}_{a_1\cdots a_\ell}\int\frac{d^Dx_0}{\lp x_{01}^2\rp^{D-\Delta}}m^{(01)\hph{b_1}c_1}_{\hph{(01)}b_1}\cdots m^{(01)\hph{b_\ell}c_\ell}_{\hph{(01)}b_\ell}\mcO_{c_1\cdots c_\ell}(x_0),
\ee
which is a non-local operator that transforms as a primary operator of weight $D-\Delta$ under conformal transformations, and under $\SO(D)$ rotations transforms in the same way as $\mcO$.  When we insert $\wtmcO(x_1)$ in a correlation function, the prescription is to insert $\mcO(x_0)$, evaluate the correlation function, and then perform the integral above.

\subsection{Mixing matrices}

We would like to compute the constants which appear in three-point functions involving shadow operators.  Since $\wtmcO$ is linearly related to $\mcO$, the constant or constants appearing in a three-point function of $\wtmcO$ with two other operators will be linear combinations of the constants in the three-point function of $\mcO$ with those same two operators.  We would like to determine the matrices which encode these linear combinations.

\subsubsection{$\langle SS\wtmcO\rangle$}

Consider first the case where $\mcO_{a_1\cdots a_\ell}$ is symmetric traceless, and the other two operators are scalars $\phi_1$ and $\phi_2$.  The three-point function with $\mcO$ is fixed up to a single constant $\la_{12\mcO}=\la_\mcO$,
\begin{multline}\label{eq:phiphiOsymmetric}
\left\langle\phi_1(x_1)\phi_2(x_2)\mcO_{a_1\cdots a_\ell}(x_3)\right\rangle\\
=\la_\mcO\lp x_{12}^2\rp^{\hlf\lp -\Delta_1-\Delta_2+\Delta_\mcO\rp}\lp x_{13}^2\rp^{\hlf\lp -\Delta_1+\Delta_2-\Delta_\mcO\rp}\lp x_{23}^2\rp^{\hlf\lp\Delta_1-\Delta_2-\Delta_\mcO\rp}\Pi^{(\ell)\,b_1\cdots b_\ell}_{a_1\cdots a_\ell}k^{(312)}_{b_1}\cdots k^{(312)}_{b_\ell},
\end{multline}
and we expect that the shadow operator will similarly have
\begin{multline}\label{eq:phiphiOshadow}
\left\langle\phi_1(x_1)\phi_2(x_2)\wtmcO_{a_1\cdots a_\ell}(x_3)\right\rangle=\la_\wtmcO\lp x_{12}^2\rp^{\hlf\lp -\Delta_1-\Delta_2+\Delta_\wtmcO\rp}\lp x_{13}^2\rp^{\hlf\lp -\Delta_1+\Delta_2-\Delta_\wtmcO\rp}\\
\times\lp x_{23}^2\rp^{\hlf\lp\Delta_1-\Delta_2-\Delta_\wtmcO\rp}\Pi^{(\ell)\,b_1\cdots b_\ell}_{a_1\cdots a_\ell}k^{(312)}_{b_1}\cdots k^{(312)}_{b_\ell},
\end{multline}
where $\Delta_\wtmcO=D-\Delta_\mcO$. Inserting the definition of the shadow operator (\ref{eq:shadowopDef}) and performing the integral leads to \footnote{Details on the computation of these integrals are given in appendix \ref{app:mixingmatrices}}
\begin{multline}
\label{eq:lambdaOtilde}
\la_\wtmcO=\pi^{D/2}\frac{\G(\Delta_\mcO-\frac{D}{2})\G(\Delta_\mcO+\ell-1)}{\G(\Delta_\mcO-1)\G(D-\Delta_\mcO+\ell)}\\
\times\frac{\G(\hlf\lp D+\Delta_1-\Delta_2-\Delta_\mcO+\ell\rp)\G(\hlf\lp D-\Delta_1+\Delta_2-\Delta_\mcO+\ell\rp)}{\G(\hlf\lp\Delta_1-\Delta_2+\Delta_\mcO+\ell\rp)\G(\hlf\lp -\Delta_1+\Delta_2+\Delta_\mcO+\ell\rp)}\la_\mcO.
\end{multline}

\subsubsection{$\langle SV\wtmcO\rangle$}

Next, we consider symmetric traceless $\mcO(x_3)$, but in a three-point function with a scalar $\phi(x_1)$ and a vector $v_a(x_2)$. In this case, both OPE contribute:
\be
\label{eq:AlphaTilde}
\begin{split}
\al_\wtmcO= & \pi^{D/2}\frac{\G(\hlf (D+\Delta_\phi-\Delta_v-\Delta_\mcO+\ell+1))\G(\hlf (D-\Delta_\phi+\Delta_v-\Delta_\mcO+\ell-1))\G(\Delta_\mcO-\frac{D}{2})}{\G(\hlf (\Delta_\phi-\Delta_v+\Delta_\mcO+\ell+1))\G(\hlf (-\Delta_\phi+\Delta_v+\Delta_\mcO+\ell+1))\G(\Delta_\mcO)}\\
& \quad\times\frac{\G(\Delta_\mcO+\ell-1)}{\G(D-\Delta_\mcO+\ell)}\ls
\hlf\lp\lp\Delta_\mcO+\ell-1\rp\lp D-\Delta_\mcO-1\rp-\lp\Delta_\mcO-1\rp\lp\Delta_\phi-\Delta_v\rp\rp\al_\mcO\right.\\
& \qquad\left.-\lp\Delta_\mcO-\frac{D}{2}\rp\lp\Delta_\phi-\Delta_v+\Delta_\mcO+\ell-1\rp\beta_\mcO\rs,
\end{split}
\ee
\be
\label{eq:BetaTilde}
\begin{split}
\beta_\wtmcO= & \pi^{D/2}\frac{\G(\hlf (D+\Delta_\phi-\Delta_v-\Delta_\mcO+\ell-1))\G(\hlf (D-\Delta_\phi+\Delta_v-\Delta_\mcO+\ell-1))\G(\Delta_\mcO-\frac{D}{2})}{\G(\hlf (\Delta_\phi-\Delta_v+\Delta_\mcO+\ell+1))\G(\hlf (-\Delta_\phi+\Delta_v+\Delta_\mcO+\ell+1))\G(\Delta_\mcO)}\\
& \quad\times\frac{\G(\Delta_\mcO+\ell-1)}{\G(D-\Delta_\mcO+\ell)}\ls\frac{\ell}{2}\lp\Delta_\mcO-\frac{D}{2}\rp\lp\Delta_\phi-\Delta_v\rp\al_\mcO+\frac{1}{4}\lp\Delta_\phi-\Delta_v+\Delta_\mcO+\ell-1\rp\right.\\
& \qquad\left.\times\lp\lp\Delta_\mcO-1\rp\lp D-\Delta_\mcO+\ell-1\rp-\lp D-\Delta_\mcO-1\rp\lp\Delta_\phi-\Delta_v\rp\rp\beta_\mcO\rs.
\end{split}
\ee

As a nice check on this result, we can show that in the case that $v_a$ is conserved, so $\al_\mcO$ and $\beta_\mcO$ obey (\ref{eq:ConservedVec3pt}) for $\ell>0$ (the $\ell=0$ case has a subtlety but can also be shown to be consistent), then $\al_\wtmcO$ and $\beta_\wtmcO$ obey the corresponding equation with $\Delta_\wtmcO$.  For future reference we will write
\be
\label{eq:MDef}
\al_{\phi v\wtmcO}=M_\al^{\hph{\al}\al}\al_{\phi v\mcO}+M_\al^{\hph{\al}\beta}\beta_{\phi v\mcO},\qquad\beta_{\phi v\wtmcO}=M_\beta^{\hph{\beta}\al}\al_{\phi v\mcO}+M_\beta^{\hph{\beta}\beta}\beta_{\phi v\mcO},
\ee
where the constants $M_r^{\hph{r}s}$ can be read off from (\ref{eq:AlphaTilde}) and (\ref{eq:BetaTilde}).

\subsubsection{$\langle SV\wtmcA\rangle$}

Finally, we turn to the mixed symmetry operator $\mcA_{b_1b_2c_1\cdots c_k}(x_3)$ and its shadow
\begin{multline}
\wtmcA_{b_1b_2c_1\cdots c_k}(x_3)\\
=\wtPi^{(k)\,d_1d_2e_1\cdots e_k}_{b_1b_2c_1\cdots c_k}\int\frac{d^Dx_0}{\lp x_{03}^2\rp^{D-\Delta_\mcA}}m^{(03)\hph{d_1}f_1}_{\hph{(03)}d_1}m^{(03)\hph{d_2}f_2}_{\hph{(03)}d_2}m^{(03)\hph{e_1}g_1}_{\hph{(03)}e_1}\cdots m^{(03)\hph{e_k}g_k}_{\hph{(03)}e_k}\mcA_{f_1f_2g_1\cdots g_k}(x_0),
\end{multline}
which can appear in a three-point function with a scalar $\phi(x_1)$ and a vector $v_a(x_2)$.

Similar techniques to those employed above give the relation between the three-point coefficients,
\begin{multline}
\g_{\wtmcA}=\pi^{D/2}\frac{\G(\Delta_\mcA+k)\G(\Delta_\mcA-\frac{D}{2})\G(\hlf\lp D+\Delta_\phi-\Delta_v-\Delta_\mcA+k+1\rp)}{\G(\Delta_\mcA)\G(D-\Delta_\mcA+k+1)\G(\hlf\lp\Delta_\phi-\Delta_v+\Delta_\mcA+k+1\rp)}\\
\times\frac{\G(\hlf\lp D-\Delta_\phi+\Delta_v-\Delta_\mcA+k+1\rp)}{\G(\hlf\lp -\Delta_\phi+\Delta_v+\Delta_\mcA+k+1\rp)}\lp\Delta_\mcA-2\rp\g_\mcA.
\end{multline}

\subsection{Shadow projectors}

Given a primary operator $\mcO$, define a shadow projector
\be
P_\mcO=\mcN_\mcO\int d^Dx_0\left|\mcO_{a_1\cdots a_\ell}(x_0)\right\rangle\left\langle\wtmcO^{a_1\cdots a_\ell}(x_0)\right|.
\ee
This should be interpreted as an operator that gets inserted into a correlation function, separating it into two correlation functions with an integral.  When inserted into a given channel in a correlation function, it is designed to pick out the contribution of $\mcO$ and its descendants.  $\mcN_\mcO$ is a normalization constant that we fix by demanding
\be
 \left\langle\vp_1(x_1)\vp_2(x_2)\mcO_{a_1\cdots a_\ell}(x_3)\right\rangle=\left\langle\mcO_{a_1\cdots a_\ell}(x_3)P_\mcO\vp_1(x_1)\vp_2(x_2)\right\rangle
\ee
and so we need to take (see appendix \ref{app:mixingmatrices})
\be
\label{eq:NO}
\mcN_\mcO=\pi^{-D}\frac{\lp \Delta_\mcO+\ell-1\rp\lp D-\Delta_\mcO+\ell-1\rp\G(\Delta_\mcO-1)\G(D-\Delta_\mcO-1)}{\G(\Delta_\mcO-\frac{D}{2})\G(\frac{D}{2}-\Delta_\mcO)}.
\ee
Note that $\mcN_\mcO$ is independent of $\Delta_1$ and $\Delta_2$, as it should be.

Similarly, for the mixed symmetry case we can define a projector,
\be
P_\mcA=\mcN_\mcA\int d^Dx_0\left|\mcA_{a_1a_2b_1\cdots b_k}(x_0)\right\rangle\left\langle\wtmcA^{a_1a_2b_1\cdots b_k}\right|,
\ee
and $\mcN_\mcA$ can be computed to be
\be
\mcN_\mcA=\pi^{-D}\frac{\lp\Delta_\mcA+k\rp\lp D-\Delta_\mcA+k\rp\G(\Delta_\mcA)\G(D-\Delta_\mcA)}{\lp\Delta_\mcA-2\rp\lp D-\Delta_\mcA-2\rp\G(\Delta_\mcA-\frac{D}{2})\G(\frac{D}{2}-\Delta_\mcA)}.
\ee

\section{Conformal blocks}
\label{sec:ConformalBlocks}

We next turn to four-point functions.  These can be evaluated by first performing operator product expansions (OPEs) of the first two operators and the last two operators, and then evaluating the remaining two-point functions.  Consider first a general OPE.  Let's use notation where $\bar{a}$ represents a multi-index, transforming as some representation of $\SO(D)$.  Then the OPE of two arbitrary operators has the form
\be
\phi_{1\,\bar{a}}(x_1)\phi_{2\,\bar{b}}(x_2)=\sum_{\mathcal{U}}f_{12\mathcal{U}\,\bar{a}\bar{b}}^{\bar{c}}(x_{12})\mathcal{U}_{\bar{c}}(x_2),
\ee
where the sum is in principle over all local operators $\mathcal{U}_{\bar{c}}(x)$ in the theory, and the coefficients $f_{12\mathcal{U}\,\bar{a}\bar{b}}^{\bar{c}}$ are functions of $x_{12}$.  Actually, for fixed representations only a finite number of tensor structures are compatible with the symmetries, so we can write this as a sum over tensor structures labeled by $r$,
\be
f_{12\mathcal{U}\,\bar{a}\bar{b}}^{\bar{c}}(x_{12})=\sum_r\la_{12\mathcal{U}\,r}s^{r\,\bar{c}}_{\bar{a}\bar{b}}(x_{12}),
\ee
where the three-point tensor structures $s^{r\,\bar{c}}_{\bar{a}\bar{b}}(x_{12})$ are universal quantities which depend on the conformal representations (meaning both the $\SO(D)$ representations and the conformal weights) involved, but are otherwise independent of the theory or the particular operators.  That dependence is entirely contained in the constants $\la_{12\mathcal{U}\,r}$.  Finally, there is one further simplification, which is that when $\mathcal{U}$ is a descendent of a primary operator $\mcO$ (and thus corresponds to some differential operator acting on $\mcO$), then its coefficients in the $\phi_1\times\phi_2$ OPE is determined linearly in terms of the coefficients of $\mcO$ in the OPE.  Thus the OPE can in fact be written as a sum over primary operators $\mcO$,
\be
\phi_{1\,\bar{a}}(x_1)\phi_{2\,\bar{b}}(x_2)=\sum_{\mcO}\sum_r\la_{12\mcO\,r}C^{r\,\bar{c}}_{\bar{a}\bar{b}}(x_{12},\p_2)\mcO_{\bar{c}}(x_2).
\ee
Again the differential operators $C^{r\,\bar{c}}_{\bar{a}\bar{b}}(x_{12},\p_2)$ are universal in the same sense as above.

Now inserting this form of the OPE into the four-point function, we can write
\be
\label{eq:General4pt}
\left\langle\phi_{1\,\bar{a}}(x_1)\phi_{2\,\bar{b}}(x_2)\phi_{3\,\bar{c}}(x_3)\phi_{4\,\bar{d}}(x_4)\right\rangle=\sum_\mcO\sum_{r,s}\la_{12\mcO\,r}\la_{34\mcO\,s}W^{rs}_{\bar{a}\bar{b}\bar{c}\bar{d}}(x_1,x_2,x_3,x_4).
\ee
The functions $W^{rs}_{\bar{a}\bar{b}\bar{c}\bar{d}}$ depend only on the $\SO(D)$ representations and conformal weights of the $\phi_i$ and of $\mcO$.  These functions are often called conformal partial waves (though this nomenclature is not universal).  Conformal invariance can actually be used to further restrict the form of the $W$'s, so that we can write
\be
\label{eq:WaveBlockRelation}
W^{rs}_{\bar{a}\bar{b}\bar{c}\bar{d}}(x_1,x_2,x_3,x_4)=X_{\Delta_i}
\sum_p g^{rs}_p(u,v)t^p_{\bar{a}\bar{b}\bar{c}\bar{d}}(x_i).
\ee
Here the sum $p$ runs over allowed tensor structures, and the four-point tensor structures $t^p_{\bar{a}\bar{b}\bar{c}\bar{d}}(x_i)$ depend only on the $\SO(D)$ representations of the external operators, not the conformal dimensions, while the functions $g^{rs}_p(u,v)$ depend on the full conformal representations (i.e.\ both $\SO(D)$ representations and conformal weights), but are themselves scalar functions of the cross-ratios $u$ and $v$.  These $g^{rs}_p(u,v)$ are called conformal blocks, and our task in the rest of this section is to compute them for the scalar and vector four-point functions of interest.

\subsection{General discussion}

Our primary purpose in this paper involves specific examples of four-point functions, but let us first have a very brief general discussion.  Roughly, the idea is that by inserting the projector $P_\mcO$ into the correlator (\ref{eq:General4pt}), we should pick out only the contribution from the primary $\mcO$ and its descendants.  This is not quite correct, as explained in~\cite{Simmons-Duffin2014} and elsewhere; rather we must insert the projector and then pick out only the terms of the result which transform with a phase $e^{2\pi i(\Delta_\mcO-\Delta_1-\Delta_2)}$ as we send $x_{12}^2\rr e^{4\pi i}x_{12}^2$.  The remaining terms will transform with a phase $e^{-2\pi i(\Delta_\mcO+\Delta_1+\Delta_2)}$ under this rotation and these terms should be thrown away.  This procedure is called monodromy projection.  In practice, we can write the result before the monodromy projection as a certain double integral over Feynman-Schwinger parameters, and then the monodromy projection can be implemented simply as a modification of the integration region, along with some insertions of signs in the integrand.  Once we have successfully picked out the contributions from $\mcO$ and its descendants, we can read off $g^{rs}_p(u,v)$ from the terms proportional to $\la_{12\mcO\,r}\la_{34\mcO\,s}t^p_{\bar{a}\bar{b}\bar{c}\bar{d}}$.

If we write the general four-point function as
\be
\left\langle\phi_{1\,\bar{a}}(x_1)\phi_{2\,\bar{b}}(x_2)\phi_{3\,\bar{c}}(x_3)\phi_{4\,\bar{d}}(x_4)\right\rangle=X_{\Delta_i}
\sum_pq_p(u,v)t^p_{\bar{a}\bar{b}\bar{c}\bar{d}}(x_i),
\ee
then we have
\be
q_p(u,v)=\sum_\mcO\sum_{r,s}\la_{12\mcO\,r}\la_{34\mcO\,s}g^{rs}_p(u,v).
\ee

\subsection{Scalars and vectors}

Let's understand what we should then be computing for our examples of interest.  First we will review the case of four scalars.  In this case the exchanged primary must be traceless symmetric, with its representation labeled by a spin $\ell$ and dimension $\Delta_\mcO$.  There is a unique three-point tensor structure for each $\ell$,
\be
s_{a_1\cdots a_\ell}=\Pi^{(\ell)\,b_1\cdots b_\ell}_{a_1\cdots a_\ell}k^{(012)}_{b_1}\cdots k^{(012)}_{b_\ell},
\ee
and a unique four-point tensor structure $t=1$.  In other words, the correlator should take the form
\begin{multline}
\left\langle\phi_1(x_1)\phi_2(x_2)\phi_3(x_3)\phi_4(x_4)\right\rangle=\sum_\mcO\la_{12\mcO}\la_{34\mcO}W(x_i;\Delta_i;\ell,\Delta_\mcO)\\
=X_{\Delta_i}
\sum_\mcO\la_{12\mcO}\la_{34\mcO}\,g(u,v;\Delta_i;\ell,\Delta_\mcO).
\end{multline}
Here we have shown on which parameters the conformal partial waves $W$ or conformal blocks $g$ can depend; often we will not indicate this explicitly.  In terms of the function $q(u,v)$ introduced in (\ref{eq:SSSSstructs}) we have
\be
q(u,v)=\sum_\mcO\la_{12\mcO}\la_{34\mcO}\,g(u,v).
\ee

Next suppose we have three scalars and a vector $v$ in the second position.  The only possibility for exchanged operators are again traceless symmetric $\mcO$ of spin $\ell$.  There is only one tensor structure which can appear in the $\langle\phi_3\phi_4\mcO\rangle$ three-point function, with coefficient $\la_{34\mcO}$, but there are two possible tensor structures in the $\langle\phi_1(x_1)v(x_2)\mcO(x_0)\rangle$ three-point function,
\be
s^\al_{ab_1\cdots b_\ell}=\Pi^{(\ell)\,c_1\cdots c_\ell}_{b_1\cdots b_\ell}k^{(210)}_ak^{(012)}_{c_1}\cdots k^{(012)}_{c_\ell},\quad s^\beta_{ab_1\cdots b_\ell}=\Pi^{(\ell)\,c_1\cdots c_\ell}_{b_1\cdots b_\ell}m^{(20)}_{ac_1}k^{(012)}_{c_1}\cdots k^{(012)}_{c_\ell},
\ee
whose coefficients we will label $\al_{12\mcO}$ and $\beta_{12\mcO}$.  There are also two four-point tensor structures\footnote{From the point of view of this correlator these might not be the first choice of tensor structures; we might prefer $k^{(213)}$ and $k^{(214)}$ to be more symmetric between $x_3$ and $x_4$.  However we will also being using these correlators and tensor structures as intermediate expressions in computing the SVSV conformal blocks, where the symmetry we will want to maintain is between $x_2$ and $x_4$.} $t^1_a=k^{(214)}_a$ and $t^2_a=k^{(234)}_a$, with coefficient functions $q_1(u,v)$ and $q_2(u,v)$ respectively, and these are related to the conformal blocks $g^{\al\la}_1$, $g^{\al\la}_2$, $g^{\beta\la}_1$, and $g^{\beta\la}_2$ by
\be
q_i(u,v)=\sum_\mcO\la_{34\mcO}\lp\al_{12\mcO}g^{\al\la}_i(u,v)+\beta_{12\mcO}g^{\beta\la}_i(u,v)\rp,
\ee
for $i=1,2$.  Each of the four conformal block functions will depend on the conformal weights $\Delta_1$, $\Delta_2=\Delta_v$, $\Delta_3$, $\Delta_4$, and $\Delta_\mcO$, as well as the spin $\ell$ of $\mcO$.

Having the vector in the fourth position is essentially the same upon interchanging $(12)\leftrightarrow(34)$, with $t^1_a=k^{(412)}_a$, $t^2_a=k^{(432)}_a$, and
\be
q'_i(u,v)=\sum_\mcO\la_{12\mcO}\lp\al_{34\mcO}g^{\la\al}_i(u,v)+\beta_{34\mcO}g^{\la\beta}_i(u,v)\rp.
\ee

Finally, for the case of two scalars and two vectors we found in (\ref{eq:SVSVstructs}) five four-point tensor structures,
\be
t^0_{ab}=m^{(24)}_{ab},\quad t^{11}_{ab}=k^{(214)}_ak^{(412)}_b,\ t^{12}_{ab}=k^{(214)}_ak^{(432)}_b,\ t^{21}_{ab}=k^{(234)}_ak^{(412)}_b,\ t^{22}_{ab}=k^{(234)}_ak^{(432)}_b,
\ee
with associated coefficient functions $q_0$ and $q_{ij}$.  In this case the exchanged operator can either be traceless symmetric $\mcO$ of spin $\ell$, or it can be a mixed-symmetry operator $\mcA$ whose representation is labeled by $k$.  In the former case, each of the three point function has two tensor structures $s^\al$ and $s^\beta$, while in the latter case there is a unique three-point tensor structure
\be
s^\g_{ab_1b_2c_1\cdots c_k}=\wtPi^{(k)\,d_1d_2e_1\cdots e_k}_{b_1b_2c_1\cdots c_k}m^{(20)}_{ad_1}k^{(012)}_{d_2}k^{(012)}_{e_1}\cdots k^{(012)}_{e_k}.
\ee
Hence, for generic (not necessarily identical) scalars and vectors, we have
\be
q_0=\sum_\mcO\lp\al_{12\mcO}\al_{34\mcO}g^{\al\al}_0+\al_{12\mcO}\beta_{34\mcO}g^{\al\beta}_0+\beta_{12\mcO}\al_{34\mcO}g^{\beta\al}_0+\beta_{12\mcO}\beta_{34\mcO}g^{\beta\beta}_0\rp+\sum_\mcA\g_{12\mcA}\g_{34\mcA}g^{\g\g}_0,
\ee
\be
q_{ij}=\sum_\mcO\lp\al_{12\mcO}\al_{34\mcO}g^{\al\al}_{ij}+\al_{12\mcO}\beta_{34\mcO}g^{\al\beta}_{ij}+\beta_{12\mcO}\al_{34\mcO}g^{\beta\al}_{ij}+\beta_{12\mcO}\beta_{34\mcO}g^{\beta\beta}_{ij}\rp+\sum_\mcA\g_{12\mcA}\g_{34\mcA}g^{\g\g}_{ij}.
\ee
Altogether there are twenty-five conformal block functions.  $2\times 2\times 5=20$ of them are associated with symmetric traceless exchange and will depend on the spin $\ell$ as well as the conformal weights $\Delta_i$ and $\Delta_\mcO$, while the other five are associated to mixed symmetry exchange, and will depend on $\Delta_i$, $\Delta_\mcA$ and $k$, which labels the mixed symmetry representation.

\subsection{Exchange symmetries}
\label{subsec:ExchangeSymmetries}

As in the classification of tensor structures, the structure of conformal blocks can simplify significantly when some of the operators are identical, so that we have extra symmetry from exchanging those operators.  Note however, that since the conformal block decomposition picks out a particular exchange channel, not all exchanges will give us constraints on individual conformal block functions.  An exchange that results in a different exchange channel is called a crossing symmetry, and will constrain only the full sum of conformal blocks, not the individual conformal blocks themselves.  Crossing symmetry is the subject of the next section, when we will set up the bootstrap.  In the current subsection, however, we will consider the exchanges which don't mix channels, and so can constrain the blocks themselves.  These can involve exchange of operator $1$ with operator $2$, of operator $3$ with operator $4$, or exchanging the pair $(12)$ with the pair $(34)$.

For example, consider the case of four scalars, with its unique conformal block function $g(u,v;\Delta_1,\Delta_2,\Delta_3,\Delta_4;\ell,\Delta_\mcO)$, where for this section we will show explicit dependence on parameters.  The four-point function will be invariant if we simultaneously exchange $x_1$ with $x_2$ and $\Delta_1$ with $\Delta_2$.  This leads to a constraint on the conformal blocks,
\be
g(u,v;\Delta_1,\Delta_2,\Delta_3,\Delta_4;\ell,\Delta_\mcO)=v^{-\hlf(\Delta_3-\Delta_4)}g(u/v,1/v;\Delta_2,\Delta_1,\Delta_3,\Delta_4;\ell,\Delta_\mcO).
\ee
Similarly, for $3\leftrightarrow 4$ exchange,
\be
g(u,v;\Delta_1,\Delta_2,\Delta_3,\Delta_4;\ell,\Delta_\mcO)=v^{\hlf(\Delta_1-\Delta_2)}g(u/v,1/v;\Delta_1,\Delta_2,\Delta_4,\Delta_3;\ell,\Delta_\mcO),
\ee
and for $(12)\leftrightarrow(34)$ exchange,
\be
g(u,v;\Delta_1,\Delta_2,\Delta_3,\Delta_4;\ell,\Delta_\mcO)=v^{\hlf(\Delta_1-\Delta_2-\Delta_3+\Delta_4)}g(u,v;\Delta_3,\Delta_4,\Delta_1,\Delta_2;\ell,\Delta_\mcO).
\ee
These relations are most useful when some of the scalars are really identical.  For example, if all four scalars are identical with weight $\Delta$, then we have
\be
g(u,v;\Delta;\ell,\Delta_\mcO)=g(u/v,1/v;\Delta;\ell,\Delta_\mcO).
\ee

In the case of three scalars and a vector, we have a couple of options.  If the vector is in the second position, then we have the $3\leftrightarrow 4$ exchange of scalars, which tells us that
\be
g_1^{r\la}(u,v;\Delta_1,\Delta_v,\Delta_3,\Delta_4;\ell,\Delta_\mcO)=v^{\hlf\lp\Delta_1-\Delta_v+1\rp}g_1^{r\la}(u/v,1/v;\Delta_1,\Delta_v,\Delta_4,\Delta_3;\ell,\Delta_\mcO),
\ee
and
\begin{multline}
g_2^{r\la}(u,v;\Delta_1,\Delta_v,\Delta_3,\Delta_4;\ell,\Delta_\mcO)\\
=-v^{\hlf\lp\Delta_1-\Delta_v\rp}\lp u^\hlf g_1^{r\la}(u/v,1/v;\Delta_1,\Delta_v,\Delta_4,\Delta_3;\ell,\Delta_\mcO)+g_2^{r\la}(u/v,1/v;\Delta_1,\Delta_v,\Delta_4,\Delta_3)\rp,
\end{multline}
where $r$ is either $\al$ or $\beta$.  Note that in deriving these relations, we needed to transform the $t^p_a$ under this exchange and then re\"express the result in terms of our basis $t^p_a$ again.  Since our chosen basis $t^1_a=k^{(214)}_a$, $t^2_a=k^{(234)}$ does not behave particularly nicely (rather we chose it to make later computations with two scalars and two vectors slightly nicer), the resulting expressions are slightly messier than they would be in a basis like $t^{\prime\,1}_a=k^{(213)}_a$, $t^{\prime\,2}_a=k^{(214)}_a$ which simply gets exchanged under $3\leftrightarrow 4$.  Performing a $(12)\leftrightarrow(34)$ exchange relates the SVSS conformal blocks to the SSSV conformal blocks,
\be
g_1^{r\la}(u,v;\Delta_1,\Delta_v,\Delta_3,\Delta_4;\ell,\Delta_\mcO)=v^{\hlf\lp\Delta_1-\Delta_v-\Delta_3+\Delta_4\rp} g_2^{\la r}(u,v;\Delta_3,\Delta_4,\Delta_1,\Delta_v;\ell,\Delta_\mcO),
\ee
and
\be
g_2^{r\la}(u,v;\Delta_1,\Delta_v,\Delta_3,\Delta_4;\ell,\Delta_\mcO)=v^{\hlf\lp\Delta_1-\Delta_v-\Delta_3+\Delta_4\rp} g_1^{\la r}(u,v;\Delta_3,\Delta_4,\Delta_1,\Delta_v;\ell,\Delta_\mcO).
\ee


Finally, for the SVSV case, the only useful exchange is $(12)\leftrightarrow(34)$, which tells us
\be
g_{22}^{rs}(u,v;\Delta_1,\Delta_2,\Delta_3,\Delta_4;\ell,\Delta_\mcO)=v^{\hlf(\Delta_1-\Delta_2-\Delta_3+\Delta_4)}g_{11}^{sr}(u,v;\Delta_3,\Delta_4,\Delta_1,\Delta_2;\ell,\Delta_\mcO),
\ee
and
\be
g_p^{rs}(u,v;\Delta_1,\Delta_2,\Delta_3,\Delta_4;\ell,\Delta_\mcO)=v^{\hlf(\Delta_1-\Delta_2-\Delta_3+\Delta_4)}g_p^{sr}(u,v;\Delta_3,\Delta_4,\Delta_1,\Delta_2;\ell,\Delta_\mcO),
\ee
for $r$ and $s$ being $\al$ or $\beta$, and for $p$ being $0$, $12$, or $21$.  Similarly
\be
g_{22}^{\g\g}(u,v;\Delta_1,\Delta_2,\Delta_3,\Delta_4;k,\Delta_\mcA)=v^{\hlf(\Delta_1-\Delta_2-\Delta_3+\Delta_4)}g_{11}^{\g\g}(u,v;\Delta_3,\Delta_4,\Delta_1,\Delta_2;k,\Delta_\mcA),
\ee
\be
g_p^{\g\g}(u,v;\Delta_1,\Delta_2,\Delta_3,\Delta_4;k,\Delta_\mcA)=v^{\hlf(\Delta_1-\Delta_2-\Delta_3+\Delta_4)}g_p^{\g\g}(u,v;\Delta_3,\Delta_4,\Delta_1,\Delta_2;k,\Delta_\mcA).
\ee
In particular, if we have identical scalars and identical vectors, then the $g_0$, $g_{12}$, and $g_{21}$ are only constrained to be symmetric in their upper indices (i.e.\ $g^{\al\beta}_p=g^{\beta\al}_p$), while the $g_{22}$ functions are determined by the $g_{11}$'s,
\be
g_{22}^{rs}(u,v)=g_{11}^{sr}(u,v).
\ee

\subsection{Computing the blocks}

At the risk of cluttering notation, we will add a hat to the conformal block functions to denote the result obtained from insertion of the shadow projector,
\be
\left\langle\phi_{1\,\bar{a}}(x_1)\phi_{2\,\bar{b}}(x_2)P_\mcO\phi_{3\,\bar{c}}(x_3)\phi_{4\,\bar{d}}(x_4)\right\rangle=X_{\Delta_i}
\la_{12\mcO\,r}\la_{34\mcO\,s}\,\whg^{rs}_p(u,v)t^p_{\bar{a}\bar{b}\bar{c}\bar{d}}.
\ee
The actual conformal blocks $g^{rs}_p(u,v)$ themselves are then obtained from the $\whg^{rs}_p(u,v)$ by a monodromy projection, which now picks out the terms in $\whg^{rs}_p(u,v)$ which transform with a phase $e^{2\pi i\Delta_\mcO}$ as $u\rr e^{4\pi i}u$, and throws away the terms which transform as $e^{-2\pi i\Delta_\mcO}$.

We will call the process of implementing the monodromy projection, going from $\whg^{rs}_p$ to $g^{rs}_p$ (i.e.\ removing the hat), {\it{doffing}}.

\subsubsection{$\langle SSSS\rangle$}
\label{subsubsec:ComputingSSSS}

We'll start by reviewing the computation of the conformal blocks for four scalar operators.  Here, on insertion of the shadow projector we have
\be
\begin{split}
& \left\langle\phi_1(x_1)\phi_2(x_2)P_\mcO\phi_3(x_3)\phi_4(x_4)\right\rangle\\
& \quad =\mcN_\mcO\int d^Dx_0\left\langle\phi_1(x_1)\phi_2(x_2)\mcO_{a_1\cdots a_\ell}(x_0)\right\rangle\left\langle\wtmcO^{a_1\cdots a_\ell}(x_0)\phi_3(x_3)\phi_4(x_4)\right\rangle\\
& \quad =\mcN_\mcO\Pi^{(\ell)\,a_1\cdots a_\ell}_{b_1\cdots b_\ell}\int d^Dx_0\ls\la_{12\mcO}\lp x_{01}^2\rp^{\hlf\lp -\Delta_1+\Delta_2-\Delta_\mcO\rp}\lp x_{02}^2\rp^{\hlf\lp\Delta_1-\Delta_2-\Delta_\mcO\rp}\lp x_{12}^2\rp^{\hlf\lp -\Delta_1-\Delta_2+\Delta_\mcO\rp}\right.\\
& \qquad\left.\times k^{(012)}_{a_1}\cdots k^{(012)}_{a_\ell}\vphantom{\lp x_{01}^2\rp^{\hlf\lp -\Delta_1+\Delta_2-\Delta_\mcO\rp}}\rs\ls\la_{34\wtmcO}\lp x_{03}^2\rp^{\hlf\lp -\Delta_3+\Delta_4+\Delta_\mcO-D\rp}\lp x_{04}^2\rp^{\hlf\lp\Delta_3-\Delta_4+\Delta_\mcO-D\rp}\right.\\
& \qquad\left.\times\lp x_{34}^2\rp^{\hlf\lp D-\Delta_3-\Delta_4-\Delta_\mcO\rp}k^{(034)\,b_1}\cdots k^{(034)\,b_\ell}\rs\\
& \quad =
X_{\Delta_i}\ls\mcN_\mcO\la_{12\mcO}\la_{34\wtmcO}\lp x_{12}^2\rp^{\hlf\Delta_\mcO}\lp x_{13}^2\rp^{\hlf\lp\Delta_3-\Delta_4\rp}\lp x_{14}^2\rp^{\hlf\lp\Delta_1-\Delta_2-\Delta_3+\Delta_4\rp}\lp x_{24}^2\rp^{\hlf\lp -\Delta_1+\Delta_2\rp}\right.\\
& \qquad\left.\times\lp x_{34}^2\rp^{\hlf\lp D-\Delta_\mcO\rp}\int d^Dx_0\lp x_{01}^2\rp^{\hlf\lp -\Delta_1+\Delta_2-\Delta_\mcO\rp}\lp x_{02}^2\rp^{\hlf\lp\Delta_1-\Delta_2-\Delta_\mcO\rp}\lp x_{03}^2\rp^{\hlf\lp -\Delta_3+\Delta_4+\Delta_\mcO-D\rp}\right.\\
& \qquad\left.\times\lp x_{04}^2\rp^{\hlf\lp\Delta_3-\Delta_4+\Delta_\mcO-D\rp}\lp k^{(012)}_{a_1}\cdots k^{(012)}_{a_\ell}\Pi^{(\ell)\,a_1\cdots a_\ell}_{b_1\cdots b_\ell}k^{(034)\,b_1}\cdots k^{(034)\,b_\ell}\rp\rs.
\end{split}
\ee
from which we can identify $\la_{12\mcO}\la_{34\mcO}\whg(u,v)$ with the quantity in square brackets.

As shown in Appendix \ref{subapp:PiO}, we can write
\be
\label{eq:LambdaAlphaStruct}
k^{(012)}_{a_1}\cdots k^{(012)}_{a_\ell}\Pi^{(\ell)\,a_1\cdots a_\ell}_{b_1\cdots b_\ell}k^{(034)\,b_1}\cdots k^{(034)\,b_\ell}=p_{D,\ell}(t),
\ee
where $p_{D,\ell}(t)$ is a polynomial of degree $\ell$ whose properties are explained in the appendix, and
\begin{multline}
\label{eq:tExpansion}
t=k^{(012)}\cdot k^{(034)}\\
=\hlf\lp x_{01}^2x_{02}^2x_{03}^2x_{04}^2x_{12}^2x_{34}^2\rp^{-1/2}\lp -x_{01}^2x_{03}^2x_{24}^2+x_{01}^2x_{04}^2x_{23}^2+x_{02}^2x_{03}^2x_{14}^2-x_{02}^2x_{04}^2x_{13}^2\rp.
\end{multline}

Let us now define integrals
\be
\label{eq:IlDef}
I_{\al,\beta,\g,\d}^{(\ell)}=\int\frac{d^Dx_0\,p_{D,\ell}(t)}{\lp x_{01}^2\rp^\al\lp x_{02}^2\rp^\beta\lp x_{03}^2\rp^\g\lp x_{04}^2\rp^\d}.
\ee
For $\ell=0$ this integral is evaluated in (\ref{eq:Basic4ptScalarIntegral}).  

With this definition and the expressions (\ref{eq:NO}) and (\ref{eq:lambdaOtilde}), we have
\be
\label{eq:SSSSgIRelation}
\begin{split}
& \whg(u,v;\Delta_i;\ell,\Delta_\mcO)\\
& \quad =\pi^{-D/2}\frac{\G(\hlf\lp D+\Delta_3-\Delta_4-\Delta_\mcO+\ell\rp)\G(\hlf\lp D-\Delta_3+\Delta_4-\Delta_\mcO+\ell\rp)\G(\Delta_\mcO+\ell)}{\G(\hlf\lp\Delta_3-\Delta_4+\Delta_\mcO+\ell\rp)\G(\hlf\lp -\Delta_3+\Delta_4+\Delta_\mcO+\ell\rp)\G(D-\Delta_\mcO+\ell-1)}\\
& \qquad\times\frac{\G(D-\Delta_\mcO-1)}{\G(\frac{D}{2}-\Delta_\mcO)}\lp x_{12}^2\rp^{\hlf\Delta_\mcO}\lp x_{13}^2\rp^{\hlf\lp\Delta_3-\Delta_4\rp}\lp x_{14}^2\rp^{\hlf\lp\Delta_1-\Delta_2-\Delta_3+\Delta_4\rp}\lp x_{24}^2\rp^{\hlf\lp -\Delta_1+\Delta_2\rp}\\
& \qquad\times\lp x_{34}^2\rp^{\hlf\lp D-\Delta_\mcO\rp}I^{(\ell)}_{\hlf\lp\Delta_1-\Delta_2+\Delta_\mcO\rp,\hlf\lp -\Delta_1+\Delta_2+\Delta_\mcO\rp,\hlf\lp D+\Delta_3-\Delta_4-\Delta_\mcO\rp,\hlf\lp D-\Delta_3+\Delta_4-\Delta_\mcO\rp}.
\end{split}
\ee
Note that 
the prefactor $(x_{12}^2)^{\Delta_\mcO/2}$ already has the desired behavior under the monodromy projection, so we will want to pick out the terms from the integral which are invariant under the monodromy.

Note that if we expand the polynomial $p_{D,\ell}(t)$ using the explicit formulae in Appendix \ref{subapp:PiO} then the integral is simply a sum of terms of a form computed in Appendix \ref{subapp:4ptIntegrals}.  For example, in the case $\ell=0$, then $p_{D,0}(t)=1$, and we have (restoring the explicit parameter dependence)
\begin{multline}
\label{eq:ScalarSSSSCB}
\whg(u,v;\Delta_i;0,\Delta_\mcO)\\
=\frac{\G(\Delta_\mcO)\G(\hlf\lp D-\Delta_3+\Delta_4-\Delta_\mcO\rp)}{\G(\frac{D}{2}-\Delta_\mcO)\G(\hlf\lp\Delta_1-\Delta_2+\Delta_\mcO\rp)\G(\hlf\lp -\Delta_1+\Delta_2+\Delta_\mcO\rp)\G(\hlf\lp -\Delta_3+\Delta_4+\Delta_\mcO\rp)}u^{\hlf\Delta_\mcO}\\
\times v^{\hlf\lp -\Delta_3+\Delta_4-\Delta_\mcO\rp}\whf_{\hlf\lp\Delta_1-\Delta_2+\Delta_\mcO\rp,\hlf\lp -\Delta_1+\Delta_2+\Delta_\mcO\rp,\hlf\lp D+\Delta_3-\Delta_4-\Delta_\mcO\rp,\hlf\lp D-\Delta_3+\Delta_4-\Delta_\mcO\rp}(uv^{-1},v^{-1}),
\end{multline}
where $\whf$ is defined in (\ref{eq:whfDef}).  Since the $u^{\Delta_\mcO/2}$ factor already behaves correctly under the monodromy projection, then to obtain the conformal block $g(u,v)$ we must restrict to the monodromy invariant piece of $\whf$, and this is given simply by a function $f$ defined in (\ref{eq:fDef}).  Then $g(u,v;\Delta_i;0,\Delta_\mcO)$ is given by doffing the expression (\ref{eq:ScalarSSSSCB}), replacing $\whf$ by $f$.  Note also that this formula shows explicity that $g(u,v;\Delta_i;\ell,\Delta_\mcO)$ doesn't depend on all four of the $\Delta_i$ individually, but only on the differences $\Delta_1-\Delta_2$ and $\Delta_3-\Delta_4$.  Because of this we can adopt some condensed notation that will be useful below, defining functions that are related to the standard blocks by shifting these two differences by integer amounts $P$ and $Q$,
\be
\label{eq:SCondensed}
g_{\ell;P,Q}(u,v)=g(u,v;\Delta_1+P,\Delta_2,\Delta_3+Q,\Delta_4;\ell,\Delta_\mcO).
\ee
In this notation (which can also be used for $\whg$) the dependence on the $\Delta_i$ and $\Delta_\mcO$ is left implicit.

In even dimensions\footnote{In arbitrary dimensions there exists a closed form for the $\ell=0$ conformal block in terms of Appel functions~\cite{Dolan2001}, but for even dimensions the result can be expressed using the much more familiar $\vphantom{F_1}_2F_1$ hypergeometric functions.} we can evaluate the integrals in $f$ explicitly, with the result
\be
\begin{split}
g_{0;0,0}(u,v) & =\frac{\G(\Delta_\mcO)\G(\hlf\lp -\Delta_1+\Delta_2+\Delta_\mcO-D+2\rp)\G(\hlf\lp\Delta_3-\Delta_4+\Delta_\mcO-D+2\rp)}{\G(\Delta_\mcO-\frac{D}{2}+1)\G(\hlf\lp -\Delta_1+\Delta_2+\Delta_\mcO\rp)\G(\hlf\lp\Delta_3-\Delta_4+\Delta_\mcO\rp)}\\
& \qquad\times\lp x\bar{x}\rp^{\hlf\Delta_\mcO}\lp\frac{1}{x-\bar{x}}\lp x\p_x-\bar{x}\p_{\bar{x}}\rp\rp^{\frac{D}{2}-1}\\
& \quad\cdot\ls\vphantom{F_1}_2F_1(\frac{-\Delta_1+\Delta_2+\Delta_\mcO-D+2}{2},\frac{\Delta_3-\Delta_4+\Delta_\mcO-D+2}{2},\Delta_\mcO-\frac{D}{2}+1;x)\right.\\
& \quad\left.\times\vphantom{F_1}_2F_1(\frac{-\Delta_1+\Delta_2+\Delta_\mcO-D+2}{2},\frac{\Delta_3-\Delta_4+\Delta_\mcO-D+2}{2},\Delta_\mcO-\frac{D}{2}+1;\bar{x})\rs,
\end{split}
\ee
where the variables $x$ and $\bar{x}$ are related to $u$ and $v$ via
\be
u=x\bar{x},\qquad v=\lp 1-x\rp\lp 1-\bar{x}\rp.
\ee

What about $\ell>0$?  As indicated, for any fixed small $\ell$ we can of course expand $p_{D,\ell}(t)$ into monomials and proceed as above.  But in fact we can be a bit more clever than that and exploit the recursion relations (\ref{eq:plRecursion}) to expand the numerator of the integrand in (\ref{eq:IlDef}).  In the recursion relation we also need to expand $t$ according to (\ref{eq:tExpansion}), and reabsorb the powers of $(x_{0i}^2)$ as shifts of the external operator dimensions.  Finally, passing to the monodromy-projected answer, the result is~\cite{Dolan2001}
\be
\begin{split}
& g_{\ell;0,0}(u,v)\\
& \quad =\frac{\Delta_\mcO+\ell-1}{D-\Delta_\mcO+\ell-2}\ls\hlf\frac{D+\Delta_3-\Delta_4-\Delta_\mcO+\ell-2}{\Delta_3-\Delta_4+\Delta_\mcO+\ell-2}u^{-1/2}\lp g_{\ell-1;1,-1}(u,v)-g_{\ell-1;-1,-1}(u,v)\rp\right.\\
& \qquad\left. +\hlf\frac{D-\Delta_3+\Delta_4-\Delta_\mcO+\ell-2}{-\Delta_3+\Delta_4+\Delta_\mcO+\ell-2}u^{-1/2}\lp vg_{\ell-1;-1,1}(u,v)-g_{\ell-1;1,1}(u,v)\rp\right.\\
& \qquad\left. -\frac{\lp\Delta_\mcO+\ell-2\rp\lp D+\Delta_3-\Delta_4-\Delta_\mcO+\ell-2\rp\lp D-\Delta_3+\Delta_4-\Delta_\mcO+\ell-2\rp}{\lp D-\Delta_\mcO+\ell-3\rp\lp\Delta_3-\Delta_4+\Delta_\mcO+\ell-2\rp\lp -\Delta_3+\Delta_4+\Delta_\mcO+\ell-2\rp}\right.\\
& \qquad\quad\left.\times\frac{\lp\ell-1\rp\lp D+\ell-4\rp}{\lp D+2\ell-4\rp\lp D+2\ell-6\rp}g_{\ell-2;0,0}(u,v)\rs
\end{split}
\ee
This recursion holds in any dimension.  In $D=2$ the recursion can actually be solved explicitly to get a closed form expression for $g(u,v;\Delta_i;\ell,\Delta_\mcO)$ in terms of elementary hypergeometric functions, and in higher even dimensions solutions can also be constructed (by using a relation between the blocks in $D+2$ dimensions and those in $D$ dimensions).  For example, in $D=4$,
\be
g(u,v;\Delta_i;\ell,\Delta_\mcO)=\lp -\hlf\rp^\ell\frac{x\bar{x}}{x-\bar{x}}\ls k_{\Delta_\mcO+\ell}(x)k_{\Delta_\mcO-\ell-2}(\bar{x})-k_{\Delta_\mcO-\ell-2}(x)k_{\Delta_\mcO+\ell}(\bar{x})\rs,
\ee
where
\be
\label{eq:kbetaDef}
k_\beta(x)=x^{\beta/2}\vphantom{F_1}_2F_1(\frac{\beta-\Delta_1+\Delta_2}{2},\frac{\beta+\Delta_3-\Delta_4}{2},\beta;x).
\ee
This is a good moment to make a point about normalizations.  The Casimir differential equation implies that in the limit $\bar{x},x\rightarrow 0$ the conformal block should behave approximately as $c_\ell x^{\hlf(\Delta_\mcO+\ell)}\bar{x}^{\hlf(\Delta_\mcO-\ell)}$ for some constants $c_\ell$.  Our blocks are defined according to (\ref{eq:General4pt}) and (\ref{eq:WaveBlockRelation}), and this turns out to imply $c_\ell=(-1/2)^\ell$.  Some authors prefer different normalizations, say with $c_\ell=1$.  It is always easy to go back and forth between conventions, as long as one is aware of them.

An alternative approach is to expand the polynomials $p_{D,\ell}(t)$ in the integrals $I^{(\ell)}$ to obtain an expression for conformal blocks with $\ell>0$ as a sum of $\ell=0$ blocks.  This result (with or without hats) is
\begin{multline}
\whg_{\ell;0,0}(u,v)=2^{-\ell}\frac{\G(\hlf\lp D+\Delta_3-\Delta_4-\Delta_\mcO+\ell\rp)\G(\hlf\lp D-\Delta_3+\Delta_4-\Delta_\mcO+\ell\rp)}{\G(\hlf\lp\Delta_3-\Delta_4+\Delta_\mcO+\ell\rp)\G(\hlf\lp -\Delta_3+\Delta_4+\Delta_\mcO+\ell\rp)}\frac{\G(\Delta_\mcO+\ell)}{\G(\Delta_\mcO)}\\
\times\frac{\G(D-\Delta_\mcO-1)}{\G(D-\Delta_\mcO+\ell-1)}\sum_{i=0}^{\lfloor\ell/2\rfloor}\sum_{A=0}^{\ell-2i}\sum_{B=0}^{\ell-2i-A}\sum_{C=0}^{\ell-2i-A-B}\lp -1\rp^{\ell+i+B+C}\frac{\ell!}{i!A!B!C!\lp \ell-2i-A-B-C\rp!}\\
\times\frac{\G(\hlf\lp\Delta_3-\Delta_4+\Delta_\mcO+\ell\rp-i-A-C)\G(\hlf\lp -\Delta_3+\Delta_4+\Delta_\mcO-\ell\rp+i+A+C)}{\G(\hlf\lp D+\Delta_3-\Delta_4-\Delta_\mcO+\ell\rp-i-A-C)\G(\hlf\lp D-\Delta_3+\Delta_4-\Delta_\mcO-\ell\rp+i+A+C)}\\
\times\frac{\G(\frac{D}{2}+\ell-i-1)}{\G(\frac{D}{2}+\ell-1)}u^{i-\frac{\ell}{2}}v^B
\whg_{0;\ell-2(i+A+B),\ell-2(i+A+C)}(u,v)
\end{multline}

At any rate, in subsequent sections we will assume that these SSSS conformal blocks are some known functions, and we will endeavor to compute the new conformal blocks in terms of these.  


\subsubsection{$\langle SVSS\rangle$}

Let's now move to the case with one vector.  The most efficient way to proceed is to first note that we can relate the three-point function of a scalar, a vector, and a symmetric traceless tensor to the three-point function of two scalars and a symmetric traceless tensor~\cite{Costa2011}.  Explicitly, we can define
\begin{multline}
S^\la_{a_1\cdots a_\ell}(x_i;\Delta_i)\\
=\lp x_{12}^2\rp^{\hlf\lp -\Delta_1-\Delta_2+\Delta_\mcO\rp}\lp x_{13}^2\rp^{\hlf\lp -\Delta_1+\Delta_2-\Delta_\mcO\rp}\lp x_{23}^2\rp^{\hlf\lp\Delta_1-\Delta_2-\Delta_\mcO\rp}\Pi^{(\ell)\,b_1\cdots b_\ell}_{a_1\cdots a_\ell}k^{(312)}_{b_1}\cdots k^{(312)}_{b_\ell},
\end{multline}
so that
\be
\left\langle\phi_1(x_1)\phi_2(x_2)\mcO_{a_1\cdots a_\ell}(x_3)\right\rangle=\la_{12\mcO}S^\la_{a_1\cdots a_\ell}(x_i;\Delta_i).
\ee
Then we can write
\be
\left\langle\phi_1(x_1)v_a(x_2)\mcO_{b_1\cdots b_\ell}(x_3)\right\rangle=\al_{12\mcO}S^\al_{a\,b_1\cdots b_\ell}(x_i;\Delta_i)+\beta_{12\mcO}S^\beta_{a\,b_1\cdots b_\ell}(x_i;\Delta_i),
\ee
where
\be\label{eq:Salpha}
\begin{split}
& S^\al_{a\,b_1\cdots b_\ell}(x_i;\Delta_\phi,\Delta_v,\Delta_\mcO)\\
& \quad =\lp x_{12}^2\rp^{\hlf\lp -\Delta_\phi-\Delta_v+\Delta_\mcO\rp}\lp x_{13}^2\rp^{\hlf\lp -\Delta_\phi+\Delta_v-\Delta_\mcO\rp}\lp x_{23}^2\rp^{\hlf\lp\Delta_\phi-\Delta_v-\Delta_\mcO\rp}\Pi^{(\ell)\,c_1\cdots c_\ell}_{b_1\cdots b_\ell}k^{(213)}_ak^{(312)}_{c_1}\cdots k^{(312)}_{c_\ell}\\
& \quad =\frac{1}{2\lp 1-\Delta_\mcO\rp}\ls m^{(12)\hph{a}c}_{\hph{(12)}a}\lp\frac{\p}{\p x_1^c}+2\lp\Delta_\phi-1\rp\frac{\lp x_{12}\rp_c}{x_{12}^2}\rp S^\la_{b_1\cdots b_\ell}(x_i;\Delta_\phi-1,\Delta_v,\Delta_\mcO)\right.\\
& \qquad\left.+\lp\frac{\p}{\p x_2^a}-2\lp\Delta_v-1\rp\frac{\lp x_{12}\rp_a}{x_{12}^2}\rp S^\la_{b_1\cdots b_\ell}(x_i;\Delta_\phi,\Delta_v-1,\Delta_\mcO)\rs,
\end{split}
\ee
and
\be\label{eq:Sbeta}
\begin{split}
& S^\beta_{a\,b_1\cdots b_\ell}(x_i;\Delta_\phi,\Delta_v,\Delta_\mcO)\\
& \quad =\lp x_{12}^2\rp^{\hlf\lp -\Delta_\phi-\Delta_v+\Delta_\mcO\rp}\lp x_{13}^2\rp^{\hlf\lp -\Delta_\phi+\Delta_v-\Delta_\mcO\rp}\lp x_{23}^2\rp^{\hlf\lp\Delta_\phi-\Delta_v-\Delta_\mcO\rp}\Pi^{(\ell)\,c_1\cdots c_\ell}_{b_1\cdots b_\ell}m^{(23)}_{ac_1}k^{(312)}_{c_2}\cdots k^{(312)}_{c_\ell}\\
& \quad =\frac{\Delta_\phi-\Delta_v-\Delta_\mcO+\ell+1}{\ell}S^\al_{a\,b_1\cdots b_\ell}(x_i;\Delta_\phi,\Delta_v,\Delta_\mcO)\\
& \qquad -\frac{1}{\ell}\lp\frac{\p}{\p x_2^a}-2\lp\Delta_v-1\rp\frac{\lp x_{12}\rp_a}{x_{12}^2}\rp S^\la_{b_1\cdots b_\ell}(x_i;\Delta_\phi,\Delta_v-1,\Delta_\mcO),
\end{split}
\ee
as can be verified by explicit computation.

The conformal blocks will be computed by the expression
\be
\label{eq:SVSSDeriv1}
\begin{split}
& X_{\Delta_1,\Delta_v,\Delta_3,\Delta_4}\la_{34\mcO}\ls\lp\al_{12\mcO}\whg_1^{\al\la}+\beta_{12\mcO}\whg_1^{\beta\la}\rp k^{(214)}_a+\lp\al_{12\mcO}\whg_2^{\al\la}+\beta_{12\mcO}\whg_2^{\beta\la}\rp k^{(234)}_a\rs\\
& \quad =\mcN_\mcO\int d^Dx_0\left\langle\phi_1(x_1)v_a(x_2)\mcO_{b_1\cdots b_\ell}(x_0)\right\rangle\left\langle\phi_3(x_3)\phi_4(x_4)\wtmcO^{b_1\cdots b_\ell}(x_0)\right\rangle\\
& \quad =\mcN_\mcO\int d^Dx_0\lp\al_{12\mcO}S^\al_{a\,b_1\cdots b_\ell}(x_1,x_2,x_0;\Delta_1,\Delta_v,\Delta_\mcO)+\beta_{12\mcO}S^\beta_{a\,b_1\cdots b_\ell}(x_1,x_2,x_0;\Delta_1,\Delta_v,\Delta_\mcO)\rp\\
& \qquad\times\la_{34\wtmcO} S^{\la\,b_1\cdots b_\ell}(x_3,x_4,x_0;\Delta_3,\Delta_4,D-\Delta_\mcO).
\end{split}
\ee
On the other hand, we have
\begin{multline}
X_{\Delta_1,\Delta_2,\Delta_3,\Delta_4}\la_{12\mcO}\la_{34\mcO}\whg(u,v;\Delta_i;\ell,\Delta_\mcO)\\
=\mcN_\mcO\la_{12\mcO}\la_{34\wtmcO}\int d^Dx_0S^\la_{a_1\cdots a_\ell}(x_1,x_2,x_0;\Delta_1,\Delta_2,\Delta_\mcO)S^{\la\,a_1\cdots a_\ell}(x_3,x_4,x_0;\Delta_3,\Delta_4,D-\Delta_\mcO).
\end{multline}
By expressing $S^\al$ and $S^\beta$ in terms of $S^\la$, and pulling the differential operators outside of the integral, we can express $\whg^{r\la}_i$ in terms of differential operators acting on $\whg$.  

For example, to compute $\whg^{\al\la}_i$ we get
\begin{multline}
\whg^{\al\la}_1k^{(214)}_a+\whg^{\al\la}_2k^{(234)}_a=\frac{1}{2\lp 1-\Delta_\mcO\rp}X_{\Delta_1,\Delta_v,\Delta_3,\Delta_4}^{-1}\\
\times\ls m^{(12)\hph{a}c}_{\hph{(12)}a}\lp\frac{\p}{\p x_1^c}+2\lp\Delta_1-1\rp\frac{\lp x_{12}\rp_c}{x_{12}^2}\rp\lp X_{\Delta_1-1,\Delta_v,\Delta_3,\Delta_4}\whg(x_i;\Delta_\phi-1,\Delta_v,\Delta_3,\Delta_4;\ell,\Delta_\mcO)\rp\right.\\
\left. +\lp\frac{\p}{\p x_2^a}-2\lp\Delta_v-1\rp\frac{\lp x_{12}\rp_a}{x_{12}^2}\rp\lp X_{\Delta_1,\Delta_v-1,\Delta_3,\Delta_4}\whg(x_i;\Delta_1,\Delta_v-1,\Delta_3,\Delta_4;\ell,\Delta_\mcO)\rp\rs.
\end{multline}
This leads to
\begin{multline}
\whg^{\al\la}_1=\frac{1}{2\lp 1-\Delta_\mcO\rp}\ls\lp 1-\Delta_1+\Delta_v+\lp 1-v\rp\lp\Delta_3-\Delta_4\rp+2v\lp 1-v\rp\p_v-2uv\p_u\rp\right.\\
\left. \times\whg(u,v;\Delta_1-1,\Delta_v,\Delta_3,\Delta_4;\ell,\Delta_\mcO)\right.\\
\left. +\lp 1+\Delta_1-\Delta_v-2u\p_u\rp\whg(u,v;\Delta_1+1,\Delta_v,\Delta_3,\Delta_4;\ell,\Delta_\mcO)\rs,
\end{multline}
\begin{multline}
\whg^{\al\la}_2=\frac{\sqrt{uv}}{2\lp 1-\Delta_\mcO\rp}\ls\lp\Delta_3-\Delta_4+2u\p_u+2v\p_v\rp\whg(u,v;\Delta_1-1,\Delta_v,\Delta_3,\Delta_4;\ell,\Delta_\mcO)\right.\\
\left. -2\p_v\whg(u,v;\Delta_1+1,\Delta_v,\Delta_3,\Delta_4;\ell,\Delta_\mcO)\rs,
\end{multline}

Similarly,
\begin{multline}
\whg^{\beta\la}_1=\frac{\Delta_1-\Delta_v-\Delta_\mcO+\ell+1}{\ell}\whg^{\al\la}_1(u,v;\Delta_1,\Delta_v,\Delta_3,\Delta_4;\ell,\Delta_\mcO)\\
-\frac{1}{\ell}\lp 1+\Delta_1-\Delta_v-2u\p_u\rp\whg(u,v;\Delta_1+1,\Delta_v,\Delta_3,\Delta_4;\ell,\Delta_\mcO),
\end{multline}
\begin{multline}
\whg^{\beta\la}_2=\frac{\Delta_1-\Delta_v-\Delta_\mcO+\ell+1}{\ell}\whg^{\al\la}_2(u,v;\Delta_1,\Delta_v,\Delta_3,\Delta_4;\ell,\Delta_\mcO)\\
+\frac{2\sqrt{uv}}{\ell}\p_v\whg(u,v;\Delta_1+1,\Delta_v,\Delta_3,\Delta_4;\ell,\Delta_\mcO).
\end{multline}

Note that as with the scalar blocks, the expressions only depend on the difference $\Delta_1-\Delta_v$ and $\Delta_3-\Delta_4$, not on the weights individually.  The other crucial property of these expressions is that the operators which act on the $\whg$ on the right hand side involve only integer powers of $\sqrt{u}$, so in particular they are all invariant under the monodromy projection.  This means that when we implement the monodromy projection, all we have to do is remove the hats from the scalar blocks on the right hand side and from the new blocks on the left hand side.  After these expressions have been thus doffed, we have relations between the full $g^{r\la}_i$ blocks and the scalar blocks $g$.

\subsubsection{$\langle SSSV\rangle$}

The case when the vector is in the fourth position is very similar.  We have
\be
\label{eq:SSSVDeriv1}
\begin{split}
& X_{\Delta_1,\Delta_2,\Delta_3,\Delta_v}\la_{12\mcO}\ls\lp\al_{34\mcO}\whg_1^{\la\al}+\beta_{34\mcO}\whg_1^{\la\beta}\rp k^{(412)}_a+\lp\al_{34\mcO}\whg_2^{\la\al}+\beta_{34\mcO}\whg_2^{\la\beta}\rp k^{(432)}_a\rs\\
& \quad =\mcN_\mcO\int d^Dx_0\left\langle\phi_1(x_1)\phi_2(x_2)\mcO_{b_1\cdots b_\ell}(x_0)\right\rangle\left\langle\phi_3(x_3)v_a(x_4)\wtmcO^{b_1\cdots b_\ell}(x_0)\right\rangle\\
& \quad =\mcN_\mcO\int d^Dx_0\la_{12\mcO}S^\la_{b_1\cdots b_\ell}(x_1,x_2,x_0;\Delta_1,\Delta_2,\Delta_\mcO)\\
& \quad\times\lp\al_{34\wtmcO}S^{\al\hph{a}b_1\cdots b_\ell}_{\hph{\al}a}(x_3,x_4,x_0;\Delta_3,\Delta_v,D-\Delta_\mcO)+\beta_{34\wtmcO}S^{\beta\hph{a}b_1\cdots b_\ell}_{\hph{\beta}a}(x_3,x_4,x_0;\Delta_3,\Delta_v,D-\Delta_\mcO)\rp.
\end{split}
\ee
Note that because $\al_{34\wtmcO}$ is not simply proportional to $\al_{34\mcO}$ (we should expand it using (\ref{eq:AlphaTilde})), and similarly for the $\beta$'s, it will now be the case (unlike for SVSS) that each conformal block will get contributions from both terms on the right-hand side.

The results (after also doffing the expressions) are
\begin{multline}
g^{\la\al}_1=\frac{\sqrt{u}}{2\lp\Delta_\mcO-1\rp}\ls\lp 1-\Delta_3+\Delta_v-2u\p_u-2v\p_v\rp g(u,v;\Delta_1,\Delta_2,\Delta_3-1,\Delta_v;\ell,\Delta_\mcO)\right.\\
\left. +\lp 1-\Delta_1+\Delta_2+\Delta_3-\Delta_v+2v\p_v\rp g(u,v;\Delta_1,\Delta_2,\Delta_3+1,\Delta_v;\ell,\Delta_\mcO)\rs,
\end{multline}
\begin{multline}
g^{\la\al}_2=\frac{\sqrt{v}}{2\lp 1-\Delta_\mcO\rp}\ls\lp 1-\Delta_3+\Delta_v-2u\p_u+2\lp 1-v\rp\p_v\rp g(u,v;\Delta_1,\Delta_2,\Delta_3-1,\Delta_v;\ell,\Delta_\mcO)\right.\\
\left. +\lp 1+\Delta_3-\Delta_v-2u\p_u\rp g(u,v;\Delta_1,\Delta_2,\Delta_3+1,\Delta_v;\ell,\Delta_\mcO)\rs,
\end{multline}
\begin{multline}
g^{\la\beta}_1=\frac{\Delta_3-\Delta_v-\Delta_\mcO+\ell+1}{\ell} g^{\la\al}_1(u,v;\Delta_1,\Delta_2,\Delta_3,\Delta_v;\ell,\Delta_\mcO)\\
+\frac{\sqrt{u}}{\ell}\lp 1-\Delta_1+\Delta_2+\Delta_3-\Delta_v+2v\p_v\rp g(u,v;\Delta_1,\Delta_2,\Delta_3+1,\Delta_v;\ell,\Delta_\mcO),
\end{multline}
\begin{multline}
g^{\la\beta}_2=\frac{\Delta_3-\Delta_v-\Delta_\mcO+\ell+1}{\ell} g^{\la\al}_2(u,v;\Delta_1,\Delta_2,\Delta_3,\Delta_v;\ell,\Delta_\mcO)\\
-\frac{\sqrt{v}}{\ell}\lp 1+\Delta_3-\Delta_v-2u\p_u\rp g(u,v;\Delta_1,\Delta_2,\Delta_3+1,\Delta_v;\ell,\Delta_\mcO),
\end{multline}
again only depending on the differences $\Delta_1-\Delta_2$ and $\Delta_3-\Delta_v$.

\subsubsection{$\langle SVSV\rangle$}
\label{subsubsec:SVSVBlocks}

At last we turn to the case of primary interest; two scalars and two vectors.  We will label the scalars $1$ and $3$, and the vectors $2$ and $4$.  As we have seen, the exchange operator can be either traceless symmetric $\mcO$ or a mixed symmetry operator $\mcA$.  

We'll start with the symmetric exchange.  There are twenty different conformal blocks which can arise, $g^{rs}_p$, where $r$ and $s$ run over $\al$ and $\beta$ (for $\ell>0$, or only $\al$ for $\ell=0$) and $p$ runs over the five tensor structures of the four-point function, which we have labeled $0$, $11$, $12$, $21$, and $22$.  By inserting the shadow projector, we can get all the symmetric exchange blocks as contractions of $S^\al$'s and $S^\beta$'s, which we can in turn write as differential operators acting on $S^\la$'s.  Finally, we impose the monodromy projection by doffing all expressions.  In fact, the resulting expressions are more compact if we write them in terms of either $g^{\al\la}_i$ or $g^{\la\al}_i$.  For the $\al\al$ blocks we'll use the former representation, and we will further introduce shorthand
\be
g^{\al\la}_{i;\ell;P,Q}=g^{\al\la}_i(u,v;\Delta_1+P,\Delta_2,\Delta_3+Q,\Delta_4;\ell,\Delta_\mcO).
\ee

The final expressions are
\bea
g^{\al\al}_0 &=& \frac{1}{2\lp 1-\Delta_\mcO\rp}\ls\frac{1}{\sqrt{v}}g^{\al\la}_{2;\ell;0,-1}-\sqrt{u}g^{\al\la}_{1;\ell;0,1}-\sqrt{v}g^{\al\la}_{2;\ell;0,1}\rs,\\
g^{\al\al}_{11} &=& \frac{\sqrt{u}}{2\lp 1-\Delta_\mcO\rp}\ls -\lp 1-\Delta_3+\Delta_4-2u\p_u-2v\p_v\rp g^{\al\la}_{1;\ell;0,-1}\right.\non\\
&& \left. +\lp\Delta_1-\Delta_2-\Delta_3+\Delta_4-2v\p_v\rp g^{\al\la}_{1;\ell;0,1}\rs,\\
g^{\al\al}_{12} &=& \frac{\sqrt{v}}{2\lp 1-\Delta_\mcO\rp}\ls\lp 1-\Delta_3+\Delta_4-2u\p_u+2\lp 1-v\rp\p_v\rp g^{\al\la}_{1;\ell;0,-1}\right.\non\\
&& \left. +\lp 1+\Delta_3-\Delta_4-2u\p_u\rp g^{\al\la}_{1;\ell;0,1}\rs,\\
g^{\al\al}_{21} &=& \frac{\sqrt{u}}{2\lp 1-\Delta_\mcO\rp}\ls -\lp 1-\Delta_3+\Delta_4-2u\p_u-2v\p_v\rp g^{\al\la}_{2;\ell;0,-1}\right.\non\\
&& \left. -\lp 1-\Delta_1+\Delta_2+\Delta_3-\Delta_4+2v\p_v\rp g^{\al\la}_{2;\ell;0,1}\rs,\\
g^{\al\al}_{22} &=& \frac{\sqrt{v}}{2\lp 1-\Delta_\mcO\rp}\ls\lp 1-\frac{1}{v}-\Delta_3+\Delta_4-2u\p_u+2\lp 1-v\rp\p_v\rp g^{\al\la}_{2;\ell;0,-1}\right.\non\\
&& \left. +\lp 2+\Delta_3-\Delta_4-2u\p_u\rp g^{\al\la}_{2;\ell;0,1}\rs.
\eea

For $\ell=0$, this is the entire answer.  For $\ell>0$, we can proceed similarly with the other blocks, obtaining the $\alpha\beta$, $\beta\alpha$, and $\beta\beta$ components given in appendix \ref{app:otherblockcomponents}.
To save space, we have omitted the arguments of the conformal blocks appearing above.  For the SVSV blocks (i.e.\ the expressions on the left-hand-sides above) the arguments are unshifted, $g^{rs}_p(u,v;\Delta_1,\Delta_2,\Delta_3,\Delta_4;\ell,\Delta_\mcO)$, while for the 
others we use the previously adopted condensed notation, along with
\be
g^{\la\al}_{i;\ell;P,Q}=g^{\la\al}_i(u,v;\Delta_1+P,\Delta_2,\Delta_3+Q,\Delta_4;\ell,\Delta_\mcO).
\ee

The combination which occurs in the four-point function is
\be
\al_{12\mcO}\al_{34\mcO}g^{\al\al}_p+\al_{12\mcO}\beta_{34\mcO}g^{\al\beta}_p+\beta_{12\mcO}\al_{34\mcO}g^{\beta\al}_p+\beta_{12\mcO}\beta_{34\mcO}g^{\beta\beta}_p.
\ee
It turns that if we write this combination in terms of scalar conformal blocks it has the remarkably simple form
\be
\label{eq:SymmExchangeGeneralExpression}
A_1A_2\mcD^{--}_pg_{\ell;-1,-1}+A_1B_2\mcD^{-+}_pg_{\ell;-1,1}+B_1A_2\mcD^{+-}_pg_{\ell;1,-1}+B_1B_2\mcD^{++}_pg_{\ell;1,1},
\ee
where
\bea
A_1 &=& \frac{1}{2\lp\Delta_\mcO-1\rp}\lp\al_{12\mcO}+\lp\Delta_1-\Delta_2-\Delta_\mcO+\ell+1\rp\frac{\beta_{12\mcO}}{\ell}\rp,\\
A_2 &=& \frac{1}{2\lp\Delta_\mcO-1\rp}\lp\al_{34\mcO}+\lp\Delta_3-\Delta_4-\Delta_\mcO+\ell+1\rp\frac{\beta_{34\mcO}}{\ell}\rp,\\
B_1 &=& \frac{1}{2\lp\Delta_\mcO-1\rp}\lp\al_{12\mcO}+\lp\Delta_1-\Delta_2+\Delta_\mcO+\ell-1\rp\frac{\beta_{12\mcO}}{\ell}\rp,\\
B_2 &=& \frac{1}{2\lp\Delta_\mcO-1\rp}\lp\al_{34\mcO}+\lp\Delta_3-\Delta_4+\Delta_\mcO+\ell-1\rp\frac{\beta_{34\mcO}}{\ell}\rp,
\eea
and $\mcD^{\pm\pm}_p$ are fairly simple differential operators whose explicit forms are given in a table in appendix \ref{app:DpOps}.

We turn next to the evaluation of blocks for exchanged of a mixed symmetry tensor $\mcA_{a_1a_2b_1\cdots b_k}$ whose representation is labeled by a non-negative integer $k$ ($k=0$ corresponds to an antisymmetric two-index tensor).  The contraction which we need is
\be
\lp\g_{12\mcA}m^{(20)}_{ac_1}k^{(012)}_{c_2}k^{(012)}_{d_1}\cdots k^{(012)}_{d_k}\rp\wtPi^{(k)\,c_1c_2d_1\cdots d_k}_{e_1e_2f_1\cdots f_k}\lp\g_{34\wtmcA}m^{(40)\hph{b}e_1}_{\hph{(40)}b}k^{(034)\,e_2}k^{(034)\,f_1}\cdots k^{(034)\,f_k}\rp.
\ee
For fixed $k$, given knowledge of the projector $\widetilde{\Pi}^{(k)}$ as detailed in appendix \ref{subapp:PiA}, we could just expand this contraction by brute force into a sum of monomials in the $x_{ij}^2$ and $x_{0i}^2$, with the free indices being carried by $(x_{ij})_a$, in which case we can pull it ouside of the integral, or $(x_{0i})_a$, in which case we can rewrite the corresponding integral as an $x_i$ derivative acting on a scalar integral.  Then each term in this collection can be evaluated using the integrals in appendix \ref{subapp:4ptIntegrals}, and the result could be written in terms of the $\ell=0$ scalar blocks.  However, this approach is impractical for several reasons, most notably that the number of monomials in such an expansion grows exponentially in $k$.  We need a cleaner expression.

Indeed, we can use (\ref{eq:mixedpartialcontraction}), (\ref{eq:fkExpression}), and (\ref{eq:gkExpression}), as well as other identities from appendices \ref{app:BuildingBlocks} and \ref{app:Projectors}, to rewrite the contraction as
\be\label{eq:mixedcontraction}
\begin{split}
& \lp\g_{12\mcA}m^{(20)}_{ac_1}k^{(012)}_{c_2}k^{(012)}_{d_1}\cdots k^{(012)}_{d_k}\rp\wtPi^{(k)\,c_1c_2d_1\cdots d_k}_{e_1e_2f_1\cdots f_k}\lp\g_{34\wtmcA}m^{(40)\hph{b}e_1}_{\hph{(40)}b}k^{(034)\,e_2}k^{(034)\,f_1}\cdots k^{(034)\,f_k}\rp\\
& \quad =\hlf\g_{12\mcA}\g_{34\wtmcA}\sqrt{\frac{x_{02}^2x_{04}^2x_{12}^2x_{34}^2}{x_{01}^2x_{03}^2}}\left\{\frac{k+2}{k+1}p_{D,k+1}(t)\frac{\p^2t}{\p x_2^a\p x_4^b}-\frac{\p^2}{\p x_2^a\p x_4^b}\ls\frac{1}{\lp k+1\rp\lp k+2\rp}p_{D,k+2}(t)\right.\right.\\
& \qquad \left.\left.+\frac{k+2}{\lp D+2k\rp\lp D+2k-2\rp\lp D+k-2\rp}p_{D,k}(t)\rs\right\},
\end{split}
\ee
In appendix \ref{app:contractiondetails} we give more details and motivation for how we arrive at this expression.

This is the main formula that will allow us to relate mixed-symmetric conformal blocks to the ones for symmetric-traceless exchange.
Notice that derivatives of the polynomials $p(t)$ will always produce terms that appear in the conformal blocks of traceless-symmetric operators since, from  (\ref{eq:SymmetricPartialContraction}), they are related to symmetric contractions where not all of the indices are contracted. Furthermore derivatives of $t$, i.e.
\begin{align}
 \frac{\p t}{\p x_{2}^{a}}=\lp \frac{\p}{\p x_{2}^{a}}k^{(012)\,b}\rp k_{b}^{(034)}
 =-\sqrt{\frac{x_{01}^{2}}{x_{02}^{2}x_{12}^{2}}}\ls \lp m^{20}\cdot k^{(034)}\rp_{a}
 +t k^{(201)}_{a}\rs
\end{align}
and similar expressions for other $x_{i}$\footnote{When there is more than one derivative with respect to the same variable, one has to include extra factors, say proportional to $(x_{12})_{a}$ in the derivative above, in order to obtain covariant structures ($k$'s and $m$'s).} produce the tensor structures from three-point functions of operators with spin (see (\ref{eq:Salpha}), (\ref{eq:Sbeta}) for example). Therefore if one could write the contractions of more general mixed symmetries in the form (\ref{eq:mixedcontraction})---i.e. total derivatives of the symmetric contraction $p_{D,k}$ and  undifferentiated polynomials times derivatives of $t$---then the relation of these expressions to those from symmetric exchanges would follow in analogy to our case. We believe that it will be possible to do this in more general situations, but this has not been definitively established.  However, to support this conjecture, we present the contraction for $[k+1,1,1]$ in appendix \ref{app:contractiondetails}, which has an analogous form.

From these arguments, we obtain
\be
\begin{split}
\frac{\p^2t}{\p x_2^a\p x_4^b} & =\sqrt{\frac{x_{01}^2x_{03}^2}{x_{02}^2x_{04}^2x_{12}^2x_{34}^2}}\lp m^{(24)}_{ab}-2\sqrt{\frac{v}{u}}k^{(214)}_ak^{(432)}_b\rp\\
& \quad -\hlf\frac{x_{01}^2x_{03}^2x_{24}^2+x_{01}^2x_{04}^2x_{23}^2+x_{02}^2x_{03}^2x_{14}^2+x_{02}^2x_{04}^2x_{13}^2}{x_{02}^2x_{04}^2x_{12}^2x_{34}^2}k^{(201)}_ak^{(403)}_b\\
& \quad +\sqrt{\frac{x_{03}^2x_{14}^2}{x_{04}^2x_{12}^2x_{24}^2x_{34}^2}}k^{(201)}_ak^{(412)}_b-\sqrt{\frac{x_{03}^2x_{23}^2}{x_{04}^2x_{24}^2}}\frac{x_{01}^2x_{24}^2+x_{02}^2x_{14}^2}{x_{02}^2x_{12}^2x_{34}^2}k^{(201)}_ak^{(432)}_b\\
& \quad -\sqrt{\frac{x_{01}^2x_{14}^2}{x_{02}^2x_{24}^2}}\frac{x_{03}^2x_{24}^2+x_{04}^2x_{23}^2}{x_{04}^2x_{12}^2x_{34}^2}k^{(214)}_ak^{(403)}_b+\sqrt{\frac{x_{01}^2x_{23}^2}{x_{02}^2x_{12}^2x_{24}^2x_{34}^2}}k^{(234)}_ak^{(403)}_b,
\end{split}
\ee
and also
\begin{multline}
\frac{\p^2}{\p x_2^a\p x_4^b}p_{D,\ell}(t)=\ell^2\sqrt{\frac{x_{01}^2x_{03}^2}{x_{02}^2x_{04}^2x_{12}^2x_{34}^2}}\lp m^{(20)}_{ac_1}+k^{(201)}_ak^{(012)}_{c_1}\rp\lp m^{(40)\hph{b}d_1}_{\hph{(40)}b}+k^{(403)}_bk^{(034)\,d_1}\rp\\
\times k^{(012)}_{c_2}\cdots k^{(012)}_{c_\ell}\Pi^{(\ell)\,c_1\cdots c_\ell}_{d_1\cdots d_\ell}k^{(034)\,d_2}\cdots k^{(034)\,d_\ell}.
\end{multline}
Thus if we define
\begin{multline}
\lp S^\al_{a\,PQ}\circ_\ell S^\al_{b\,RS}\rp\\
=\int d^Dx_0S^\al_{a\,c_1\cdots c_\ell}(x_1,x_2,x_0;\Delta_1+P,\Delta_2+Q,\Delta_\mcA)S^{\al\hph{b}c_1\cdots c_\ell}_{\hph{\al}b}(x_3,x_4,x_0;\Delta_3+R,\Delta_4+S,D-\Delta_\mcA),
\end{multline}
and similarly for $(S^\al_{a\,PQ}\circ_\ell S^\la_{RS})$, etc., we can compute (recall that (\ref{eq:mixedcontraction}) appears inside a conformal integral like (\ref{eq:SVSSDeriv1}))
\be
\label{eq:IntermediategA}
\begin{split}
& X_{\Delta_i}\g_{12\mcA}\g_{34\mcA}
\whg^{\g\g}_p(u,v)t^p_{ab}\\
& \quad =\mcN_\mcA\g_{12\mcA}\g_{34\wtmcA}\left\{\hlf\frac{k+2}{k+1}\ls\lp m^{(24)}_{ab}-2\sqrt{\frac{v}{u}}k^{(214)}_ak^{(432)}_b\rp\lp S^\la_{00}\circ_{k+1}S^\la_{00}\rp\right.\right.\\
& \quad\left.\left.-\hlf\frac{1}{\sqrt{x_{12}^2x_{34}^2}}\lp x_{24}^2\lp S^\al_{a\,-\hlf\,\hlf}\circ_{k+1}S^\al_{b\,-\hlf\,\hlf}\rp+x_{23}^2\lp S^\al_{a\,-\hlf\,\hlf}\circ_{k+1}S^\al_{b\,\hlf\,-\hlf}\rp\right.\right.\right.\\
& \quad\left.\left.\left. +x_{14}^2\lp S^\al_{a\,\hlf\,-\hlf}\circ_{k+1}S^\al_{b\,-\hlf\,\hlf}\rp+x_{13}^2\lp S^\al_{a\,\hlf\,-\hlf}\circ_{k+1}S^\al_{b\,\hlf\,-\hlf}\rp\rp-\sqrt{\frac{x_{14}^2}{x_{24}^2}}k^{(412)}_b\lp S^\al_{a\,\hlf\,-\hlf}\circ_{k+1}S^\la_{00}\rp\right.\right.\\
& \quad\left.\left. +\sqrt{\frac{x_{23}^2x_{24}^2}{x_{12}^2x_{34}^2}}k^{(432)}_b\lp S^\al_{a\,-\hlf\,\hlf}\circ_{k+1}S^\la_{00}\rp+x_{14}^2\sqrt{\frac{x_{23}^2}{x_{12}^2x_{24}^2x_{34}^2}}k^{(432)}_b\lp S^\al_{a\,\hlf\,-\hlf}\circ_{k+1}S^\la_{00}\rp\right.\right.\\
& \quad\left.\left. +\sqrt{\frac{x_{14}^2x_{24}^2}{x_{12}^2x_{34}^2}}k^{(214)}_a\lp S^\la_{00}\circ_{k+1}S^\al_{b\,-\hlf\,\hlf}\rp+x_{23}^2\sqrt{\frac{x_{14}^2}{x_{12}^2x_{24}^2x_{34}^2}}k^{(214)}_a\lp S^\la_{00}\circ_{k+1}S^\al_{b\,\hlf\,-\hlf}\rp\right.\right.\\
& \quad\left.\left. -\sqrt{\frac{x_{23}^2}{x_{24}^2}}k^{(234)}_a\lp S^\la_{00}\circ_{k+1}S^\al_{b\,\hlf\,-\hlf}\rp\rs-\hlf\frac{k+2}{k+1}\ls\lp S^\al_{a\,00}\circ_{k+2}S^\al_{b\,00}\rp-\lp S^\al_{a\,00}\circ_{k+2}S^\beta_{b\,00}\rp\right.\right.\\
& \quad\left.\left. -\lp S^\beta_{a\,00}\circ_{k+2}S^\al_{b\,00}\rp+\lp S^\beta_{a\,00}\circ_{k+2}S^\beta_{b\,00}\rp\rs-\hlf\frac{k^2\lp k+2\rp}{\lp D+2k\rp\lp D+2k-2\rp\lp D+k-2\rp}\right.\\
& \quad\left.\times\ls\lp S^\al_{a\,00}\circ_kS^\al_{b\,00}\rp-\lp S^\al_{a\,00}\circ_kS^\beta_{b\,00}\rp-\lp S^\beta_{a\,00}\circ_kS^\al_{b\,00}\rp+\lp S^\beta_{a\,00}\circ_kS^\beta_{b\,00}\rp\rs
\right\}.
\end{split}
\ee

To evaluate these we make use of relations like
\be
\lp S^\la_{00}\circ_\ell S^\la_{00}\rp=\frac{X_{\Delta_1,\Delta_2,\Delta_3,\Delta_4}}{\mcN_\mcO\lp\la_{34\wtmcO}/\la_{34\mcO}\rp}\whg_{\ell;0,0}(u,v)
\ee
where $\mcN_\mcO$, $\la_{34\wtmcO}$, and $\whg_{\ell;0,0}(u,v)$ are given by (\ref{eq:NO}), (\ref{eq:lambdaOtilde}), and (\ref{eq:SCondensed}) respectively, where we make the substitution $\Delta_\mcO\rr\Delta_\mcA$ in all three definitions (also we of course substitute $\Delta_1\rr\Delta_3$ and $\Delta_2\rr\Delta_4$ in the definition of $\la$).  Similarly,
\be
\lp S^\al_{a\,PQ}\circ_\ell S^\la_{00}\rp=\frac{X_{\Delta_1+P,\Delta_2+Q,\Delta_3,\Delta_4}}{\mcN_\mcO\lp\la_{34\wtmcO}/\la_{34\mcO}\rp}\ls
\whg^{\al\la}_{1;\ell;P-Q,0}(u,v)
k^{(214)}_a
+
\whg^{\al\la}_{2;\ell;P-Q,0}(u,v)
k^{(234)}_a\rs.
\ee
For the other expressions we need to invert the matrix $M_r^{\hph{r}s}$ introduced in (\ref{eq:MDef}) (substituting $\Delta_\mcO\rr\Delta_\mcA$, as well as $\Delta_\phi\rr\Delta_3+R$ and $\Delta_v\rr\Delta_4+S$).  Then we have
\be
\lp S^\la_{00}\circ_\ell S^\al_{a\,RS}\rp=\frac{X_{\Delta_1,\Delta_2,\Delta_3+R,\Delta_4+S}}{\mcN_\mcO}\lp M^{-1}\rp_r^{\hph{r}\al}
\whg^{\la r}_{p;\ell;0,R-S}(u,v)
t^p_a,
\ee
where we sum $r$ over $\al$ and $\beta$, sum $p$ over $1$ and $2$, and where $t^1_a=k^{(412)}_a$, $t^2_a=k^{(432)}_a$.  Similarly,
\be
\lp S^r_{a\,PQ}\circ_\ell S^s_{b\,RS}\rp
=\frac{X_{\Delta_1+P,\Delta_2+Q,\Delta_3+R,\Delta_4+S}}{\mcN_\mcO}\lp M^{-1}\rp_t^{\hph{t}s}
\whg^{rt}_{p;\ell;P-Q,R-S}(u,v)
t^p_{ab}.
\ee
The necessary combination of coefficients involves
\begin{multline}
\mcN_\mcO^{-1}M^{-1}=\pi^{D/2}\frac{\G(\frac{D}{2}-\Delta_\mcO)\G(D-\Delta_\mcO+\ell-1)}{\G(D-\Delta_\mcO)\G(\Delta_\mcO+\ell)}\\
\times\frac{\G(\hlf\lp\Delta_3-\Delta_4+\Delta_\mcO+\ell+1+R-S\rp)\G(\hlf\lp -\Delta_3+\Delta_4+\Delta_\mcO+\ell-1-R+S\rp)}{\G(\hlf\lp D+\Delta_3-\Delta_4-\Delta_\mcO+\ell-1+R-S\rp)\G(\hlf\lp D-\Delta_3+\Delta_4-\Delta_\mcO+\ell+1-R+S\rp)}\\
\times\lp\begin{matrix}\frac{\lp\Delta_\mcO-1\rp\lp D-\Delta_\mcO+\ell-1\rp-\lp D-\Delta_\mcO-1\rp\lp\Delta_3-\Delta_4+R-S\rp}{D+\Delta_3-\Delta_4-\Delta_\mcO+\ell-1+R-S} & 2\Delta_\mcO-D \\ -\frac{\ell\lp 2\Delta_\mcO-D\rp\lp\Delta_3-\Delta_4+R-S\rp}{\lp\Delta_3-\Delta_4+\Delta_\mcO+\ell-1+R-S\rp\lp D+\Delta_3-\Delta_4-\Delta_\mcO+\ell-1+R-S\rp} & \frac{\lp\Delta_\mcO+\ell-1\rp\lp D-\Delta_\mcO-1\rp-\lp\Delta_\mcO-1\rp\lp\Delta_3-\Delta_4+R-S\rp}{\Delta_3-\Delta_4+\Delta_\mcO+\ell-1+R-S}\end{matrix}\rp
\end{multline}

Finally we can plug in these expressions to (\ref{eq:IntermediategA}) and collect the different tensor structures to obtain
\be
\begin{split}
& g^{\g\g}_0(u,v;\Delta_1,\Delta_2,\Delta_3,\Delta_4;k,\Delta_\mcA)\\
& \quad =\hlf\frac{k+2}{k+1}\ls C_1g_{k+1;0,0}-\frac{1}{2\sqrt{u}}\lp C_2\lp g^{\al\al}_{0;k+1;-1,-1}+g^{\al\al}_{0;k+1;1,-1}\rp+C_3\lp g^{\al\beta}_{0;k+1;-1,-1}+g^{\al\beta}_{0;k+1;1,-1}\rp\right.\right.\\
& \qquad\quad\left.\left. +C_4\lp vg^{\al\al}_{0;k+1;-1,1}+g^{\al\al}_{0;k+1;1,1}\rp+C_5\lp vg^{\al\beta}_{0;k+1;-1,1}+g^{\al\beta}_{0;k+1;1,1}\rp\rp\vphantom{\frac{1}{2\sqrt{u}}}\rs\\
& \quad -\hlf\frac{k+2}{k+1}\ls C_6\lp g^{\al\al}_{0;k+2;0,0}-g^{\beta\al}_{0;k+2;0,0}\rp+C_7\lp g^{\beta\beta}_{0;k+2;0,0}-g^{\al\beta}_{0;k+2;0,0}\rp\rs\\
& \quad -\hlf\frac{k^2\lp k+2\rp}{\lp D+2k\rp\lp D+2k-2\rp\lp D+k-2\rp}\ls C_8\lp g^{\al\al}_{0;k;0,0}-g^{\beta\al}_{0;k;0,0}\rp+C_9\lp g^{\beta\beta}_{0;k;0,0}-g^{\al\beta}_{0;k;0,0}\rp\rs,
\end{split}
\ee
\be
\begin{split}
& g^{\g\g}_{11}(u,v;\Delta_1,\Delta_2,\Delta_3,\Delta_4;k,\Delta_\mcA)\\
& \quad =\hlf\frac{k+2}{k+1}\ls -\frac{1}{2\sqrt{u}}\lp C_2\lp g^{\al\al}_{11;k+1;-1,-1}+g^{\al\al}_{11;k+1;1,-1}\rp+C_3\lp g^{\al\beta}_{11;k+1;-1,-1}+g^{\al\beta}_{11;k+1;1,-1}\rp\right.\right.\\
& \qquad\quad\left.\left. +C_4\lp vg^{\al\al}_{11;k+1;-1,1}+g^{\al\al}_{11;k+1;1,1}\rp+C_5\lp vg^{\al\beta}_{11;k+1;-1,1}+g^{\al\beta}_{11;k+1;1,1}\rp\rp-C_1g^{\al\la}_{1;k+1;1,0}\right.\\
& \qquad\left. +\frac{C_2}{\sqrt{u}}g^{\la\al}_{1;k+1;0,-1}+\frac{C_3}{\sqrt{u}}g^{\la\beta}_{1;k+1;0,-1}+\frac{C_4v}{\sqrt{u}}g^{\la\al}_{1;k+1;0,1}+\frac{C_5v}{\sqrt{u}}g^{\la\beta}_{1;k+1;0,1}\rs\\
& \quad-\hlf\frac{k+2}{k+1}\ls C_6\lp g^{\al\al}_{11;k+2;0,0}-g^{\beta\al}_{11;k+2;0,0}\rp+C_7\lp g^{\beta\beta}_{11;k+2;0,0}-g^{\al\beta}_{11;k+2;0,0}\rp\rs\\
& \quad -\hlf\frac{k^2\lp k+2\rp}{\lp D+2k\rp\lp D+2k-2\rp\lp D+k-2\rp}\ls C_8\lp g^{\al\al}_{11;k;0,0}-g^{\beta\al}_{11;k;0,0}\rp+C_9\lp g^{\beta\beta}_{11;k;0,0}-g^{\al\beta}_{11;k;0,0}\rp\rs,
\end{split}
\ee
\be
\begin{split}
& g^{\g\g}_{12}(u,v;\Delta_1,\Delta_2,\Delta_3,\Delta_4;k,\Delta_\mcA)\\
& \quad =\hlf\frac{k+2}{k+1}\ls -2\sqrt{\frac{v}{u}}C_1g_{k+1;0,0}-\frac{1}{2\sqrt{u}}\lp C_2\vphantom{g^{\al\beta}_{12;k+1;-1,-1}}\lp g^{\al\al}_{12;k+1;-1,-1}+g^{\al\al}_{12;k+1;1,-1}\rp\right.\right.\\
& \qquad\quad\left.\left. +C_3\lp g^{\al\beta}_{12;k+1;-1,-1}+g^{\al\beta}_{12;k+1;1,-1}\rp+C_4\lp vg^{\al\al}_{12;k+1;-1,1}+g^{\al\al}_{12;k+1;1,1}\rp\right.\right.\\
& \qquad\quad\left.\left. +C_5\lp vg^{\al\beta}_{12;k+1;-1,1}+g^{\al\beta}_{12;k+1;1,1}\rp\rp+C_1\sqrt{\frac{v}{u}}\lp g^{\al\la}_{1;k+1;-1,0}+g^{\al\la}_{1;k+1;1,0}\rp+\frac{C_2}{\sqrt{u}}g^{\la\al}_{2;k+1;0,-1}\right.\\
& \qquad\left. +\frac{C_3}{\sqrt{u}}g^{\la\beta}_{2;k+1;0,-1}+\frac{C_4v}{\sqrt{u}}g^{\la\al}_{2;k+1;0,1}+\frac{C_5v}{\sqrt{u}}g^{\la\beta}_{2;k+1;0,1}\rs\\
& \quad -\hlf\frac{k+2}{k+1}\ls C_6\lp g^{\al\al}_{12;k+2;0,0}-g^{\beta\al}_{12;k+2;0,0}\rp+C_7\lp g^{\beta\beta}_{12;k+2;0,0}-g^{\al\beta}_{12;k+2;0,0}\rp\rs\\
& \quad -\hlf\frac{k^2\lp k+2\rp}{\lp D+2k\rp\lp D+2k-2\rp\lp D+k-2\rp}\ls C_8\lp g^{\al\al}_{12;k;0,0}-g^{\beta\al}_{12;k;0,0}\rp+C_9\lp g^{\beta\beta}_{12;k;0,0}-g^{\al\beta}_{12;k;0,0}\rp\rs,
\end{split}
\ee
\be
\begin{split}
& g^{\g\g}_{21}(u,v;\Delta_1,\Delta_2,\Delta_3,\Delta_4;k,\Delta_\mcA)\\
& \quad =\hlf\frac{k+2}{k+1}\ls -\frac{1}{2\sqrt{u}}\lp C_2\lp g^{\al\al}_{21;k+1;-1,-1}+g^{\al\al}_{21;k+1;1,-1}\rp+C_3\lp g^{\al\beta}_{21;k+1;-1,-1}+g^{\al\beta}_{21;k+1;1,-1}\rp\right.\right.\\
& \qquad\quad\left.\left. +C_4\lp vg^{\al\al}_{21;k+1;-1,1}+g^{\al\al}_{21;k+1;1,1}\rp+C_5\lp vg^{\al\beta}_{21;k+1;-1,1}+g^{\al\beta}_{21;k+1;1,1}\rp\rp\right.\\
& \qquad\left. -C_1g^{\al\la}_{2;k+1;1,0}-C_4\sqrt{v}g^{\la\al}_{1;k+1;0,1}-C_5\sqrt{v}g^{\la\beta}_{1;k+1;0,1}\vphantom{\frac{1}{2\sqrt{u}}}\rs\\
& \quad -\hlf\frac{k+2}{k+1}\ls C_6\lp g^{\al\al}_{21;k+2;0,0}-g^{\beta\al}_{21;k+2;0,0}\rp+C_7\lp g^{\beta\beta}_{21;k+2;0,0}-g^{\al\beta}_{21;k+2;0,0}\rp\rs\\
& \quad -\hlf\frac{k^2\lp k+2\rp}{\lp D+2k\rp\lp D+2k-2\rp\lp D+k-2\rp}\ls C_8\lp g^{\al\al}_{21;k;0,0}-g^{\beta\al}_{21;k;0,0}\rp+C_9\lp g^{\beta\beta}_{21;k;0,0}-g^{\al\beta}_{21;k;0,0}\rp\rs,
\end{split}
\ee
\be
\begin{split}
& g^{\g\g}_{22}(u,v;\Delta_1,\Delta_2,\Delta_3,\Delta_4;k,\Delta_\mcA)\\
& \quad =\hlf\frac{k+2}{k+1}\ls -\frac{1}{2\sqrt{u}}\lp C_2\lp g^{\al\al}_{22;k+1;-1,-1}+g^{\al\al}_{22;k+1;1,-1}\rp+C_3\lp g^{\al\beta}_{22;k+1;-1,-1}+g^{\al\beta}_{22;k+1;1,-1}\rp\right.\right.\\
& \qquad\quad\left.\left. +C_4\lp vg^{\al\al}_{22;k+1;-1,1}+g^{\al\al}_{22;k+1;1,1}\rp+C_5\lp vg^{\al\beta}_{22;k+1;-1,1}+g^{\al\beta}_{22;k+1;1,1}\rp\rp\right.\\
& \qquad\left. +C_1\sqrt{\frac{v}{u}}\lp g^{\al\la}_{2;k+1;-1,0}+g^{\al\la}_{2;k+1;1,0}\rp-C_4\sqrt{v}g^{\la\al}_{2;k+1;0,1}-C_5\sqrt{v}g^{\la\beta}_{2;k+1;0,1}\rs\\
& \quad -\hlf\frac{k+2}{k+1}\ls C_6\lp g^{\al\al}_{22;k+2;0,0}-g^{\beta\al}_{22;k+2;0,0}\rp+C_7\lp g^{\beta\beta}_{22;k+2;0,0}-g^{\al\beta}_{22;k+2;0,0}\rp\rs\\
& \quad -\hlf\frac{k^2\lp k+2\rp}{\lp D+2k\rp\lp D+2k-2\rp\lp D+k-2\rp}\ls C_8\lp g^{\al\al}_{22;k;0,0}-g^{\beta\al}_{22;k;0,0}\rp+C_9\lp g^{\beta\beta}_{22;k;0,0}-g^{\al\beta}_{22;k;0,0}\rp\rs.
\end{split}
\ee
The constants appearing above are written in appendix \ref{app:mixedsymmetricconstants}.  For $k=0$ the $k^2$ numerator in the last line of each conformal block kills those terms, and so we don't need to worry about the fact that for $k=0$ only $g^{\al\al}$ is defined.  One detail that we do need to worry about is the $\ell$-dependent normalization of the scalar conformal blocks, mentioned below equation (\ref{eq:kbetaDef}).  Since the expressions used to compute the mixed symmetry blocks involve adding contributions from scalar conformal blocks of different spins, it is important to use our normalization for the scalar blocks.  Otherwise, some of the relative coefficients will be off (e.g.\ by powers of two relative to another common normalization)\footnote{We have checked that our expressions for $k=0,1$ are consistent with the latest version of \cite{Costa2014}.}.

\section{Setting up the bootstrap}
\label{sec:Bootstrap}

\subsection{General discussion}

The picture now is that we are given explicit expressions for the conformal blocks, which depend only on the weights $\Delta_i$ and\footnote{In sections discussing very general four-point functions, such as this one, $\mcO$ will stand for all possible primary exchange operators, of arbitrary $\SO(D)$ representations, while in sections discussing particular assignments of representations, such as the following two subsections, $\mcO$ will refer only to traceless symmetric exchanges.  In that case we will also have exchange operators $\mcA$ of mixed symmetry.} $\Delta_\mcO$, and the $\SO(D)$ representations of the five operators in question.  The blocks are otherwise theory-independent.  Indeed, a CFT is specified by the spectrum of primary operators, i.e. a list of the $\phi_{i\,\bar{a}}$, characterized by their weights $\Delta_i$ and representations, and their OPE coefficients $\la_{ijk\,r}$ (a finite list of constants enumerated by $r$ for each fixed triple of operators $\phi_{i\,\bar{a}}$, $\phi_{j\,\bar{b}}$, $\phi_{k\,\bar{c}}$).  Then from this data we can compute any four-point function by
\be
\left\langle\phi_{1\,\bar{a}}(x_1)\phi_{2\,\bar{b}}(x_2)\phi_{3\,\bar{c}}(x_3)\phi_{4\,\bar{d}}(x_4)\right\rangle=X_{\Delta_i}
\sum_p\lp\sum_\mcO\sum_{r,s}\la_{12\mcO\,r}\la_{34\mcO\,s}g^{rs}_p(u,v)\rp t^p_{\bar{a}\bar{b}\bar{c}\bar{d}}(x_i),
\ee

However, in deriving this expression we made a choice to first perform the OPEs of $\phi_{1\,\bar{a}}$ with $\phi_{2\,\bar{b}}$ and $\phi_{3\,\bar{c}}$ with $\phi_{4\,\bar{d}}$, then evaluating the resulting two-point function.  Starting with the same correlation function and the same CFT data, we could have evaluated instead the OPE of $\phi_{1\,\bar{a}}$ with $\phi_{4\,\bar{d}}$ and $\phi_{2\,\bar{b}}$ with $\phi_{3\,\bar{c}}$, or equivalently, we could have performed a $2\leftrightarrow 4$ crossing symmetry exchange before performing our OPEs.  This should be an equivalent path to the same four-point correlator, and by comparing the two results we obtain a non-trivial constraint on the defining data of our CFT.

Let's recall how this works for four scalars operators $\phi_i$.  We have
\begin{multline}
\left\langle\phi_1(x_1)\phi_2(x_2)\phi_3(x_3)\phi_4(x_4)\right\rangle=\lp\frac{x_{14}^2}{x_{13}^2}\rp^{\hlf\lp\Delta_3-\Delta_4\rp}\lp\frac{x_{24}^2}{x_{14}^2}\rp^{\hlf\lp\Delta_1-\Delta_2\rp}\\
\times\lp x_{12}^2\rp^{-\hlf\lp\Delta_1+\Delta_2\rp}\lp x_{34}^2\rp^{-\hlf\lp\Delta_3+\Delta_4\rp}\sum_\mcO\la_{12\mcO}\la_{34\mcO}g(u,v;\Delta_1,\Delta_2,\Delta_3,\Delta_4;\ell,\Delta_\mcO),
\end{multline}
while in the other channel we have
\begin{multline}
\left\langle\phi_1(x_1)\phi_2(x_2)\phi_3(x_3)\phi_4(x_4)\right\rangle=\lp\frac{x_{12}^2}{x_{13}^2}\rp^{\hlf\lp\Delta_3-\Delta_2\rp}\lp\frac{x_{24}^2}{x_{12}^2}\rp^{\hlf\lp\Delta_1-\Delta_4\rp}\\
\times\lp x_{14}^2\rp^{-\hlf\lp\Delta_1+\Delta_4\rp}\lp x_{23}^2\rp^{-\hlf\lp\Delta_3+\Delta_2\rp}\sum_\mcO\la_{14\mcO}\la_{32\mcO}g(v,u;\Delta_1,\Delta_4,\Delta_3,\Delta_2;\ell,\Delta_\mcO).
\end{multline}
Comparing the two we learn that
\begin{multline}
0=\sum_\mcO\left\{\vphantom{u^{\hlf\lp\Delta_2+\Delta_4\rp}}\la_{12\mcO}\la_{34\mcO}g(u,v;\Delta_1,\Delta_2,\Delta_3,\Delta_4;\ell,\Delta_\mcO)\right.\\
\left. -\la_{14\mcO}\la_{32\mcO}u^{\hlf\lp\Delta_3+\Delta_4\rp}v^{-\hlf\lp\Delta_2+\Delta_3\rp}g(v,u;\Delta_1,\Delta_4,\Delta_3,\Delta_2;\ell,\Delta_\mcO)\right\}
\end{multline}

This constraint is of limited usefulness when the scalars are all distinct.  A somewhat better case is when $\phi_2$ and $\phi_4$ are identical scalars, in which case we get
\be
0=\sum_\mcO\la_{12\mcO}\la_{32\mcO}\ls g_{\ell;0,0}(u,v)-\lp\frac{u}{v}\rp^{\hlf\lp\Delta_2+\Delta_3\rp}g_{\ell;0,0}(v,u)\rs.
\ee
In practice, it is not easy to extract information from this form either.  Rather, the comparison becomes most powerful (at least in the absence of other information) when the first and third scalars are also identical and the theory is unitary.  In this case we get
\be
0=\sum_\mcO\la_{12\mcO}^2\ls g_{\ell;0,0}(u,v)-\lp\frac{u}{v}\rp^{\hlf\lp\Delta_1+\Delta_2\rp}g_{\ell;0,0}(v,u)\rs.
\ee
Now the coefficients are all positive (since the $\la$s are real in a unitary theory).  For fixed $\Delta_1$ and $\Delta_2$, we can view the quantity in square brackets as a family of functions of $u$ and $v$, parameterized by $\ell$ and $\Delta_\mcO$.  If we choose a functional $F$ (taking functions of $u$ and $v$ and returning a real number; for instance we can act by an arbitrary differential operator and then evaluate at some choice of fixed $u$ and $v$), and apply it to the functions in square brackets, then we get a set of real numbers $F_{\ell,\Delta_\mcO}$, and our constraint simply looks like
\be
0=\sum_\mcO\la_{12\mcO}^2F_{\ell,\Delta_\mcO}.
\ee
A necessary condition for this to have a solution is that the spectrum must include operators $\mcO_1$ and $\mcO_2$, both with nonzero OPE coefficients $\la_{12\mcO}$, such that $F_{\ell_1,\Delta_{\mcO_1}}$ and $F_{\ell_2,\Delta_{\mcO_2}}$ have opposite sign.  The art now is to choose a functional (or better a set of functionals) (given fixed $\Delta_1$ and $\Delta_2$) that splits the space of operators in a useful way.  For example, in the case of all four scalars being identical, it is possible to choose functionals such that $F_{\ell,\Delta_\mcO}\ge 0$ for $\ell>0$ and $\Delta_\mcO$ above the unitarity bound ($\Delta_\mcO>D+\ell-2$), $F_{0,0}=1$ (this correseponds to the identity operator and excludes a trivial spectrum) and $F_{0,\Delta_\mcO}\ge 0$ for $\Delta_\mcO\ge \Delta_c$, with $\Delta_c$ some critical value (that depends on $D$ and $\Delta_\phi$).  In this case, we can conclude that the spectrum must include a scalar operator $\mcO$ which appears in the $\phi_1\phi_2$ OPE and satisfies $(D-2)/2\le\Delta_\mcO\le\Delta_c$ (the first inequality is the unitarity bound for scalar operators other than the identity).

For three scalars and one vector, there is no configuration which is quite as powerful.  For SVSS, we can consider $1\leftrightarrow 3$ exchange, which leads to
\be
\begin{split}
0 & =\sum_\mcO\left\{\la_{34\mcO}\lp\lp\al_{12\mcO}g^{\al\la}_1(u,v;\Delta_1,\Delta_v,\Delta_3,\Delta_4;\ell,\Delta_\mcO)\vphantom{\beta_{12\mcO}g^{\beta\la}_1}\right.\right.\right.\\
& \qquad\left.\left.\left. +\beta_{12\mcO}g^{\beta\la}_1(u,v;\Delta_1,\Delta_v,\Delta_3,\Delta_4;\ell,\Delta_\mcO)\rp k^{(214)}_a+\lp\al_{12\mcO}g^{\al\la}_2(u,v;\Delta_1,\Delta_v,\Delta_3,\Delta_4;\ell,\Delta_\mcO)\vphantom{\beta_{12\mcO}g^{\beta\la}_2}\right.\right.\right.\\
& \qquad\left.\left.\left. +\beta_{12\mcO}g^{\beta\la}_2(u,v;\Delta_1,\Delta_v,\Delta_3,\Delta_4;\ell,\Delta_\mcO)\rp k^{(234)}_a\rp\right.\\
& \qquad\left. -\la_{14\mcO}u^{\hlf\lp\Delta_3+\Delta_4\rp}v^{-\hlf\lp\Delta_v+\Delta_3\rp}\lp\lp\al_{32\mcO}g^{\al\la}_1(v,u;\Delta_3,\Delta_v,\Delta_1,\Delta_4;\ell,\Delta_\mcO)\vphantom{\beta_{32\mcO}g^{\beta\la}_2}\right.\right.\right.\\
& \qquad\left.\left.\left. +\beta_{32\mcO}g^{\beta\la}_1(v,u;\Delta_3,\Delta_v,\Delta_1,\Delta_4;\ell,\Delta_\mcO)\rp k^{(234)}_a+\lp\al_{32\mcO}g^{\al\la}_2(v,u;\Delta_3,\Delta_v,\Delta_1,\Delta_4;\ell,\Delta_\mcO)\vphantom{\beta_{32\mcO}g^{\beta\la}_2}\right.\right.\right.\\
& \qquad\left.\left.\left. +\beta_{32\mcO}g^{\beta\la}_2(v,u;\Delta_3,\Delta_v,\Delta_1,\Delta_4;\ell,\Delta_\mcO)\rp k^{(214)}_a\rp\right\}.
\end{split}
\ee
By grouping the two tensor structures, we get two scalar equations.  Let's write them out just for the case that $\phi_1$ and $\phi_3$ are identical.  We get
\begin{multline}
0=\sum_\mcO\la_{14\mcO}\left\{\al_{12\mcO}\lp g^{\al\la}_1(u,v)-u^{\hlf\lp\Delta_1+\Delta_4\rp}v^{-\hlf\lp\Delta_1+\Delta_v\rp}g^{\al\la}_2(v,u)\rp\right.\\
\left. +\beta_{12\mcO}\lp g^{\beta\la}_1(u,v)-u^{\hlf\lp\Delta_1+\Delta_4\rp}v^{-\hlf\lp\Delta_1+\Delta_v\rp}g^{\beta\la}_2(v,u)\rp\right\},
\end{multline}
and a similar equation where we act on the subscript $p$ of
\be
g^{r\la}_p(u,v)=g^{r\la}_p(u,v;\Delta_1,\Delta_v,\Delta_1,\Delta_4;\ell,\Delta_\mcO),
\ee
by exchanging $1\leftrightarrow 2$.

By considering $2\leftrightarrow 4$ exchange, we would obtain equations relating a sum over $g^{r\la}_p$ blocks of the SVSS correllator with the $g^{\la r}_p$ blocks of SSSV, or we could obtain an equation by considering $1\leftrightarrow 3$ exchange in the SSSV case.

Next we turn to our primary interest in this paper - the case of two scalars and two vectors.

\subsection{SVSV case with generic vectors}

Finally we consider the case of two scalars, in the $1$ and $3$ positions, and two vectors, in the $2$ and $4$ positions.  By comparing the four-point function obtained by taking the OPEs of $\phi_1$ with $v_2$ and $\phi_3$ with $v_4$ to the result obtained by taking $\phi_1$ with $v_4$ and $\phi_3$ with $v_2$ (obtained from the former by a $2\leftrightarrow 4$ exchange), we get
\begin{multline}
\left\{\sum_\mcO\ls\al_{12\mcO}\al_{34\mcO}g^{\al\al}_p+\al_{12\mcO}\beta_{34\mcO}g^{\al\beta}_p+\beta_{12\mcO}\al_{34\mcO}g^{\beta\al}_p+\beta_{12\mcO}\beta_{34\mcO}g^{\beta\beta}_p\rs+\sum_{\mcA}\g_{12\mcA}\g_{34\mcA}g^{\g\g}_p\right\}t^p_{ab}\\
=u^{\hlf\lp\Delta_3+\Delta_4\rp}v^{-\hlf\lp\Delta_2+\Delta_3\rp}\left\{\sum_\mcO\ls\al_{14\mcO}\al_{32\mcO}g^{\al\al}_p+\al_{14\mcO}\beta_{32\mcO}g^{\al\beta}_p+\beta_{14\mcO}\al_{32\mcO}g^{\beta\al}_p+\beta_{14\mcO}\beta_{32\mcO}g^{\beta\beta}_p\rs\right.\\
\left. +\sum_\mcA\g_{14\mcA}\g_{32\mcA}g^{\g\g}_p\right\}t^{\prime\,p}_{ab}.
\end{multline}
Here we have abbreviated all the blocks on the left-hand side as
\be
g^{rs}_p=g^{rs}_p(u,v;\Delta_1,\Delta_2,\Delta_3,\Delta_4;\ell,\Delta_\mcO),\quad g^{\g\g}_p=g^{\g\g}_p(u,v;\Delta_1,\Delta_2,\Delta_3,\Delta_4;k,\Delta_\mcA),
\ee
while on the right-hand side we expand as
\be
g^{rs}_p=g^{rs}_p(v,u;\Delta_1,\Delta_4,\Delta_3,\Delta_2;\ell,\Delta_\mcO),\quad g^{\g\g}_p=g^{\g\g}_p(v,u;\Delta_1,\Delta_4,\Delta_3,\Delta_2;k,\Delta_\mcA).
\ee
we also recall that we use tensor structures
\be
t^0_{ab}=m^{(24)}_{ab},\ t^{11}_{ab}=k^{(214)}_ak^{(412)}_b,\ t^{12}_{ab}=k^{(214)}_ak^{(432)}_b,\ t^{21}_{ab}=k^{(234)}_ak^{(412)}_b,\ t^{22}_{ab}=k^{(234)}_ak^{(432)}_b,
\ee
and the primed tensor structures are obtained by exchanging $2\leftrightarrow 4$ and $a\leftrightarrow b$,
\be
t^{\prime\,0}_{ab}=m^{(24)}_{ab},\ t^{\prime\,11}_{ab}=k^{(214)}_ak^{(412)}_b,\ t^{\prime\,12}_{ab}=k^{(234)}_ak^{(412)}_b,\ t^{\prime\,21}_{ab}=k^{(214)}_ak^{(432)}_b,\ t^{\prime\,22}_{ab}=k^{(234)}_ak^{(432)}_b.
\ee
In other words, $t^{\prime\,0}_{ab}=t^0_{ab}$ and $t^{\prime\,ij}_{ab}=t^{ji}_{ab}$.

Grouping like tensor structures together now gives us five equations on the underlying data of the CFT.  Note that by using (\ref{eq:SymmExchangeGeneralExpression}), we can rewrite the symmetric exchange summands in terms of scalar blocks.


As with the case of four-scalars, the equations are much more constraining for the case where we have two identical scalars and two identical vectors.  In this case, we saw in section \ref{subsec:ExchangeSymmetries} that $g^{rs}_{22}(u,v)=g^{sr}_{11}(u,v)$ and $g^{rs}_p=g^{sr}_p$ for the other $p$.  This results in only three independent bootstrap constraints,
\be
0=\sum_\mcO\ls G_0(u,v)-\lp\frac{u}{v}\rp^{\hlf\lp\Delta_\phi+\Delta_v\rp}G_0(v,u)\rs+\sum_\mcA\g_{\phi v\mcA}^2\ls g_0^{\g\g}(u,v)-\lp\frac{u}{v}\rp^{\hlf\lp\Delta_\phi+\Delta_v\rp}g_0^{\g\g}(v,u)\rs,
\ee
\be
0=\sum_\mcO \ls G_{11}(u,v)-\lp\frac{u}{v}\rp^{\hlf\lp\Delta_\phi+\Delta_v\rp}G_{11}(v,u)\rs+\sum_\mcA\g_{\phi v\mcA}^2\ls g_{11}^{\g\g}(u,v)-\lp\frac{u}{v}\rp^{\hlf\lp\Delta_\phi+\Delta_v\rp}g_{11}^{\g\g}(v,u)\rs,
\ee
and
\be
0=\sum_\mcO \ls G_{12}(u,v)-\lp\frac{u}{v}\rp^{\hlf\lp\Delta_\phi+\Delta_v\rp}G_{21}(v,u)\rs+\sum_\mcA\g_{\phi v\mcA}^2\ls g_{12}^{\g\g}(u,v)-\lp\frac{u}{v}\rp^{\hlf\lp\Delta_\phi+\Delta_v\rp}g_{21}^{\g\g}(v,u)\rs,
\ee
where we have defined
\be
G_p(u,v)=A_\mcO^2\mcD^{--}_pg_{\ell;-1,-1}+A_\mcO B_\mcO\lp\mcD^{-+}_pg_{\ell;-1,1}+\mcD^{+-}_pg_{\ell;1,-1}\rp+B_\mcO^2\mcD^{++}_pg_{\ell;1,1},
\ee
with
\bea
\label{eq:AODef}
A_\mcO &=& \frac{1}{2\lp\Delta_\mcO-1\rp}\lp\al_{\phi v\mcO}+\lp\Delta_\phi-\Delta_v-\Delta_\mcO+\ell+1\rp\frac{\beta_{\phi v\mcO}}{\ell}\rp,\\
\label{eq:BODef}
B_\mcO &=& \frac{1}{2\lp\Delta_\mcO-1\rp}\lp\al_{\phi v\mcO}+\lp\Delta_\phi-\Delta_v+\Delta_\mcO+\ell-1\rp\frac{\beta_{\phi v\mcO}}{\ell}\rp,
\eea
(or $A_\mcO=B_\mcO=\al_{\phi v\mcO}/2(\Delta_\mcO-1)$ for $\ell=0$) and using the notation of (\ref{eq:SymmExchangeGeneralExpression}).

\subsection{SVSV with conserved vectors}

The situation is even more tractable in the case that the identical vectors are in fact conserved currents.  In this case we have $\Delta_1=\Delta_3=\Delta_\phi$ and $\Delta_2=\Delta_4=D-1$.  There are also restrictions on the data of symmetric operator exchange, i.e.\ on which operators $\mcO$ can appear in the $\langle\phi v\mcO\rangle$ three-point function, which were discussed in section \ref{subsec:ConservedVectors}.  Let's split into the cases $\ell=0$ and $\ell>0$.

For $\ell=0$, as reviewed in section \ref{subsec:ConservedVectors}, we can assume that either $\phi$ is neutral under the symmetry, or that is the real part of a complex scalar operator of charge $Q$.  We'll focus on the latter case, and the former case can be recovered by setting $Q=0$.  Thus, there will be a unique scalar operator $\mcO$ that can be exchanged, with $\Delta_\mcO=\Delta_\phi$ and $\al_{\phi v\mcO}=-Q$.  In this case we can split this piece out of the bootstrap constraints, much in the same way that the contribution from exchange of the identity operator is typically split off for the case of identical scalars in the SSSS bootstrap, and move it to the right-hand-side of the constraint equations.

For $\ell>0$ we also have another important result - the relation between $\al_{\phi v\mcO}$ and $\beta_{\phi v\mcO}$ given in (\ref{eq:ConservedVec3pt}).  Since we could in principle have either $\al_{\phi v\mcO}$ or $\beta_{\phi v\mcO}$ vanishing, it is more useful to define a non-vanishing constant $c_\mcO$ related to them by
\be
\al_{\phi v\mcO}=\lp\Delta_\phi-\Delta_\mcO+D+\ell-2\rp c_\mcO,\qquad\beta_{\phi v\mcO}=\lp\Delta_\mcO-\Delta_\phi\rp c_\mcO,
\ee
which ensures that (\ref{eq:ConservedVec3pt}) holds.  In terms of this, (\ref{eq:AODef}) and (\ref{eq:BODef}) become
\bea
A_\mcO &=& -\frac{\lp\Delta_\mcO-\Delta_\phi-\ell\rp\lp\Delta_\mcO-\Delta_\phi+D+\ell-2\rp}{2\ell\lp\Delta_\mcO-1\rp}c_\mcO=:a_\mcO c_\mcO,\\
B_\mcO &=& \ls\frac{\Delta_\mcO-\Delta_\phi}{\ell}-\frac{\lp\Delta_\mcO-\Delta_\phi-\ell\rp\lp\Delta_\mcO-\Delta_\phi+D+\ell-2\rp}{2\ell\lp\Delta_\mcO-1\rp}\rs c_\mcO=:b_\mcO c_\mcO,
\eea
which defines constants $a_\mcO$ and $b_\mcO$ that depend only on the dimensions $\Delta_\phi$ and $\Delta_\mcO$, and then we have $G_p(u,v)=c_\mcO^2\widehat{G}_p(u,v)$, with
\be
\widehat{G}_p(u,v)=a_\mcO^2\mcD^{--}_pg_{\ell;-1,-1}+a_\mcO b_\mcO\lp\mcD^{-+}_pg_{\ell;-1,1}+\mcD^{+-}_pg_{\ell;1,-1}\rp+b_\mcO^2\mcD^{++}_pg_{\ell;1,1}.
\ee

Now we can rewrite the bootstrap constraints as follows,
\begin{multline}
\sum_{\mcO,\ell>0}c_\mcO^2\ls\widehat{G}_0(u,v)-\lp\frac{u}{v}\rp^{\hlf\lp\Delta_\phi+D-1\rp}\widehat{G}_0(v,u)\rs\\
+\sum_\mcA\g_{\phi v\mcA}^2\ls g^{\g\g}_0(u,v)-\lp\frac{u}{v}\rp^{\hlf\lp\Delta_\phi+D-1\rp}g^{\g\g}_0(v,u)\rs\\
=-\frac{Q^2}{4\lp\Delta_\phi-1\rp^2}\ls H_0(u,v)-\lp\frac{u}{v}\rp^{\hlf\lp\Delta_\phi+D-1\rp}H_0(v,u)\rs,
\end{multline}
\begin{multline}
\sum_{\mcO,\ell>0}c_\mcO^2\ls\widehat{G}_{11}(u,v)-\lp\frac{u}{v}\rp^{\hlf\lp\Delta_\phi+D-1\rp}\widehat{G}_{11}(v,u)\rs\\
+\sum_\mcA\g_{\phi v\mcA}^2\ls g^{\g\g}_{11}(u,v)-\lp\frac{u}{v}\rp^{\hlf\lp\Delta_\phi+D-1\rp}g^{\g\g}_{11}(v,u)\rs\\
=-\frac{Q^2}{4\lp\Delta_\phi-1\rp^2}\ls H_{11}(u,v)-\lp\frac{u}{v}\rp^{\hlf\lp\Delta_\phi+D-1\rp}H_{11}(v,u)\rs,
\end{multline}
\begin{multline}
\sum_{\mcO,\ell>0}c_\mcO^2\ls\widehat{G}_{12}(u,v)-\lp\frac{u}{v}\rp^{\hlf\lp\Delta_\phi+D-1\rp}\widehat{G}_{21}(v,u)\rs\\
+\sum_\mcA\g_{\phi v\mcA}^2\ls g^{\g\g}_{12}(u,v)-\lp\frac{u}{v}\rp^{\hlf\lp\Delta_\phi+D-1\rp}g^{\g\g}_{21}(v,u)\rs\\
=-\frac{Q^2}{4\lp\Delta_\phi-1\rp^2}\ls H_{12}(u,v)-\lp\frac{u}{v}\rp^{\hlf\lp\Delta_\phi+D-1\rp}H_{21}(v,u)\rs,
\end{multline}
where we have defined
\be
H_p(u,v)=\mcD^{--}_pg_{0;-1,-1}+\mcD^{-+}_pg_{0;-1,1}+\mcD^{+-}_pg_{0;1,-1}+\mcD^{++}_pg_{0;1,1},
\ee
for the $\ell=0$ exchange, and the scalar blocks in this last expression are evaluated with $\Delta_1=\Delta_3=\Delta_\mcO=\Delta_\phi$ and $\Delta_2=\Delta_4=D-1$.  In these equations the left-hand sides are written as sums of functions which depend only on the conformal representations (i.e.\ only on $D$, $\Delta_\phi$, and either $\Delta_\mcO$ and $\ell$ or $\Delta_\mcA$ and $k$) but not on the three-point function coefficients, multiplied by manifestly positive real numbers (for unitary theories).  This is a situation in which we can profitably use the techniques developed in the scalar bootstrap literature to obtain bounds on the spectrum of our theory. The (numerical) analysis of the consistent space of theories with conserved currents will be carried out in future work. 

\section{Conclusions}
\label{sec:Conclusions}

Our primary goal in this paper was to set up, in explicit detail, the bootstrap equations for a four-point function of two scalars and two vectors (abbreviated $\langle SVSV\rangle$) in a general CFT in arbitrary dimension.  To compute this four-point function, one performs OPEs between a scalar and a vector, producing either traceless symmetric tensors $\mcO$ or mixed symmetry tensors $\mcA$ (indeed this is the simplest four-point function that includes exchange of a mixed symmetry operator), and evaluates the resulting two-point functions.  The bootstrap compares the two different channels for these OPEs.  The contribution of a primary operator, either $\mcO$ or $\mcA$, and all its descendants to the four-point function is captured by a conformal block $g^{rs}_p$, where the indices $r$ and $s$ refer to possible tensor structures in either the three-point function $\langle SV\mcO\rangle$ or $\langle SV\mcA\rangle$, while $p$ refers to $\langle SVSV\rangle$.  For general scalars and vectors, $\langle SV\mcO\rangle$ has two structures and $\langle SV\mcA\rangle$ has one, while $\langle SVSV\rangle$ has five structures, so there are really $(2^2+1^2)\times 5=25$ different conformal block functions, and we have computed them all in section \ref{subsubsec:SVSVBlocks}.  We presented the explicit form of conformal blocks associated to the exchange of symmetric traceless operators $\mcO$, by applying suitable differential operators to scalar blocks using ideas from \cite{Costa2011}.  Furthermore we showed that the conformal blocks of the mixed operator $\mcA$ can also be written as differential operators acting on scalar blocks, if we allow them to have shifted spins $k$,$k+1$,$k+2$. We also found that writing the SVSV blocks in terms of lower spin ones (SVSS,SSSV) makes the expressions simpler. This could potentially be important for making the computation of higher spin blocks (four vectors,four stress tensors) more tractable.  For identical scalars and/or identical vectors there are many relations between the blocks which we have spelled out.  

Next we set up the bootstrap equations, starting with the most general $\langle SVSV\rangle$ (or $\langle SSSV\rangle$ or $\langle SVSS\rangle$) four-point function, and then specializing to identical scalars and identical vectors.  Finally we restrict to the particularly simple case of conserved vectors, for which a unique scalar $\mcO$ can contribute to $\langle SV\mcO\rangle$, while for $\mcO$ with $\ell>0$ the two possible structures in $\langle SV\mcO\rangle$ collapse to one.  This latter property ensured that the bootstrap equations resemble those of identical scalars with global symmetry \cite{Rattazzi2011a}, i.e. a sum of vector functions with positive coefficients. Thus one can exploit the already developed techniques for bootstrapping identical scalars\footnote{Bootstrapping non-conserved vectors could in principle be done using semi-definite techniques\cite{Polanda,Kos2014a,Kos2014}.}.

As a check of our results, we have verified numerically that our blocks (as well as the SSSV and SVSS blocks which we construct as intermediate steps) exhibit the correct behavior under exchanges, though we have omitted the details in the present work, preferring to defer all numerical details to a follow up paper.  It would be worthwhile to develop further checks on our results, by comparing with cases (such as generalized free CFTs) where the full four-point function can be computed directly.  It would also be interesting to understand our results in the context of holography, and particularly to match our results with the bulk geometric quantities studied in~\cite{Hijano:2015zsa}.

Going forward, there are two natural extensions to this work.  The first is to actually apply our formalism to seek, both numerically and analytically, bounds on the data of some general class of CFTs.  It would be particularly interesting to derive results for conserved vectors, which could constrain theories with continuous global symmetries and their spectra of scalars charged under the symmetry.

The second direction heading forward is to use the techniques developed in this work to set up the bootstrap for even more complicated four-point functions.  The next one to attempt is probably four vectors, either conserved or not, and this case should be tractable by hand.  Another possibility would be two scalars and two spin-two tensors, especially for the case where the tensors are conserved (e.g.\ stress-energy tensors).  Finally, the most ambitious goal would be to bootstrap the four-point function of conserved stress-tensors (see for example the discussion in~\cite{Dymarsky:2013wla}).  This is probably not feasible using our current techniques ``by hand", but might be possible if we can computerize the necessary steps.


\acknowledgments

D.~R.\ was supported by the grants PHY-1214344 and NSF Focused Research Grant DMS-1159404 and by the George P. and Cynthia W. Mitchell Institute for Fundamental Physics and Astronomy. F. R.-B.\ was supported by the Mexican Consejo Nacional de Ciencia y Tecnolog\'{i}a (CONACyT scholarship 382219).
We thank S.~El-Showk, S.~Hellerman, D.~Hofman, H.~Osborn, E.~Perlmutter, S.~Rychkov, and S.~Sethi for useful discussions.  D.~R.\ would like to thank the Simons Center for Geometry and Physics for hospitality during the final days of preparation of this work.

\appendix

\section{Building blocks and identities}
\label{app:BuildingBlocks}

In this paper, the physical space is flat $\R^D$ with Euclidean signature.  Indices $a$, $b$, etc., are raised and lowered with the Kronecker delta $\d_{ab}$.  For two vectors $x_i^a$ and $x_j^a$, we define
\be
x_{ij}^a=x_i^a-x_j^a.
\ee

Two particular structures play a significant role in constructing correlators in a conformal field theory,
\be
k^{(ijk)}_a=\frac{x_{ij}^2\lp x_{ik}\rp_a-x_{ik}^2\lp x_{ij}\rp_a}{\lp x_{ij}^2x_{ik}^2x_{jk}^2\rp^{1/2}},
\ee
where $x_i$, $x_j$, and $x_k$ are assumed to be distinct points, and
\be
m^{(ij)}_{ab}=\d_{ab}-\frac{2}{x_{ij}^2}\lp x_{ij}\rp_a\lp x_{ij}\rp_b,
\ee
where again $x_i$ and $x_j$ are distinct.

From the fact that $x_{ij}+x_{jk}=x_{ik}$, we can show that
\be
\label{eq:kLinComb}
k^{(ik\ell)}_a=-\sqrt{\frac{x_{i\ell}^2x_{jk}^2}{x_{ij}^2x_{k\ell}^2}}k^{(ijk)}_a+\sqrt{\frac{x_{ik}^2x_{j\ell}^2}{x_{ij}^2x_{k\ell}^2}}k^{(ij\ell)}_a
\ee

Using the basic identity that
\be
x_{ij}\cdot x_{k\ell}=\hlf\lp -x_{ik}^2+x_{i\ell}^2+x_{jk}^2-x_{j\ell}^2\rp,
\ee
we can prove identities
\be
k^{(ijk)}\cdot k^{(i\ell m)}=\hlf\lp x_{ij}^2x_{ik}^2x_{i\ell}^2x_{im}^2x_{jk}^2x_{\ell m}^2\rp^{-1/2}\lp -x_{ij}^2x_{i\ell}^2x_{km}^2+x_{ij}^2x_{im}^2x_{k\ell}^2+x_{ik}^2x_{i\ell}^2x_{jm}^2-x_{ik}^2x_{im}^2x_{j\ell}^2\rp,
\ee
\be
m^{(ij)}_{ab}k^{(jk\ell)\,b}=-\sqrt{\frac{x_{ik}^2x_{j\ell}^2}{x_{ij}^2x_{k\ell}^2}}k^{(ijk)}_a+\sqrt{\frac{x_{i\ell}^2x_{jk}^2}{x_{ij}^2x_{k\ell}^2}}k^{(ij\ell)}_a,
\ee
and
\be
\d^{cd}m^{(ik)}_{ac}m^{(kj)}_{db}=m^{(ij)}_{ab}-2k^{(ijk)}_ak^{(jik)}_b.
\ee
As special cases of these fomulae, we have
\be
\lp k^{(ijk)}\rp^2=1,\qquad m^{(ij)}_{ab}k^{(jik)\,b}=k^{(ijk)}_a,\qquad \d^{cd}m^{(ij)}_{ac}m^{(ji)}_{db}=\d_{ab}.
\ee

One more useful identity is
\be
\label{eq:pkIdentity}
\frac{\p}{\p x_k^a}k^{(ijk)}_b=-\sqrt{\frac{x_{ij}^2}{x_{ik}^2x_{jk}^2}}\lp m^{(ki)}_{ab}+k^{(kij)}_ak^{(ijk)}_b\rp.
\ee

\section{Lorentz representation projectors}
\label{app:Projectors}

We will be grouping tensor operators by their representations under $\SO(D)$.  There is a large body of work on irreducible representations of $\SO(D)$ (for instance see the nice discussion in~\cite{Costa2014} and references therein), but we really don't need the full power of this theory for the current work.

Consider a tensor with $n$ indices.  It must transform as a sub-representation of the tensor product ${\mathbf{D}}^{\otimes n}$ of $n$ copies of the $D$-dimensional vector representation.  To distinguish the different irreducible representations $I$ which appear in the decomposition of ${\mathbf{D}}^{\otimes n}$, we can use projectors, $\Pi^{I\,a_1\cdots a_n}_{b_1\cdots b_n}$.  Being projectors, these must satisfy
\be
\Pi^{I\,a_1\cdots a_n}_{c_1\cdots c_n}\Pi^{J\,c_1\cdots c_n}_{b_1\cdots b_n}=\d^{IJ}\Pi^{I\,a_1\cdots a_n}_{b_1\cdots b_n}.
\ee
The projectors are built exclusively with Kronecker deltas $\d^{a_i}_{b_j}$, $\d^{a_ia_j}$, or $\d_{b_ib_j}$.

Below, we will need the projectors for the totally symmetric traceless representation of spin $\ell$ (i.e.\ with $\ell$ indices), and also for a mixed symmetry representation with $k+2$ indices which we will describe below.

\subsection{Totally symmetric}
\label{subapp:PiO}

Consider first the projector onto totally symmetric traceless representations, $\Pi^{(\ell)\,a_1\cdots a_\ell}_{b_1\cdots b_\ell}$.  By the symmetries of the problem, it must have the form
\be
\Pi^{(\ell)\,a_1\cdots a_\ell}_{b_1\cdots b_\ell}=A_0\d^{(a_1}_{(b_1}\cdots\d^{a_\ell)}_{b_\ell)}+\sum_{i=1}^{\lfloor\ell/2\rfloor}A_i\d^{(a_1a_2}\cdots\d^{a_{2i-1}a_{2i}}\d_{(b_1b_2}\cdots\d_{b_{2i-1}b_{2i}}\d^{a_{2i+1}}_{b_{2i+1}}\cdots\d^{a_\ell)}_{b_\ell)},
\ee
where the $A_i$ are constants.  For $\ell\ge 2$, taking the trace with $\d^{b_{\ell-1}b_\ell}$ we get
\begin{multline}
A_0\d^{(a_1}_{(b_1}\cdots\d^{a_{\ell-2}}_{b_{\ell-2})}\d^{a_{\ell-1}a_\ell)}\\
+\sum_{i=1}^{\lfloor\ell/2\rfloor}A_i\left\{\frac{\lp\ell-2i\rp\lp\ell-1-2i\rp}{\ell\lp\ell-1\rp}\d^{(a_1a_2}\cdots\d^{a_{2i+1}a_{2i+2}}\d_{(b_1b_2}\cdots\d_{b_{2i-1}b_{2i}}\d^{a_{2i+3}}_{b_{2i+1}}\cdots\d^{a_\ell)}_{b_{\ell-2})}\right.\\
\left. +\frac{2i\lp D+2\ell-2i-2\rp}{\ell\lp\ell-1\rp}\d^{(a_1a_2}\cdots\d^{a_{2i-1}a_{2i}}\d_{(b_1b_2}\cdots\d_{b_{2i-3}b_{2i-2}}\d^{a_{2i+1}}_{b_{2i-1}}\cdots\d^{a_\ell)}_{b_{\ell-2})}\right\}.
\end{multline}
Thus tracelessness requires
\be
A_i=-\frac{\lp\ell+2-2i\rp\lp\ell+1-2i\rp}{2i\lp D+2\ell-2-2i\rp}A_{i-1},\qquad\mathrm{for\ }1\le i\le\lfloor\ell/2\rfloor,
\ee
or
\be
A_i=\lp -1\rp^i\frac{\ell!\G(\frac{D}{2}+\ell-i-1)}{2^{2i}\lp\ell-2i\rp!i!\G(\frac{D}{2}+\ell-1)}A_0.
\ee
Finally, we can fix $A_0$ by the condition that $\Pi^2=\Pi$, i.e.
\be
\Pi^{(\ell)\,a_1\cdots a_\ell}_{c_1\cdots c_\ell}\Pi^{(\ell)\,c_1\cdots c_\ell}_{b_1\cdots b_\ell}=\Pi^{(\ell)\,a_1\cdots a_\ell}_{b_1\cdots b_\ell}.
\ee
In fact we only need to check the leading terms, not the subleading traceless terms, because the latter can't contribute to the former when we square.  Then since
\be
\d^{(a_1}_{(c_1}\cdots \d^{a_\ell)}_{c_\ell)}\d^{(c_1}_{(b_1}\cdots\d^{c_\ell)}_{b_\ell)}=\d^{(a_1}_{(b_1}\cdots\d^{a_\ell)}_{b_\ell)},
\ee
we require $A_0^2=A_0$, and hence we should take $A_0=1$, and we can write
\begin{multline}
\label{eq:STPi}
\Pi^{(\ell)\,a_1\cdots a_\ell}_{b_1\cdots b_\ell}=\sum_{i=0}^{\lfloor\ell/2\rfloor}\lp -\frac{1}{4}\rp^i\frac{\ell!\,\G(\frac{D}{2}+\ell-i-1)}{i!\lp\ell-2i\rp!\,\G(\frac{D}{2}+\ell-1)}\\
\times\d^{(a_1a_2}\cdots\d^{a_{2i-1}a_{2i}}\d_{(b_1b_2}\cdots\d_{b_{2i-1}b_{2i}}\d^{a_{2i+1}}_{b_{2i+1}}\cdots\d^{a_\ell)}_{b_\ell)}.
\end{multline}

These projectors obey certain recursion relations.  With the explicit expressions for coefficients above, one can show that
\be
\label{eq:STPiRecursion}
\Pi^{(\ell)\,a_1\cdots a_\ell}_{b_1\cdots b_\ell}=\d^{(a_1}_{(b_1}\Pi^{(\ell-1)\,a_2\cdots a_\ell)}_{b_2\cdots b_\ell)}-\frac{\lp D+\ell-4\rp\lp\ell-1\rp}{\lp D+2\ell-6\rp\lp D+2\ell-4\rp}\d^{(a_1a_2}\d_{(b_1b_2}\Pi^{(\ell-2)\,a_3\cdots a_\ell)}_{b_3\cdots b_\ell)}.
\ee

Now we can define polynomials $p_{D,\ell}(t)$ by
\be
X_{a_1}\cdots X_{a_\ell}\Pi^{(\ell)\,a_1\cdots a_\ell}_{b_1\cdots b_\ell}Y^{b_1}\cdots Y^{b_\ell}=\lp X^2Y^2\rp^{\ell/2}p_{D,\ell}(t),\qquad t=\frac{X\cdot Y}{\sqrt{X^2Y^2}}.
\ee
Explicitly, using (\ref{eq:STPi}), we have
\be
p_{D,\ell}(t)=\sum_{i=0}^{\lfloor\ell/2\rfloor}\lp -\frac{1}{4}\rp^i\frac{\ell!\G(\frac{D}{2}+\ell-i-1)}{i!(\ell-2i)!\G(\frac{D}{2}+\ell-1)}t^{\ell-2i}.
\ee
These are related to the more familiar Gegenbauer polynomials by
\be
p_{D,\ell}(t)=\frac{\ell!\G(\frac{D}{2}-1)}{2^\ell\G(\frac{D}{2}+\ell-1)}C^{(\frac{D}{2}-1)}_\ell(t).
\ee
They obey a simple differential identity,
\be
\label{eq:DRecursion}
p_{D,\ell}'(t)=\ell p_{D+2,\ell-1}(t),
\ee
and also
\be
p_{D+2,\ell}(t)=p_{D,\ell}(t)+\frac{\ell\lp\ell-1\rp}{\lp D+2\ell-2\rp\lp D+2\ell-4\rp}p_{D+2,\ell-2}(t).
\ee
We can also prove a recursion relation for fixed $D$ from (\ref{eq:STPiRecursion}),
\be
\label{eq:plRecursion}
p_{D,\ell}(t)=tp_{D,\ell-1}(t)-\frac{\lp D+\ell-4\rp\lp\ell-1\rp}{\lp D+2\ell-4\rp\lp D+2\ell-6\rp}p_{D,\ell-2}(t).
\ee
The first few of these polynomials are
\be
p_0=1,\qquad p_1=t,\qquad p_2=t^2-\frac{1}{D},\qquad p_3=t^3-\frac{3}{D+2}t,\non
\ee
\be
p_4=t^4-\frac{6}{D+4}t^2+\frac{3}{(D+2)(D+4)},\quad p_5=t^5-\frac{10}{D+6}t^3+\frac{15}{(D+4)(D+6)}t.
\ee

We will also need the result of the following partial contractions of $\Pi^{(\ell)}$,
\be
\label{eq:SymmetricPartialContraction}
\begin{split}
& X_{c_1}\cdots X_{c_{\ell-1}}\Pi^{(\ell)\,ac_1\cdots c_{\ell-1}}_{bd_1\cdots d_{\ell-1}}Y^{d_1}\cdots Y^{d_{\ell-1}}=\frac{1}{\ell^2}\frac{\p}{\p X_a}\frac{\p}{\p Y^b}\ls\lp X^2Y^2\rp^{\ell/2}p_\ell(t)\rs\\
& \quad =\frac{1}{\ell^2}\lp X^2Y^2\rp^{\frac{\ell-1}{2}}\ls\d^a_b\p_t+\lp\frac{X^aX_b}{X^2}+\frac{Y^aY_b}{Y^2}\rp\lp\lp\ell-1\rp\p_t-t\p_t^2\rp\right.\\
& \qquad\left. +\frac{X^aY_b}{\sqrt{X^2Y^2}}\lp\ell^2-\lp 2\ell-1\rp t\p_t+t^2\p_t^2\rp+\frac{Y^aX_b}{\sqrt{X^2Y^2}}\p_t^2\rs p_\ell(t)\\
& \quad =\frac{1}{\ell}\lp X^2Y^2\rp^{\frac{\ell-1}{2}}\ls\d^a_bp_{D+2,\ell-1}(t)+\lp\ell-1\rp\lp\frac{X^aX_b}{X^2}+\frac{Y^aY_b}{Y^2}\rp\lp p_{D+2,\ell-1}(t)-tp_{D+4,\ell-2}(t)\rp\right.\\
& \qquad\left. +\frac{X^aY_b}{\sqrt{X^2Y^2}}\lp\ell p_{D,\ell}(t)-\lp 2\ell-1\rp tp_{D+2,\ell-1}(t)+\lp\ell-1\rp t^2p_{D+4,\ell-2}(t)\rp\right.\\
& \qquad\left. +\lp\ell-1\rp\frac{Y^aX_b}{\sqrt{X^2Y^2}}p_{D+4,\ell-2}(t)\rs.
\end{split}
\ee

\subsection{Mixed symmetry}
\label{subapp:PiA}

Now we would like to find projectors onto the mixed symmetry representations that we need for the scalar-vector bootstrap.  Recall that these tensors are antisymmetric in their first two indices, totally symmetric in their remaining $k$ indices, they vanish when antisymmetrized over any three indices (this condition is trivial unless the three are the first two indices plus one more), and are completely traceless.  We will write the corresponding projectors $\wtPi^{(k)\,a_1a_2b_1\cdots b_k}_{c_1c_2d_1\cdots d_k}$ with tildes to distinguish from the totally symmetric case considered above.

For $k=0$, the only index structure compatible with antisymmetry is
\be
\widetilde{\Pi}^{(0)\,a_1a_2}_{c_1c_2}=A_0\lp\d^{a_1}_{c_1}\d^{a_2}_{c_2}-\d^{a_1}_{c_2}\d^{a_2}_{c_1}\rp.
\ee
Imposing $\widetilde{\Pi}^2=\widetilde{\Pi}$ then implies $A_0=1/2$.

For $k=1$, there are three terms compatible with the antisymmetry in the $a_i$ and the $c_i$,
\begin{multline}
\widetilde{\Pi}^{(1)\,a_1a_2b}_{c_1c_2d}=A_0\lp\d^{a_1}_{c_1}\d^{a_2}_{c_2}-\d^{a_1}_{c_2}\d^{a_2}_{c_1}\rp\d^b_d+B_0\lp\d^{a_1}_{c_1}\d^{a_2}_d\d^b_{c_2}-\d^{a_1}_{c_2}\d^{a_2}_d\d^b_{c_1}-\d^{a_1}_d\d^{a_2}_{c_1}\d^b_{c_2}+\d^{a_1}_d\d^{a_2}_{c_2}\d^b_{c_1}\rp\\
+C_1\lp\d^{a_1b}\d_{c_1d}\d^{a_2}_{c_2}-\d^{a_1b}\d_{c_2d}\d^{a_2}_{c_1}-\d^{a_2b}\d_{c_1d}\d^{a_1}_{c_2}+\d^{a_2b}\d_{c_2d}\d^{a_1}_{c_1}\rp.
\end{multline}
Demanding that this vanish on antisymetrizing $[a_1a_2b]$ leads to the constraint
\be
2A_0-4B_0=0,
\ee
while demanding that it vanishes when we trace with $\d^{c_2d}$ gives
\be
A_0+B_0+\lp D-1\rp C_1=0.
\ee
Finally, demanding that $\widetilde{\Pi}^2=\widetilde{\Pi}$ requires
\be
2A_0^2+4B_0^2=A_0,\quad 4A_0B_0-2B_0^2=B_0,\quad\mathrm{and}\quad 4A_0C_1+4B_0C_1+2\lp D-1\rp C_1^2=C_1.
\ee
The unique non-vanishing solution to these constraints is that
\be
A_0=\frac{1}{3},\qquad B_0=\frac{1}{6},\qquad C_1=-\frac{1}{2\lp D-1\rp},
\ee
so
\begin{multline}
\widetilde{\Pi}^{(1)\,a_1a_2b}_{c_1c_2d}=\frac{1}{3}\lp\d^{a_1}_{c_1}\d^{a_2}_{c_2}-\d^{a_1}_{c_2}\d^{a_2}_{c_1}\rp\d^b_d+\frac{1}{6}\lp\d^{a_1}_{c_1}\d^{a_2}_d\d^b_{c_2}-\d^{a_1}_{c_2}\d^{a_2}_d\d^b_{c_1}-\d^{a_1}_d\d^{a_2}_{c_1}\d^b_{c_2}+\d^{a_1}_d\d^{a_2}_{c_2}\d^b_{c_1}\rp\\
-\frac{1}{2\lp D-1\rp}\lp\d^{a_1b}\d_{c_1d}\d^{a_2}_{c_2}-\d^{a_1b}\d_{c_2d}\d^{a_2}_{c_1}-\d^{a_2b}\d_{c_1d}\d^{a_1}_{c_2}+\d^{a_2b}\d_{c_2d}\d^{a_1}_{c_1}\rp.
\end{multline}

For $k>1$, the following structure is the most general consistent with antisymmetry of the $a_i$ and $c_i$, symmetry of the $b_i$ and $d_i$, and symmetry between upper and lower indices,
\be
\label{eq:GeneralMSPi}
\begin{split}
& \widetilde{\Pi}^{(k)\,a_1a_2b_1\cdots b_k}_{c_1c_2d_1\cdots d_k}\\
& \ =\sum_{i=0}^{\lfloor k/2\rfloor}A_i\lp\d^{a_1}_{c_1}\d^{a_2}_{c_2}-\d^{a_1}_{c_2}\d^{a_2}_{c_1}\rp\d^{(b_1b_2}\cdots\d^{b_{2i-1}b_{2i}}\d_{(d_1d_2}\cdots\d_{d_{2i-1}d_{2i}}\d^{b_{2i+1}}_{d_{2i+1}}\cdots\d^{b_k)}_{d_k)}\\
& \quad+\sum_{i=0}^{\lfloor(k-1)/2\rfloor}B_i\lp\d^{a_1}_{c_1}\d^{a_2}_{(d_1}\d^{(b_1}_{|c_2|}-\d^{a_1}_{c_2}\d^{a_2}_{(d_1}\d^{(b_1}_{|c_1|}-\d^{a_1}_{(d_1}\d^{a_2}_{|c_1}\d^{(b_1}_{c_2|}+\d^{a_1}_{(d_1}\d^{a_2}_{|c_2}\d^{(b_1}_{c_1|}\rp\\
& \qquad\times\d^{b_2b_3}\cdots\d^{b_{2i}b_{2i+1}}\d_{d_2d_3}\cdots\d_{d_{2i}d_{2i+1}}\d^{b_{2i+2}}_{d_{2i+2}}\cdots\d^{b_k)}_{d_k)}\\
& \quad +\sum_{i=1}^{\lfloor(k+1)/2\rfloor}C_i\lp\d^{a_1(b_1}\d_{c_1(d_1}\d^{|a_2|}_{|c_2|}-\d^{a_1(b_1}\d_{c_2(d_1}\d^{|a_2|}_{|c_1|}-\d^{a_2(b_1}\d_{c_1(d_1}\d^{|a_1|}_{|c_2|}+\d^{a_2(b_1}\d_{c_2(d_1}\d^{|a_1|}_{|c_1|}\rp\\
& \qquad\times\d^{b_2b_3}\cdots\d^{b_{2i-2}b_{2i-1}}\d_{d_2d_3}\cdots\d_{d_{2i-2}d_{2i-1}}\d^{b_{2i}}_{d_{2i}}\cdots\d^{b_k)}_{d_k)}\\
& \quad +\sum_{i=1}^{\lfloor k/2\rfloor}D_i\lp\d^{a_1(b_1}\d_{c_1(d_1}\d^{|a_2|}_{d_2}\d^{b_2}_{|c_2|}-\d^{a_1(b_1}\d_{c_2(d_1}\d^{|a_2|}_{d_2}\d^{b_2}_{|c_1|}-\d^{a_2(b_1}\d_{c_1(d_1}\d^{|a_1|}_{d_2}\d^{b_2}_{|c_2|}\right.\\
& \qquad\left.+\d^{a_2(b_1}\d_{c_2(d_1}\d^{|a_1|}_{d_2}\d^{b_2}_{|c_1|}\rp\d^{b_3b_4}\cdots\d^{b_{2i-1}b_{2i}}\d_{d_3d_4}\cdots\d_{d_{2i-1}d_{2i}}\d^{b_{2i+1}}_{d_{2i+1}}\cdots\d^{b_k)}_{d_k)}\\
& \quad +\sum_{i=1}^{\lfloor k/2\rfloor}E_i\lp\d^{a_1(b_1}\d_{(d_1d_2}\d^{|a_2|}_{|c_1}\d^{b_2}_{c_2|}-\d^{a_1(b_1}\d_{(d_1d_2}\d^{|a_2|}_{|c_2}\d^{b_2}_{c_1|}-\d^{a_2(b_1}\d_{(d_1d_2}\d^{|a_1|}_{|c_1}\d^{b_2}_{c_2|}\right.\\
& \qquad\left.+\d^{a_2(b_1}\d_{(d_1d_2}\d^{|a_1|}_{|c_2}\d^{b_2}_{c_1|}+\d^{(b_1b_2}\d_{c_1(d_1}\d^{|a_1}_{|c_2|}\d^{a_2|}_{d_2}-\d^{(b_1b_2}\d_{c_2(d_1}\d^{|a_1}_{|c_1|}\d^{a_2|}_{d_2}-\d^{(b_1b_2}\d_{c_1(d_1}\d^{|a_1}_{d_2}\d^{a_2|}_{|c_2|}\right.\\
& \qquad\left.+\d^{(b_1b_2}\d_{c_2(d_1}\d^{|a_1}_{d_2}\d^{a_2|}_{|c_1|}\rp\d^{b_3b_4}\cdots\d^{b_{2i-1}b_{2i}}\d_{d_3d_4}\cdots\d_{d_{2i-1}d_{2i}}\d^{b_{2i+1}}_{d_{2i+1}}\cdots\d^{b_k)}_{d_k)}.
\end{split}
\ee
Demanding this vanish when we antisymmetrize over $[a_1a_2b_1]$, when we trace with $\d^{d_{k-1}d_k}$, and when we trace with $\d^{c_2d_k}$ fixes everything up to one constant $A_0$ which can then be fixed by the condition that $\widetilde{\Pi}^2=\widetilde{\Pi}$.  The result is that
\bea
B_i &=& \frac{k-2i}{2}A_i,\\
C_i &=& -\frac{k-2i+2}{D+k-2}\ls\frac{i\lp D+k-1\rp}{D+2k-2i}+\hlf\rs A_{i-1},\\
D_i &=& \frac{i\lp D+2k\rp}{D+k-2}A_i,\\
E_i &=& -iA_i,
\eea
while the $A_i$ are given by
\be
A_0=\frac{1}{k+2},
\ee
and the recursion
\be
A_i=-\frac{\lp k-2i+1\rp\lp k-2i+2\rp}{2i\lp D+2k-2i\rp}A_{i-1},
\ee
solved by
\be
A_i=\lp -\frac{1}{4}\rp^i\frac{k!\,\G(\frac{D}{2}+k-i)}{\lp k+2\rp i!\lp k-2i\rp!\,\G(\frac{D}{2}+k)}.
\ee

For $D\le 4$, the story so far is not quite complete.

In $D=2$, these mixed symmetry tensors labeled by $k$ are equivalent to spin-$k$ symmetric traceless tensors, with the map
\be
\mcA_{a_1a_2b_1\cdots b_k}=\e_{a_1a_2}\mcO_{b_1\cdots b_k},\qquad\mcO_{a_1\cdots a_k}=\hlf\e^{b_1b_2}\mcA_{b_1b_2a_1\cdots a_k}.
\ee

In $D=3$ similarly, there is an isomorphism between mixed symmetry labeled by $k$ and traceless symmetric of spin $k+1$, via
\be
\mcA_{a_1a_2b_1\cdots b_k}=\e_{a_1a_2}^{\hph{a_1a_2}c}\mcO_{b_1\cdots b_kc}+\frac{k}{2}\lp\e_{a_1(b_1}^{\hph{a_1(b_1}c}\mcO_{|a_2|b_2\cdots b_k)c}-\e_{a_2(b_1}^{\hph{a_2(b_1}c}\mcO_{|a_1|b_2\cdots b_k)c}\rp,
\ee
and
\be
\mcO_{a_1\cdots a_{k+1}}=\frac{1}{k+2}\e_{(a_1}^{\hph{(a_1}c_1c_2}\mcA_{|c_1c_2|a_2\cdots a_{k+1})}.
\ee

Finally, in $D=4$ we don't have to worry about any isomorphisms of this sort, but we instead need to recognize that our mixed symmetry representations are in fact reducible.  To split the two pieces apart, we can define
\be
\Pi^{(\pm)\,a_1a_2}_{b_1b_2}=\frac{1}{4}\lp\d^{a_1}_{b_1}\d^{a_2}_{b_2}-\d^{a_1}_{b_2}\d^{a_2}_{b_1}\pm\e^{a_1a_2}_{\hph{a_1a_2}b_1b_2}\rp,
\ee
and then define
\be
\widetilde{\Pi}^{(k\pm)\,a_1a_2b_1\cdots b_k}_{c_1c_2d_1\cdots d_k}=\Pi^{(\pm)\,a_1a_2}_{e_1e_2}\Pi^{(\pm)\,f_1f_2}_{c_1c_2}\widetilde{\Pi}^{(k)\,e_1e_2b_1\cdots b_k}_{f_1f_2d_1\cdots d_k}.
\ee

As in the symmetric case, we will need to consider the result of contracting these projectors with vectors $X$ and $Y$, so we consider the expression
\be
\label{eq:wtPiContraction}
X_{c_1}\cdots X_{c_{k+1}}\wtPi^{(k)\,ac_1\cdots c_{k+1}}_{bd_1\cdots d_{k+1}}Y^{d_1}\cdots Y^{d_{k+1}}.
\ee
The free indices $a$ and $b$ can only be carried by a Kronecker delta $\d^a_b$ or by the vectors $X^a$ and $Y^a$.  Moreover, the expression must be symmetric under simultaneous interchange of $X$ with $Y$ and $a$ with $b$, and it must be identically zero when we contract with $X_a$ or with $Y^b$.  These conditions imply that it must have the form
\begin{multline}\label{eq:mixedpartialcontraction}
X_{c_1}\cdots X_{c_{k+1}}\wtPi^{(k)\,ac_1\cdots c_{k+1}}_{bd_1\cdots d_{k+1}}Y^{d_1}\cdots Y^{d_{k+1}}\\
=\lp X^2Y^2\rp^{\frac{k+1}{2}}\ls\lp -\d^a_b+\frac{X^aX_b}{X^2}+\frac{Y^aY_b}{Y^2}-\frac{X^aY_b}{\sqrt{X^2Y^2}}t\rp f_{k-1}(t)+\lp -\d^a_bt+\frac{Y^aX_b}{\sqrt{X^2Y^2}}\rp g_k(t)\rs,
\end{multline}
for some polynomials $f_{k-1}(t)$ and $g_k(t)$ of degree $k-1$ and $k$ respectively, with $t=X\cdot Y/\sqrt{X^2Y^2}$ as before.  These polynomials can be determined by explicit contraction of (\ref{eq:GeneralMSPi}) and use of the solutions for coefficients determined above.  The result is
\be
f_{k-1}(t)=\hlf\sum_{i=0}^{\lfloor\frac{k-1}{2}\rfloor}\lp -\frac{1}{4}\rp^i\frac{k!\G(\frac{D}{2}+k-i)}{i!\lp k-2i-1\rp!\lp D+k-2\rp\G(\frac{D}{2}+k)}t^{k-2i-1},
\ee
and
\be
g_k(t)=-\hlf\sum_{i=0}^{\lfloor\frac{k}{2}\rfloor}\lp -\frac{1}{4}\rp^i\frac{k!\lp D+k-2i-2\rp\G(\frac{D}{2}+k-i)}{i!\lp k-2i\rp!\lp D+k-2\rp\G(\frac{D}{2}+k)}t^{k-2i}.
\ee
Actually, these can be recast in terms of the polynomials $p_{D,\ell}(t)$ which we defined in the symmetric case (and which are related to the usual Gegenbauer polynomials),
\bea
\label{eq:fkExpression}
f_{k-1}(t) &=& \frac{1}{2\lp k+1\rp\lp D+k-2\rp}p_{D,k+1}''(t)=\frac{k}{2\lp D+k-2\rp}p_{D+4,k-1}(t),\\
g_k(t) &=& -\frac{1}{2\lp k+1\rp\lp D+k-2\rp}\lp\lp D-2\rp p_{D,k+1}'(t)+tp_{D,k+1}''(t)\rp\non\\
\label{eq:gkExpression}
&=& -\frac{1}{2\lp D+k-2\rp}\lp\lp D-2\rp p_{D+2,k}(t)+ktp_{D+4,k-1}(t)\rp,
\eea
where a prime denotes the derivative with respect to the argument $t$.
\section{Mixed symmetric contractions}\label{app:contractiondetails}
In general one expects that the contraction of the projector $\Pi^{[\lambda]}$ associated to some Young symmetry $\lambda$ is given by  
\begin{align}
 X_{f_{1}}\cdots X_{f_{k}}\Pi^{[\lambda]e_{1}\cdots e_{n}f_{1}\cdots f_{k}}_{g_{1}\cdots g_{n}h_{1}\cdots h_{k}}Y^{h_{1}}\cdots Y^{h_{k}}=\sum_{i}T_{i}(X,Y)^{e_{1}\cdots e_{n}}_{g_{1}\cdots g_{n}}\mathcal{P}_{i}(t),
\end{align}
where $T_{i}$ are tensor structures made out of combinations of $X$, $Y$, and the Kronecker delta, and $\mathcal{P}$ polynomials on $t\equiv \frac{X\cdot Y}{\sqrt{X^{2}Y^{2}}}$. In the previous section we showed this, explicitly, for $[k+1,1]$. More generally, from the work of \cite{Geyer1999,Geyer2000,Eilers2006}, one can understand this expression as the result of a particular differential operator (say in $X$) acting on the symmetric contraction of $\lambda_{1}$ indices
\begin{align}\label{eq:gencontraction}
 X_{f_{1}}\cdots X_{f_{k}}\Pi^{[\lambda]e_{1}\cdots e_{n}f_{1}\cdots f_{k}}_{g_{1}\cdots g_{n}h_{1}\cdots h_{k}}Y^{h_{1}}\cdots Y^{h_{k}}=\mathcal{D}^{[\lambda]e_{1}\cdots e_{n}}_{g_{1}\cdots g_{n}}(X)H_{\lambda_{1}}(X\cdot Y)^{\lambda_{1}},
\end{align}
where 
\begin{align}
 H_{\lambda_{1}}(X\cdot Y)^{\lambda_{1}}=X_{f_{1}}\cdots X_{f_{\lambda_{1}}}\Pi^{[\lambda_{1}]f_{1}\cdots f_{\lambda_{1}}}_{h_{1}\cdots h_{\lambda_{1}}}Y^{h_{1}}\cdots Y^{h_{\lambda_{1}}}=(X^{2}Y^{2})^{\lambda_{1}/2}p_{D,\lambda_{1}}(t),
\end{align}
and $\lambda_{1}$ is the length of the top row of the Young pattern $[\lambda]$ (in our case this is $k+1$). In the context of conformal blocks, the extra indices $e_{j}$ are contracted with $m^{(10)}$, $m^{(20)}$, and the indices $g_{j}$ with $m^{(30)}$, $m^{(40)}$. Furthermore, $X\equiv k^{(012)}$, $Y\equiv k^{(034)}$ with $X^{2}=Y^{2}=1$. Thus a generic contraction with $T_{i}$
\begin{align}\label{eq:tensorcoeffcontraction}
 m^{10}\cdots m^{10}m^{20}\cdots m^{20} \cdot T_{i}(k^{(012)},k^{(034)})\cdot m^{(30)}\cdots m^{(30)} m^{(40)}\cdots m^{(40)},
\end{align}
can include combinations of 1- and 2-index elements
\begin{align}
 m^{(i0)}_{ab}k^{(012)\,b},\quad  m^{(i0)}_{ab}k^{(034)\,b},
 \quad m^{(i0)}_{ab}m^{(0j)\,b}_{\hph{(0j)\,b}c},
\end{align}
with $i,j=1,2,3,4$. The result presented in this paper suggests that one can write the contractions (\ref{eq:tensorcoeffcontraction}) as derivatives of $t$ only. For example, for (\ref{eq:mixedpartialcontraction}) we have
\begin{multline}
m^{(20)}_{ae} T_{1\,c}^{e} m^{(04)\,c}_{\hph{(04)\,c}b}=
 m^{(20)}_{ae} (t \delta^{e}_{c}-k^{(012)}_{c}k^{(034)\,e} ) m^{(04)\,c}_{\hph{(04)\,c}b}
\\=\sqrt{\frac{x_{02}^{2}x_{04}^{2}x_{12}^{2}x_{34}^{2}}{x_{01}^{2}x_{03}^{2}}}\left(t \frac{\partial^{2}t}{\partial x_{2}^{a}\partial x_{4}^{b}}-\frac{\partial t}{\partial x_{2}^{a}}\frac{\partial t}{\partial x_{4}^{b}} \right).
\end{multline}
\begin{multline}
m^{(20)}_{ae} T_{2\,c}^{e} m^{(04)\,c}_{\hph{(04)\,c}b}\\
=
 m^{(20)}_{ae} \left((t^{2}-1)\delta^{e}_{c}+k^{(012)}_{c}k^{(012)\,e}+k^{(034)}_{c}k^{(034)\,e}-t(k^{(012)}_{c}k^{(034)\,e}+k^{(034)}_{c}k^{(012)\,e})\right) m^{(04)\,c}_{\hph{(04)\,c}b}
 \\
 =\sqrt{\frac{x_{02}^{2}x_{04}^{2}x_{12}^{2}x_{34}^{2}}{x_{01}^{2}x_{03}^{2}}}\left((t^{2}-1) \frac{\partial^{2}t}{\partial x_{2}^{a}\partial x_{4}^{b}}-t\frac{\partial t}{\partial x_{2}^{a}}\frac{\partial t}{\partial x_{4}^{b}} \right),
\end{multline}
where we have extracted $T_1$ and $T_2$ by rewriting (\ref{eq:mixedpartialcontraction}) as
\be
-T_{1\,b}^a\lp tf_{k-1}+g_k\rp+T_{2\,b}^af_{k-1}(t),
\ee
and we picked the particular combinations because $f_{k-1}$ and $tf_{k-1}+g_k$ are just constant multiples of $p_{D+4,k-1}$ and $p_{D+2,k}$ respectively.
These results rely on the fact that 
\begin{align}
 \frac{\partial k^{(0ij)}_{c}}{\partial x_{j}^{a}}=-\sqrt{\frac{x_{0i}^{2}}{x_{0j}^{2}x_{ij}^{2}}}\left( m^{(0j)}_{ac}+k^{(j0i)}_{a}k^{(0ij)}_{c} \right),
\end{align}
and the key observation is that the particular combinations of $\delta$, $k^{(012)}$, $k^{(034)}$, that appear in $T_{i}$, are such that the terms $k_{a}^{(j0i)}$ cancel out, leaving only the terms $m^{(0j)}$ that we want. This leads to 
\begin{multline}
 m^{(20)}_{ac}k^{(012)}_{d_{1}}\cdots k^{(012)}_{d_{k+1}}\widetilde{\Pi}^{(k)\,cd_{1}\cdots d_{k+1}}_{e f_{1}\cdots f_{k+1}}m^{(04)\,e}_{\hph{(04)\,e}b}k^{(034)\,f_1}\cdots k^{(034)\,f_{k+1}}
 \\
 =\sqrt{\frac{x_{02}^{2}x_{04}^{2}x_{12}^{2}x_{34}^{2}}{x_{01}^{2}x_{03}^{2}}}
 \frac{1}{2(D+k-2)}
 \left(
 k\left( (t^{2}-1)\frac{\partial^{2}t}{\partial x_{2}^{a}\partial x_{4}^{b}}-t \frac{\partial t}{\partial x_{2}^{a}}\frac{\partial t}{\partial x_{4}^{b}}\right)
 p_{D+4,k-1}\right.\\
 \left.
 +(D-2)
 \left( 
 t\frac{\partial^{2}t}{\partial x_{2}^{a}\partial x_{4}^{b}}- \frac{\partial t}{\partial x_{2}^{a}}\frac{\partial t}{\partial x_{4}^{b}}
 \right)p_{D+2,k}
 \right).
\end{multline}
Equation (\ref{eq:mixedcontraction}) then follows from the chain rule and simple Gegenbauer identities listed in appendix B.

As an extra result, we present the contraction under the projector $\Pi^{[k+1,1,1]}$ associated to the Young pattern $[k+1,1,1]$. Using techniques from \cite{Geyer1999,Geyer2000,Eilers2006} one obtains 
\begin{multline}
 \mathcal{P}^{[k+1,1,1]e_{1}e_{2}}_{c_{1}c_{2}}\equiv X_{f_{1}}\cdots X_{f_{k+1}}\Pi^{[k+1,1,1]e_{1}e_{2}f_{1}\cdots f_{k+1}}_{c_{1}c_{2}d_{1}\cdots d_{k+1}}Y^{d_{1}}\cdots Y^{d_{k+1}}
 \\\propto\frac{(k+1)(X^{2}Y^{2})^{\frac{k+1}{2}}}{(k+3)(D+k-3)}\delta_{c_{1}}^{[d}\delta_{c_{2}}^{f]}\delta_{[g}^{e_{1}}\delta_{h]}^{e_{2}} \lp 
(D-3)\delta_{d}^{g}\ls t\delta^{h}_{f}-2\frac{X_{f}Y^{h}}{\sqrt{X^{2}Y^{2}}}\rs p_{D+2,k}(t)
\right.\\
\left. 
+2k\ls 
\delta_{d}^{g}\lp 
\frac{(t^{2}-1)\delta^{h}_{f}}{2}+\frac{X_{f}X^{h}}{X^{2}}-\frac{t\lp X_{f}Y^{h}+Y_{f}X^{h} \rp}{\sqrt{X^{2}Y^{2}}}+\frac{Y_{f}Y^{h}}{Y^{2}}\rp
-\frac{X_{d}Y_{f}X^{g}Y^{h}}{X^{2}Y^{2}}
\rs p_{D+4,k-1}
 \rp.
\end{multline}
Thus from the previous discussion one finds that
\begin{multline}
 m^{(10)}_{ae_{1}}m^{(20)}_{be_{2}}\mathcal{P}^{[k+1,1,1]e_{1}e_{2}}_{c_{1}c_{2}}
 m^{(30)\, c_{1}}_{c}m^{(40)\, c_{2}}_{d}
 \\= \frac{x_{02}^{2}x_{04}^{2}x_{12}^{2}x_{34}^{2}}{x_{01}^{2}x_{03}^{2}}
 m^{(12)}_{ab'}m^{(34)}_{cd'}\left\{  \frac{\p^{2}t}{\p x_{2}^{b}\p x_{4}^{[d|}}\frac{\p^{2}t}{\p x_{2}^{b'}\p x_{4}^{|d']}}p_{D,k+1}
 -2
   \frac{\p^{2}t}{\p x_{2}^{[b|}\p x_{4}^{[d|}} \frac{\p^{2}}{\p x_{2}^{|b']}\p x_{4}^{|d']}}
 \right.\\
 \left. 
 \times\left[\frac{1}{(k+2)(k+3)}p_{D,k+2}
 +\frac{(k+1)}{(D+2k)(D+2k-2)(D+k-3)}
 p_{D,k}
 \right]
 \right\},
\end{multline}
where in the second term, the square bracket notation is indicating that $b$ ($d$) is antisymmetrized with $b'$ ($d'$).

This is one of the two new contractions that appear in the conformal blocks for the $[k+1,1,1]$ exchange in $\langle VVVV \rangle$. Those conformal blocks are left for future work.

\section{Integrals}
\label{app:Integrals}

Much of the material in this appendix follows~\cite{Simmons-Duffin2014}.

The basic building block for our integrals is
\be
\label{eq:BasicIntegral}
\int d^Dx_0\lp\sum_ia_ix_{0i}^2\rp^{-D}=\frac{2^{1-D}\pi^{\frac{D+1}{2}}}{\G(\frac{D+1}{2})}\lp\sum_{i<j}a_ia_jx_{ij}^2\rp^{-D/2},
\ee
along with the Feynman-Schwinger trick which uses the identity
\be
\label{eq:FSTrick}
\frac{1}{\prod_{i=1}^nX_i^{c_i}}=\frac{\G(\sum_{i=1}^nc_i)}{\prod_{j=1}^n\G(c_j)}\lp\prod_{k=2}^n\int_0^\infty d\m_k\,\m^{c_k-1}\rp\frac{1}{\lp X_1+\sum_{\ell=2}^n\m_\ell X_\ell\rp^{\sum_{m=1}^nc_m}}.
\ee

\subsection{Three-point integrals}
\label{app:3ptIntegrals}

Suppose $\al+\beta+\g=D$.  Then the integral
\be
I_{\al,\beta,\g}(x_1,x_2,x_3)=\int\frac{d^Dx_0}{\lp x_{01}^2\rp^\al\lp x_{02}^2\rp^\beta\lp x_{03}^2\rp^\g},
\ee
will be a conformal scalar of weight $\al$, $\beta$, and $\g$ under conformal transformations of $x_1$, $x_2$, and $x_3$ respectively.  To evaluate the integral, we first use (\ref{eq:FSTrick}) and then (\ref{eq:BasicIntegral}) to write 
\begin{multline}
I_{\al,\beta,\g}(x_1,x_2,x_3)=\frac{\G(D)}{\G(\al)\G(\beta)\G(\g)}\int d^Dx_0\int_0^\infty ds\,s^{\beta-1}\int_0^\infty dt\,t^{\g-1}\frac{1}{\lp x_{01}^2+sx_{02}^2+tx_{03}^2\rp^D}\\
=\frac{2^{1-D}\pi^{\frac{D+1}{2}}\G(D)}{\G(\al)\G(\beta)\G(\g)\G(\frac{D+1}{2})}\int_0^\infty ds\,s^{\beta-1}\int_0^\infty dt\,t^{\g-1}\lp sx_{12}^2+tx_{13}^2+stx_{23}^2\rp^{-D/2}.
\end{multline}
To perform the remaining integrals, we recall one of the representations of the beta function
\be
\int_0^\infty du\frac{u^{x-1}}{\lp 1+u\rp^{x+y}}=\frac{\G(x)\G(y)}{\G(x+y)}.
\ee
Then
\be
\begin{split}
& I_{\al,\beta,\g}(x_1,x_2,x_3)\\
& \ =\frac{2^{1-D}\pi^{\frac{D+1}{2}}\G(D)}{\G(\al)\G(\beta)\G(\g)\G(\frac{D+1}{2})}\int_0^\infty ds\,s^{\beta-1}\lp sx_{12}^2\rp^{-D/2}\int_0^\infty dt\,t^{\g-1}\lp 1+t\lp\frac{x_{13}^2+sx_{23}^2}{sx_{12}^2}\rp\rp^{-D/2}\\
& \ =\frac{2^{1-D}\pi^{\frac{D+1}{2}}\G(D)}{\G(\al)\G(\beta)\G(\g)\G(\frac{D+1}{2})}\lp x_{12}^2\rp^{\g-\frac{D}{2}}\int_0^\infty ds\,s^{\beta+\g-\frac{D}{2}-1}\lp x_{13}^2+sx_{23}^2\rp^{-\g}\int_0^\infty du\,u^{\g-1}\lp 1+u\rp^{-D/2}\\
& \ =\frac{2^{1-D}\pi^{\frac{D+1}{2}}\G(D)\G(\frac{D}{2}-\g)}{\G(\al)\G(\beta)\G(\frac{D+1}{2})\G(\frac{D}{2})}\lp x_{12}^2\rp^{\g-\frac{D}{2}}\lp x_{13}^2\rp^{-\g}\int_0^\infty ds\,s^{\beta+\g-\frac{D}{2}-1}\lp 1+s\frac{x_{23}^2}{x_{13}^2}\rp^{-\g}\\
& \ =\frac{\pi^{D/2}\G(\frac{D}{2}-\g)}{\G(\al)\G(\beta)}\lp x_{12}^2\rp^{\g-\frac{D}{2}}\lp x_{13}^2\rp^{\beta-\frac{D}{2}}\lp x_{23}^2\rp^{\frac{D}{2}-\beta-\g}\int_0^\infty dv\,v^{\beta+\g-\frac{D}{2}-1}\lp 1+v\rp^{-\g}\\
& \ =\frac{\pi^{D/2}\G(\frac{D}{2}-\g)\G(\beta+\g-\frac{D}{2})\G(\frac{D}{2}-\beta)}{\G(\al)\G(\beta)\G(\g)}\lp x_{12}^2\rp^{\g-\frac{D}{2}}\lp x_{13}^2\rp^{\beta-\frac{D}{2}}\lp x_{23}^2\rp^{\frac{D}{2}-\beta-\g}\\
& \ =\pi^{D/2}\frac{\G(\frac{D}{2}-\al)\G(\frac{D}{2}-\beta)\G(\frac{D}{2}-\g)}{\G(\al)\G(\beta)\G(\g)}\lp x_{12}^2\rp^{\g-\frac{D}{2}}\lp x_{13}^2\rp^{\beta-\frac{D}{2}}\lp x_{23}^2\rp^{\al-\frac{D}{2}},
\end{split}
\ee
where we have also made use of the duplication formula for the gamma function, which in this case tells us
\be
\G(\frac{D}{2})\G(\frac{D+1}{2})=2^{1-D}\sqrt{\pi}\G(D).
\ee

Similarly, we will need to evaluate
\be
I_{\al\beta\g;a_1\cdots a_n}(x_1,x_2,x_3)=\Pi^{(n)\,b_1\cdots b_n}_{a_1\cdots a_n}\int\frac{d^Dx_0}{\lp x_{01}^2\rp^\al\lp x_{02}^2\rp^\beta\lp x_{03}^2\rp^\g}k^{(302)}_{b_1}\cdots k^{(302)}_{b_n},
\ee
which for $\al+\beta+\g=D$ will be a conformal scalar of weight $\al$ ($\beta$) under conformal transformations of $x_1$ ($x_2$), and a traceless symmetric tensor of conformal weight $\g$ under transformations of $x_3$.  We compute by doing a binomial expansion of the $k^{(302)}$'s,
\be
\begin{split}
& I_{\al\beta\g;a_1\cdots a_n}(x_1,x_2,x_3)=\Pi^{(n)\,b_1\cdots b_n}_{a_1\cdots a_n}\sum_{k=0}^n\frac{n!}{k!\lp n-k\rp!}\lp -1\rp^k\lp x_{23}^2\rp^{\frac{n-2k}{2}}\lp x_{23}\rp_{b_1}\cdots\lp x_{23}\rp_{b_k}\\
& \quad\times\int\frac{d^Dx_0}{\lp x_{01}^2\rp^\al\lp x_{02}^2\rp^{\beta+\frac{n}{2}}\lp x_{03}^2\rp^{\g+\frac{n-2k}{2}}}\lp x_{03}\rp_{b_{k+1}}\cdots\lp x_{03}\rp_{b_n}\\
& \ =\Pi^{(n)\,b_1\cdots b_n}_{a_1\cdots a_n}\sum_{k=0}^n\frac{n!}{k!\lp n-k\rp!}\lp -1\rp^k\lp x_{23}^2\rp^{\frac{n-2k}{2}}\lp x_{23}\rp_{b_1}\cdots\lp x_{23}\rp_{b_k}\\
& \quad\times\frac{\G(\g-\frac{n}{2})}{2^{n-k}\G(\g+\frac{n}{2}-k)}\frac{\p}{\p x_3^{b_{k+1}}}\cdots\frac{\p}{\p x_3^{b_n}}I_{\al,\beta+\frac{n}{2},\g-\frac{n}{2}}(x_1,x_2,x_3)\\
& \ =\pi^{D/2}\Pi^{(n)\,b_1\cdots b_n}_{a_1\cdots a_n}\sum_{k=0}^n\sum_{m=0}^{n-k}\frac{n!}{k!m!\lp n-k-m\rp!}\lp -1\rp^k\\
& \quad\times\frac{\G(\frac{D}{2}+n-m-k-\al)\G(\frac{D}{2}-\frac{n}{2}+m-\beta)\G(\frac{D}{2}+\frac{n}{2}-\g)}{\G(\al)\G(\beta+\frac{n}{2})\G(\g+\frac{n}{2}-k)}\\
& \quad\times\lp x_{12}^2\rp^{\g-\frac{n}{2}-\frac{D}{2}}\lp x_{13}^2\rp^{\beta+\frac{n}{2}-m-\frac{D}{2}}\lp x_{23}^2\rp^{\al-\frac{n}{2}+m-\frac{D}{2}}\lp x_{13}\rp_{b_1}\cdots\lp x_{13}\rp_{b_m}\lp x_{23}\rp_{b_{m+1}}\cdots\lp x_{23}\rp_{b_n}\\
& \ =\pi^{D/2}\Pi^{(n)\,b_1\cdots b_n}_{a_1\cdots a_n}\sum_{m=0}^n\frac{n!}{m!\lp n-m\rp!}\lp -1\rp^{n-m}\frac{\G(\frac{D}{2}-\al)\G(\frac{D}{2}+\frac{n}{2}-\beta)\G(\frac{D}{2}+\frac{n}{2}-\g)}{\G(\al)\G(\beta+\frac{n}{2})\G(\g+\frac{n}{2})}\\
& \quad \times\lp x_{12}^2\rp^{\g-\frac{n}{2}-\frac{D}{2}}\lp x_{13}^2\rp^{\beta+\frac{n}{2}-m-\frac{D}{2}}\lp x_{23}^2\rp^{\al-\frac{n}{2}+m-\frac{D}{2}}\lp x_{13}\rp_{b_1}\cdots\lp x_{13}\rp_{b_m}\lp x_{23}\rp_{b_{m+1}}\cdots\lp x_{23}\rp_{b_n}\\
& \ =\pi^{D/2}\Pi^{(n)\,b_1\cdots b_n}_{a_1\cdots a_n}\frac{\G(\frac{D}{2}-\al)\G(\frac{D}{2}+\frac{n}{2}-\beta)\G(\frac{D}{2}+\frac{n}{2}-\g)}{\G(\al)\G(\beta+\frac{n}{2})\G(\g+\frac{n}{2})}\\
& \quad\times\lp x_{12}^2\rp^{\g-\frac{D}{2}}\lp x_{13}^2\rp^{\beta-\frac{D}{2}}\lp x_{23}^2\rp^{\al-\frac{D}{2}}k^{(312)}_{b_1}\cdots k^{(312)}_{b_n},
\end{split}
\ee
where we used the identity
\be
\label{eq:CombId1}
\sum_{k=0}^N\frac{N!}{k!\lp N-k\rp!}\lp -1\rp^k\frac{\G(x-k)}{\G(y-k)}=\lp -1\rp^N\frac{\G(x-N)\G(y-x+N)}{\G(y)\G(y-x)},
\ee
with $N=n-m$, $x=\frac{D}{2}+n-m-\al$, and $y=\g+\frac{n}{2}$.

We will also need one more result along these lines,
\be
\begin{split}
& I_{\al,\beta,\g;a;b_1\cdots b_n}=\Pi^{(n)\,c_1\cdots c_n}_{b_1\cdots b_n}\int\frac{d^Dx_0}{\lp x_{01}^2\rp^\al\lp x_{02}^2\rp^\beta\lp x_{03}^2\rp^\g}k^{(203)}_ak^{(302)}_{c_1}\cdots k^{(302)}_{c_n}\\
& \ =\Pi^{(n)\,c_1\cdots c_n}_{b_1\cdots b_n}\left\{\frac{\sqrt{x_{23}^2}}{2\beta+n-1}\frac{\p}{\p x_2^a}I_{\al,\beta-\hlf,\g+\hlf;c_1\cdots c_n}+\frac{2\beta-1}{2\beta+n-1}\frac{\lp x_{23}\rp_a}{\sqrt{x_{23}^2}}I_{\al,\beta-\hlf,\g+\hlf;c_1\cdots c_n}\right.\\
& \qquad\left. +\frac{nm^{(23)}_{ac_1}}{2\beta+n-1}I_{\al,\beta,\g;c_2\cdots c_n}\right\}\\
& \ =\pi^{D/2}\Pi^{(n)\,c_1\cdots c_n}_{b_1\cdots b_n}\frac{\G(\frac{D}{2}-\al)\G(\frac{D}{2}+\frac{n-1}{2}-\beta)\G(\frac{D}{2}+\frac{n-1}{2}-\g)}{\G(\al)\G(\beta+\frac{n+1}{2})\G(\g+\frac{n+1}{2})}\lp x_{12}^2\rp^{\g-\frac{D}{2}}\lp x_{13}^2\rp^{\beta-\frac{D}{2}}\lp x_{23}^2\rp^{\al-\frac{D}{2}}\\
& \quad\times\ls\lp\frac{D}{2}+\frac{n-1}{2}-\beta\rp\lp\frac{D}{2}+\frac{n-1}{2}-\g\rp k^{(213)}_ak^{(213)}_{c_1}+\frac{n}{2}\lp\frac{D}{2}-\al\rp m^{(23)}_{ac_1}\rs k^{(312)}_{c_2}\cdots k^{(312)}_{c_n}.
\end{split}
\ee
In this case we made use of (\ref{eq:pkIdentity}).

\subsection{Four-point integrals}
\label{subapp:4ptIntegrals}

As with the previous section, we start with integrals of the form
\be
\label{eq:Basic4ptScalarIntegral}
I_{\al,\beta,\g,\d}(x_1,x_2,x_3,x_4)=\int\frac{d^Dx_0}{\lp x_{01}^2\rp^\al\lp x_{02}^2\rp^\beta\lp x_{03}^2\rp^\g\lp x_{04}^2\rp^\d},
\ee
where $\al+\beta+\g+\d=D$.  Using (\ref{eq:BasicIntegral}) and (\ref{eq:FSTrick}) we can show
\begin{multline}
I_{\al,\beta,\g,\d}=\frac{2^{1-D}\pi^{\frac{D+1}{2}}\G(D)}{\G(\al)\G(\beta)\G(\g)\G(\d)\G(\frac{D+1}{2})}\int_0^\infty ds\,s^{\beta-1}\int_0^\infty dt\,t^{\g-1}\int_0^\infty dq\,q^{\d-1}\\
\times\lp sx_{12}^2+tx_{13}^2+qx_{14}^2+stx_{23}^2+sqx_{24}^2+tqx_{34}^2\rp^{-D/2}.
\end{multline}
After a change of variables we can do one of the three integrals, giving us a result
\begin{multline}
\label{eq:4ptScalarIntegral}
I_{\al,\beta,\g,\d}=\pi^{D/2}\frac{\G(\frac{D}{2}-\d)}{\G(\al)\G(\beta)\G(\g)}\lp x_{14}^2\rp^{-\al}\lp x_{23}^2\rp^{\d-\frac{D}{2}}\lp x_{24}^2\rp^{\frac{D}{2}-\beta-\d}\lp x_{34}^2\rp^{\frac{D}{2}-\g-\d}\\
\times \whf_{\al,\beta,\g,\d}(uv^{-1},v^{-1}),
\end{multline}
where we have defined
\be
\label{eq:whfDef}
\whf_{\al,\beta,\g,\d}(z_1,z_2)=\int_0^\infty ds\,s^{\beta-1}\int_0^\infty dt\,t^{\g-1}\lp sz_1+tz_2+st\rp^{\frac{\d-\al-\beta-\g}{2}}\lp 1+s+t\rp^{-\d},
\ee
and $u$ and $v$ are the usual invariant cross-ratios defined in (\ref{eq:CrossRatios}).


As explained in Section \ref{subsubsec:ComputingSSSS}, the monodromy projection requires us to keep only the terms in $\whf_{\al,\beta,\g,\d}(z_1,z_2)$ which are invariant under $z_1\rr e^{4\pi i}z_1$.  In~\cite{Simmons-Duffin2014} it is shown how to do this very elegantly using contour deformation arguments, with the result that the invariant pieces are given precisely by
\begin{multline}
\label{eq:fDef}
f_{\al,\beta,\g,\d}(z_1,z_2)=\left.\whf_{\al,\beta,\g,\d}(z_1,z_2)\right|_{\mathrm{monodromy-invariant}}\\
=\frac{\sin(\pi\d)}{\sin(\frac{\pi}{2}\lp\g+\d-\al-\beta\rp)}\int_0^\infty ds\,s^{\beta-1}\int_{s+1}^\infty dt\,t^{\g-1}\lp st+tz_2-sz_1\rp^{\frac{\d-\al-\beta-\g}{2}}\lp t-s-1\rp^{-\d}.
\end{multline}
The function $f_{\al,\beta,\g,\d}(z_1,z_2)$ obeys several easily verified identities (also $\whf$ obeys the same identities),
\bea
\label{eq:p1f}
\frac{\p}{\p z_1}f_{\al,\beta,\g,\d}(z_1,z_2) &=& \frac{\d-\al-\beta-\g}{2}f_{\al+1,\beta+1,\g,\d}(z_1,z_2),\\
\label{eq:p2f}
\frac{\p}{\p z_2}f_{\al,\beta,\g,\d}(z_1,z_2) &=& \frac{\d-\al-\beta-\g}{2}f_{\al+1,\beta,\g+1,\d}(z_1,z_2),
\eea
as well as
\be
f_{\al,\beta,\g,\d}(z_1,z_2)=f_{\al+1,\beta,\g,\d+1}(z_1,z_2)+f_{\al,\beta+1,\g,\d+1}(z_1,z_2)+f_{\al,\beta,\g+1,\d+1}(z_1,z_2),
\ee
and
\be
f_{\al,\beta,\g,\d}(z_1,z_2)=f_{\al,\beta+1,\g+1,\d}(z_1,z_2)+z_1f_{\al+1,\beta+1,\g,\d}(z_1,z_2)+z_2f_{\al+1,\beta,\g+1,\d}(z_1,z_2).
\ee

When $\al+\beta+\g+\d$ is an even integer, which we will call $2h$ (so $h=D/2$ in the four-point integral above, and this would be valid in even dimensions), then $f_{\al,\beta,\g,\d}$ can actually be evaluated explicitly in terms of hypergeometric functions.  First we change from $z_1$ and $z_2$ to a complex variable $x$ related by
\be
z_1=\frac{x\bar{x}}{\lp 1-x\rp\lp 1-\bar{x}\rp},\qquad z_2=\frac{1}{\lp 1-x\rp\lp 1-\bar{x}\rp},
\ee
and then it can be shown~\cite{Dolan2001,Simmons-Duffin2014} that
\begin{multline}
f_{\al,\beta,\g,\d}(z_1,z_2)=\frac{\G(\al)\G(1-h+\beta)\G(1-\d)\G(h-\g)\G(\g+\d-h)}{\G(\d)\G(h-\d)\G(1+h-\g-\d)}\lp\lp 1-x\rp\lp 1-\bar{x}\rp\rp^{h-\d}\\
\times\lp\frac{1}{x-\bar{x}}\lp x\p_x-\bar{x}\p_{\bar{x}}\rp\rp^{h-1}\ls\vphantom{F_1}_2F_1(1-h+\beta,1-\d,1+h-\g-\d;x)\right.\\
\left. \times\vphantom{F_1}_2F_1(1-h+\beta,1-\d,1+h-\g-\d;\bar{x})\rs.
\end{multline}

\section{Mixing matrices and normalization factors}\label{app:mixingmatrices}
For the case of two scalars and a symmetric traceless tensor, inserting (\ref{eq:shadowopDef}) into (\ref{eq:phiphiOsymmetric}) leads to
\begin{multline}
\left\langle\phi_1(x_1)\phi_2(x_2)\wtmcO_{a_1\cdots a_\ell}(x_3)\right\rangle=\Pi^{(\ell)\,b_1\cdots b_\ell}_{a_1\cdots a_\ell}\int\frac{d^Dx_0}{\lp x_{03}^2\rp^{D-\Delta_\mcO}}m^{(03)\hph{b_1}c_1}_{\hph{(03)}b_1}\cdots m^{(03)\hph{b_\ell}c_\ell}_{\hph{(03)}b_\ell}\\
\times\lp\la_\mcO\lp x_{01}^2\rp^{\hlf\lp -\Delta_1+\Delta_2-\Delta_\mcO\rp}\lp x_{02}^2\rp^{\hlf\lp\Delta_1-\Delta_2-\Delta_\mcO\rp}\lp x_{12}^2\rp^{\hlf\lp -\Delta_1-\Delta_2+\Delta_\mcO\rp}\Pi^{(\ell)\,d_1\cdots d_\ell}_{c_1\cdots c_\ell}k^{(012)}_{d_1}\cdots k^{(012)}_{d_\ell}\rp.
\end{multline}
Since (as reviewed in Appendix \ref{app:BuildingBlocks}) $m^{(03)\hph{a}c}_{\hph{(03)}a}m^{(03)}_{bc}=\d_{ab}$, and since $\Pi^{(\ell)}$ removes traces, it follows that
\be
\Pi^{(\ell)\,b_1\cdots b_\ell}_{a_1\cdots a_\ell}m^{(03)\hph{b_1}c_1}_{\hph{(03)}b_1}\cdots m^{(03)\hph{b_\ell}c_\ell}_{\hph{(03)}b_\ell}\Pi^{(\ell)\,d_1\cdots d_\ell}_{c_1\cdots c_\ell}k^{(012)}_{d_1}\cdots k^{(012)}_{d_\ell}\\
=\Pi^{(\ell)\,b_1\cdots b_\ell}_{a_1\cdots a_\ell}y_{b_1}\cdots y_{b_\ell},
\ee
where
\be
y_a=m^{(03)}_{ab}k^{(012)\,b}=\lp\sqrt{\frac{x_{01}^2x_{23}^2}{x_{03}^2x_{12}^2}}-\frac{x_{02}^2x_{13}^2}{\sqrt{x_{01}^2x_{03}^2x_{12}^2x_{23}^2}}\rp k^{(302)}_a+\sqrt{\frac{x_{02}^2x_{13}^2}{x_{01}^2x_{23}^2}}k^{(312)}_a.
\ee
Expanding in a trinomial expansion, we then obtain
\be
\begin{split}
&\left\langle\phi_1(x_1)\phi_2(x_2)\wtmcO_{a_1\cdots a_\ell}(x_3)\right\rangle\\
& \quad =\la_\mcO\Pi^{(\ell)\,b_1\cdots b_\ell}_{a_1\cdots a_\ell}\sum_{k=0}^\ell\sum_{m=0}^{\ell-k}\frac{\ell!}{k!m!\lp\ell-k-m\rp!}\lp -1\rp^m\\
& \qquad\times\lp x_{12}^2\rp^{\hlf\lp -\Delta_1-\Delta_2+\Delta_\mcO-k-m\rp}\lp x_{13}^2\rp^{\hlf\lp\ell-k+m\rp}\lp x_{23}^2\rp^{k-\frac{\ell}{2}}k^{(312)}_{b_1}\cdots k^{(312)}_{b_{\ell-k-m}}\\
& \qquad\times I_{\hlf\lp\Delta_1-\Delta_2+\Delta_\mcO+\ell-2k\rp,\hlf\lp -\Delta_1+\Delta_2+\Delta_\mcO-\ell+k-m\rp,D-\Delta_\mcO+\frac{k+m}{2};b_{\ell-k-m+1}\cdots b_\ell}(x_1,x_2,x_3)\\
& \quad =\pi^{D/2}\la_\mcO\Pi^{(\ell)\,b_1\cdots b_\ell}_{a_1\cdots a_\ell}k^{(312)}_{b_1}\cdots k^{(312)}_{b_\ell}\sum_{k=0}^\ell\sum_{m=0}^{\ell-k}\frac{\ell!}{k!m!\lp\ell-k-m\rp!}\lp -1\rp^m\\
& \qquad\times\frac{\G(\hlf\lp D-\Delta_1+\Delta_2-\Delta_\mcO-\ell\rp+k)\G(\hlf\lp D+\Delta_1-\Delta_2-\Delta_\mcO+\ell\rp+m)\G(\Delta_\mcO-\frac{D}{2})}{\G(\hlf\lp\Delta_1-\Delta_2+\Delta_\mcO+\ell\rp-k)\G(\hlf\lp -\Delta_1+\Delta_2+\Delta_\mcO-\ell\rp+k)\G(D-\Delta_\mcO+k+m)}\\
& \qquad\times\lp x_{12}^2\rp^{\hlf\lp D-\Delta_1-\Delta_2-\Delta_\mcO\rp}\lp x_{13}^2\rp^{\hlf\lp -\Delta_1+\Delta_2+\Delta_\mcO-D\rp}\lp x_{23}^2\rp^{\hlf\lp\Delta_1-\Delta_2+\Delta_\mcO-D\rp}\\
& \quad =\pi^{D/2}\Pi^{(\ell)\,b_1\cdots b_\ell}_{a_1\cdots a_\ell}k^{(312)}_{b_1}\cdots k^{(312)}_{b_\ell}\lp x_{12}^2\rp^{\hlf\lp -\Delta_1-\Delta_2+\Delta_\wtmcO\rp}\lp x_{13}^2\rp^{\hlf\lp -\Delta_1+\Delta_2-\Delta_\wtmcO\rp}\lp x_{23}^2\rp^{\hlf\lp\Delta_1-\Delta_2-\Delta_\wtmcO\rp}\\
& \qquad\times\frac{\G(\Delta_\mcO-\frac{D}{2})\G(\Delta_\mcO+\ell-1)}{\G(\Delta_\mcO-1)\G(D-\Delta_\mcO+\ell)}\\
& \qquad\times\frac{\G(\hlf\lp D+\Delta_1-\Delta_2-\Delta_\mcO+\ell\rp)\G(\hlf\lp D-\Delta_1+\Delta_2-\Delta_\mcO+\ell\rp)}{\G(\hlf\lp\Delta_1-\Delta_2+\Delta_\mcO+\ell\rp)\G(\hlf\lp -\Delta_1+\Delta_2+\Delta_\mcO+\ell\rp)}\la_\mcO,
\end{split}
\ee
where we use the notation and results for integrals defined in Appendix \ref{app:3ptIntegrals}, and we evaluated the sums, first over $m$ and then over $k$, using the identities
\be
\label{eq:CombId2}
\sum_{k=0}^N\frac{N!}{k!\lp N-k\rp!}\lp -1\rp^k\frac{\G(x+k)}{\G(y+k)}=\frac{\G(x)\G(y-x+N)}{\G(y+N)\G(y-x)},
\ee
which is equivalent to (\ref{eq:CombId1}), and
\be
\label{eq:CombId3}
\sum_{k=0}^N\frac{N!}{k!\lp N-k\rp!}\frac{1}{\G(x+k)\G(y-k)}=\frac{\G(x+y+N-1)}{\G(x+N)\G(y)\G(x+y-1)}.
\ee
Thus comparing with (\ref{eq:phiphiOshadow}) one can read off (\ref{eq:lambdaOtilde}).

Now for a scalar, a vector, and a traceless symmetric tensor we have
\be
\begin{split}
&\left\langle\phi(x_1)v_a(x_2)\wtmcO_{b_1\cdots b_\ell}(x_3)\right\rangle\\
& \quad =\Pi^{(\ell)\,c_1\cdots c_\ell}_{b_1\cdots b_\ell}\int\frac{d^Dx_0}{\lp x_{03}^2\rp^{D-\Delta_\mcO}}m^{(03)\hph{c_1}d_1}_{\hph{(03)}c_1}\cdots m^{(03)\hph{c_\ell}d_\ell}_{\hph{(03)}c_\ell}\lp x_{01}^2\rp^{\hlf\lp -\Delta_\phi+\Delta_v-\Delta_\mcO\rp}\lp x_{02}^2\rp^{\hlf\lp\Delta_\phi-\Delta_v-\Delta_\mcO\rp}\\
& \qquad\times\lp x_{12}^2\rp^{\hlf\lp -\Delta_\phi-\Delta_v+\Delta_\mcO\rp}\Pi^{(\ell)\,e_1\cdots e_\ell}_{d_1\cdots d_\ell}\ls -\al_\mcO k^{(201)}_ak^{(012)}_{e_1}+\beta_\mcO m^{(20)}_{ae_1}\rs k^{(012)}_{e_2}\cdots k^{(012)}_{e_\ell}\\
& \quad =\Pi^{(\ell)\,c_1\cdots c_\ell}_{b_1\cdots b_\ell}\lp x_{12}^2\rp^{\hlf\lp -\Delta_\phi-\Delta_v+\Delta_\mcO\rp}\int d^Dx_0\lp x_{01}^2\rp^{\hlf\lp -\Delta_\phi+\Delta_v-\Delta_\mcO\rp}\lp x_{02}^2\rp^{\hlf\lp\Delta_\phi-\Delta_v-\Delta_\mcO\rp}\lp x_{03}^2\rp^{\Delta_\mcO-D}\\
& \qquad\times\ls -\al_\mcO\lp\sqrt{\frac{x_{03}^2x_{12}^2}{x_{01}^2x_{23}^2}}k^{(203)}_a-\sqrt{\frac{x_{02}^2x_{13}^2}{x_{01}^2x_{23}^2}}k^{(213)}_a\rp y_{c_1}+\beta_\mcO\lp m^{(23)}_{ac_1}-2k^{(203)}_ak^{(302)}_{c_1}\rp\rs y_{c_2}\cdots y_{c_\ell},
\end{split}
\ee
using identities from Appendix \ref{app:BuildingBlocks}.

We then proceed as before, performing trinomial expansions on the $y_a$'s, perform the integrals using the results of Appendix \ref{app:3ptIntegrals}, and the identities (\ref{eq:CombId2}) and (\ref{eq:CombId3}). This results in (\ref{eq:AlphaTilde}) and (\ref{eq:BetaTilde}).

Related to these integration techniques is the determination of the normalization factor $\mcN_{\mcO}$ that appears in the shadow projector $P_{\mcO}$. As discussed in the main text, this is fixed by requiring
\be
\begin{split}
& \left\langle\vp_1(x_1)\vp_2(x_2)\mcO_{a_1\cdots a_\ell}(x_3)\right\rangle=\left\langle\mcO_{a_1\cdots a_\ell}(x_3)P_\mcO\vp_1(x_1)\vp_2(x_2)\right\rangle\\
& \quad =\mcN_\mcO\int d^Dx_0\left\langle\mcO_{a_1\cdots a_\ell}(x_3)\mcO_{b_1\cdots b_\ell}(x_0)\right\rangle\left\langle\wtmcO^{b_1\cdots b_\ell}(x_0)\vp_1(x_1)\vp_2(x_2)\right\rangle\\
& \quad =\mcN_\mcO\la_{12\wtmcO}\Pi^{(\ell)\,c_1\cdots c_\ell}_{b_1\cdots b_\ell}\Pi^{(\ell)\,d_1\cdots d_\ell}_{a_1\cdots a_\ell}\int d^Dx_0\lp x_{12}^2\rp^{\hlf\lp D-\Delta_1-\Delta_2-\Delta_\mcO\rp}\lp x_{01}^2\rp^{\hlf\lp -\Delta_1+\Delta_2+\Delta_\mcO-D\rp}\\
& \qquad\times\lp x_{02}^2\rp^{\hlf\lp\Delta_1-\Delta_2+\Delta_\mcO-D\rp}k^{(012)}_{c_1}\cdots k^{(012)}_{c_\ell}\lp x_{03}^2\rp^{-\Delta_\mcO}m^{(03)\hph{d_1}b_1}_{\hph{(03)}d_1}\cdots m^{(03)\hph{d_\ell}b_\ell}_{\hph{(03)}d_\ell}\\
& \quad =\mcN_\mcO\la_{12\wtmcO}\Pi^{(\ell)\,b_1\cdots b_\ell}_{a_1\cdots a_\ell}\lp x_{12}^2\rp^{\hlf\lp D-\Delta_1-\Delta_2-\Delta_\mcO\rp}\int d^Dx_0\lp x_{01}^2\rp^{\hlf\lp -\Delta_1+\Delta_2+\Delta_\mcO-D\rp}\\
& \qquad\times\lp x_{02}^2\rp^{\hlf\lp\Delta_1-\Delta_2+\Delta_\mcO-D\rp}\lp x_{03}^2\rp^{-\Delta_\mcO}y_{b_1}\cdots y_{b_\ell}\\
& \quad =\mcN_\mcO\la_{12\wtmcO}\Pi^{(\ell)\,b_1\cdots b_\ell}_{a_1\cdots a_\ell}\lp x_{12}^2\rp^{\hlf\lp -\Delta_1-\Delta_2+\Delta_\mcO\rp}\lp x_{13}^2\rp^{\hlf\lp -\Delta_1+\Delta_2-\Delta_\mcO\rp}\lp x_{23}^2\rp^{\hlf\lp\Delta_1-\Delta_2-\Delta_\mcO\rp}\\
& \qquad\times k^{(312)}_{b_1}\cdots k^{(312)}_{b_\ell}\frac{\G(\frac{D}{2}-\Delta_\mcO)}{\G(D-\Delta_\mcO-1)}\frac{\G(D-\Delta_\mcO+\ell-1)}{\G(\Delta_\mcO+\ell)}\\
& \qquad\times\frac{\G(\hlf\lp\Delta_1-\Delta_2+\Delta_\mcO+\ell\rp)}{\G(\hlf\lp D+\Delta_1-\Delta_2-\Delta_\mcO+\ell\rp)}\frac{\G(\hlf\lp -\Delta_1+\Delta_2+\Delta_\mcO+\ell\rp)}{\G(\hlf\lp D-\Delta_1+\Delta_2-\Delta_\mcO+\ell\rp)}\\
& \quad =\left\langle\vp_1(x_1)\vp_2(x_2)\mcO_{a_1\cdots a_\ell}(x_3)\right\rangle\mcN_\mcO\pi^D\\
& \qquad\times\frac{\G(\Delta_\mcO-\frac{D}{2})\G(\frac{D}{2}-\Delta_\mcO)}{\lp \Delta_\mcO+\ell-1\rp\lp D-\Delta_\mcO+\ell-1\rp\G(\Delta_\mcO-1)\G(D-\Delta_\mcO-1)},
\end{split}
\ee
where we read off (\ref{eq:NO}).
\section{$\alpha\beta$, $\beta\alpha$, and $\beta\beta$ components of the $\langle SVSV\rangle$ blocks}\label{app:otherblockcomponents}
Here we write the additional conformal block components that appear for $\ell>0$. These are expressed in condensed notation where the blocks on the LHS, as well as the $\al\al$ blocks on the right-hand-sides have unshifted arguments, $g^{rs}_p(u,v;\Delta_1,\Delta_2,\Delta_3,\Delta_4;\ell,\Delta_\mcO)$. The others follow the conventions in the main text
\bea
g^{\al\beta}_0 &=& \frac{\Delta_3-\Delta_4-\Delta_\mcO+\ell+1}{\ell}g^{\al\al}_0+\frac{1}{\ell}\ls\sqrt{u}g^{\al\la}_{1;\ell;0,1}+\sqrt{v}g^{\al\la}_{2;\ell;0,1}\rs,\non\\
g^{\al\beta}_{11} &=& \frac{\Delta_3-\Delta_4-\Delta_\mcO+\ell+1}{\ell}g^{\al\al}_{11}-\frac{1}{\ell}\sqrt{u}\lp\Delta_1-\Delta_2-\Delta_3+\Delta_4-2v\p_v\rp g^{\al\la}_{1;\ell;0,1},\non\\
\label{eq:AlphaBetaBlocks}
g^{\al\beta}_{12} &=& \frac{\Delta_3-\Delta_4-\Delta_\mcO+\ell+1}{\ell}g^{\al\al}_{12}-\frac{1}{\ell}\sqrt{v}\lp\Delta_3-\Delta_4+1-2u\p_u\rp g^{\al\la}_{1;\ell;0,1},\\
g^{\al\beta}_{21} &=& \frac{\Delta_3-\Delta_4-\Delta_\mcO+\ell+1}{\ell}g^{\al\al}_{21}-\frac{1}{\ell}\sqrt{u}\lp\Delta_1-\Delta_2-\Delta_3+\Delta_4-1-2v\p_v\rp g^{\al\la}_{2;\ell;0,1},\non\\
g^{\al\beta}_{22} &=& \frac{\Delta_3-\Delta_4-\Delta_\mcO+\ell+1}{\ell}g^{\al\al}_{22}-\frac{1}{\ell}\sqrt{v}\lp\Delta_3-\Delta_4+2-2u\p_u\rp g^{\al\la}_{2;\ell;0,1},\non
\eea
\bea
g^{\beta\al}_0 &=& \frac{\Delta_1-\Delta_2-\Delta_\mcO+\ell+1}{\ell} g^{\al\al}_0+\frac{1}{\ell}\ls g^{\la\al}_{1;\ell;1,0}+\sqrt{\frac{u}{v}}g^{\la\al}_{2;\ell;1,0}\rs,\non\\
g^{\beta\al}_{11} &=& \frac{\Delta_1-\Delta_2-\Delta_\mcO+\ell+1}{\ell}g^{\al\al}_{11}-\frac{1}{\ell}\lp\Delta_1-\Delta_2+2-2u\p_u\rp g^{\la\al}_{1;\ell;1,0},\non\\
\label{eq:BetaAlphaBlocks}
g^{\beta\al}_{12} &=& \frac{\Delta_1-\Delta_2-\Delta_\mcO+\ell+1}{\ell}g^{\al\al}_{12}-\frac{1}{\ell}\lp\Delta_1-\Delta_2+1-2u\p_u\rp g^{\la\al}_{2;\ell;1,0},\\
g^{\beta\al}_{21} &=& \frac{\Delta_1-\Delta_2-\Delta_\mcO+\ell+1}{\ell}g^{\al\al}_{21}+\frac{1}{\ell}\sqrt{\frac{u}{v}}2v\p_vg^{\la\al}_{1;\ell;1,0},\non\\
g^{\beta\al}_{22} &=& \frac{\Delta_1-\Delta_2-\Delta_\mcO+\ell+1}{\ell}g^{\al\al}_{22}-\frac{1}{\ell}\sqrt{\frac{u}{v}}\lp 1-2v\p_v\rp g^{\la\al}_{2;\ell;1,0},\non
\eea
\bea
g^{\beta\beta}_0 &=& \frac{\lp\Delta_1-\Delta_2-\Delta_\mcO+\ell+1\rp\lp\Delta_3-\Delta_4-\Delta_\mcO+\ell+1\rp}{\ell^2}g^{\al\al}_0\non\\
&& +\frac{\Delta_1-\Delta_2-\Delta_\mcO+\ell+1}{\ell^2}\ls\sqrt{u}g^{\al\la}_{1;\ell;0,1}+\sqrt{v}g^{\al\la}_{2;\ell;0,1}\rs\non\\
&& +\frac{\Delta_3-\Delta_4-\Delta_\mcO+\ell+1}{\ell^2}\ls g^{\la\al}_{1;\ell;1,0}+\sqrt{\frac{u}{v}}g^{\la\al}_{2;\ell;1,0}\rs\non\\
&& -\frac{1}{\ell^2}\sqrt{u}\lp\Delta_1-\Delta_2+1-2u\p_u-2v\p_v\rp g_{\ell;1,1},\non\\
g^{\beta\beta}_{11} &=& \frac{\lp\Delta_1-\Delta_2-\Delta_\mcO+\ell+1\rp\lp\Delta_3-\Delta_4-\Delta_\mcO+\ell+1\rp}{\ell^2}g^{\al\al}_{11}\non\\
&& -\frac{\Delta_1-\Delta_2-\Delta_\mcO+\ell+1}{\ell^2}\sqrt{u}\lp\Delta_1-\Delta_2-\Delta_3+\Delta_4-2v\p_v\rp g^{\al\la}_{1;\ell;0,1}\non\\
&& -\frac{\Delta_3-\Delta_4-\Delta_\mcO+\ell+1}{\ell^2}\lp\Delta_1-\Delta_2+2-2u\p_u\rp g^{\la\al}_{1;\ell;1,0}\non\\
&& +\frac{1}{\ell^2}\sqrt{u}\lp\Delta_1-\Delta_2+1-2u\p_u\rp\lp\Delta_1-\Delta_2-\Delta_3+\Delta_4-2v\p_v\rp g_{\ell;1,1},\non\\
g^{\beta\beta}_{12} &=& \frac{\lp\Delta_1-\Delta_2-\Delta_\mcO+\ell+1\rp\lp\Delta_3-\Delta_4-\Delta_\mcO+\ell+1\rp}{\ell^2}g^{\al\al}_{12}\non\\
&& -\frac{\Delta_1-\Delta_2-\Delta_\mcO+\ell+1}{\ell^2}\sqrt{v}\lp\Delta_3-\Delta_4+1-2u\p_u\rp g^{\al\la}_{1;\ell;0,1}\non\\
&& -\frac{\Delta_3-\Delta_4-\Delta_\mcO+\ell+1}{\ell^2}\lp\Delta_1-\Delta_2+1-2u\p_u\rp g^{\la\al}_{2;\ell;1,0}\non\\
\label{eq:BetaBetaBlocks}
&& +\frac{1}{\ell^2}\sqrt{v}\lp\Delta_1-\Delta_2+1-2u\p_u\rp\lp\Delta_3-\Delta_4+1-2u\p_u\rp g_{\ell;1,1},
\eea
\bea
g^{\beta\beta}_{21} &=& \frac{\lp\Delta_1-\Delta_2-\Delta_\mcO+\ell+1\rp\lp\Delta_3-\Delta_4-\Delta_\mcO+\ell+1\rp}{\ell^2}g^{\al\al}_{21}\non\\
&& -\frac{\Delta_1-\Delta_2-\Delta_\mcO+\ell+1}{\ell^2}\sqrt{u}\lp\Delta_1-\Delta_2-\Delta_3+\Delta_4-1-2v\p_v\rp g^{\al\la}_{2;\ell;0,1}\non\\
&& +\frac{\Delta_3-\Delta_4-\Delta_\mcO+\ell+1}{\ell^2}\sqrt{\frac{u}{v}} 2v\p_vg^{\la\al}_{1;\ell;1,0}\non\\
&& -\frac{1}{\ell^2}\frac{u}{\sqrt{v}}2v\p_v\lp\Delta_1-\Delta_2-\Delta_3+\Delta_4-2v\p_v\rp g_{\ell;1,1},\non\\
g^{\beta\beta}_{22} &=& \frac{\lp\Delta_1-\Delta_2-\Delta_\mcO+\ell+1\rp\lp\Delta_3-\Delta_4-\Delta_\mcO+\ell+1\rp}{\ell^2}g^{\al\al}_{22}\non\\
&& -\frac{\Delta_1-\Delta_2-\Delta_\mcO+\ell+1}{\ell^2}\sqrt{v}\lp\Delta_3-\Delta_4+2-2u\p_u\rp g^{\al\la}_{2;\ell;0,1}\non\\
&& -\frac{\Delta_3-\Delta_4-\Delta_\mcO+\ell+1}{\ell^2}\sqrt{\frac{u}{v}}\lp 1-2v\p_v\rp g^{\la\al}_{2;\ell;1,0}\non\\
&& -\frac{1}{\ell^2}\sqrt{u} 2v\p_v\lp\Delta_3-\Delta_4+1-2u\p_u\rp g_{\ell;1,1},\non
\eea
\section{Mixed symmetric constants}\label{app:mixedsymmetricconstants}
The constants appearing in the mixed-symmetric conformal blocks are defined by
\be
C_1=\frac{\mcN_\mcA\g_{34\wtmcA}/\g_{34\mcA}}{\lp\mcN_\mcO\la_{34\wtmcO}/\la_{34\mcO}\rp_{k+1\,00}}=\frac{D-\Delta_\mcA-1}{D-\Delta_\mcA-2},
\ee
\begin{multline}
C_2=\mcN_\mcA\g_{34\wtmcA}/\g_{34\mcA}\lp\mcN_\mcO^{-1}\lp M^{-1}\rp_\al^{\hph{\al}\al}\rp_{k+1\,-\hlf\,\hlf}\\
=\frac{\lp\Delta_\mcA-1\rp\lp D-\Delta_\mcA+k\rp-\lp D-\Delta_\mcA-1\rp\lp\Delta_3-\Delta_4-1\rp}{\lp D-\Delta_\mcA-2\rp\lp D-\Delta_3+\Delta_4-\Delta_\mcA+k+1\rp},
\end{multline}
\begin{multline}
C_3=\mcN_\mcA\g_{34\wtmcA}/\g_{34\mcA}\lp\mcN_\mcO^{-1}\lp M^{-1}\rp_\beta^{\hph{\beta}\al}\rp_{k+1\,-\hlf\,\hlf}\\
=\frac{\lp k+1\rp\lp D-2\Delta_\mcA\rp\lp\Delta_3-\Delta_4-1\rp}{\lp D-\Delta_\mcA-2\rp\lp D-\Delta_3+\Delta_4-\Delta_\mcA+k+1\rp\lp\Delta_3-\Delta_4+\Delta_\mcA+k-1\rp}
\end{multline}
\begin{multline}
C_4=\mcN_\mcA\g_{34\wtmcA}/\g_{34\mcA}\lp\mcN_\mcO^{-1}\lp M^{-1}\rp_\al^{\hph{\al}\al}\rp_{k+1\,\hlf\,-\hlf}\\
=\frac{\lp\Delta_3-\Delta_4+\Delta_\mcA+k+1\rp\lp\lp\Delta_\mcA-1\rp\lp D-\Delta_\mcA+k\rp-\lp D-\Delta_\mcA-1\rp\lp\Delta_3-\Delta_4+1\rp\rp}{\lp D-\Delta_\mcA-2\rp\lp D+\Delta_3-\Delta_4-\Delta_\mcA+k+1\rp\lp -\Delta_3+\Delta_4+\Delta_\mcA+k-1\rp},
\end{multline}
\begin{multline}
C_5=\mcN_\mcA\g_{34\wtmcA}/\g_{34\mcA}\lp\mcN_\mcO^{-1}\lp M^{-1}\rp_\beta^{\hph{\beta}\al}\rp_{k+1\,\hlf\,-\hlf}\\
=\frac{\lp k+1\rp\lp D-2\Delta_\mcA\rp\lp\Delta_3-\Delta_4+1\rp}{\lp D-\Delta_\mcA-2\rp\lp D+\Delta_3-\Delta_4-\Delta_\mcA+k+1\rp\lp -\Delta_3+\Delta_4+\Delta_\mcA+k-1\rp},
\end{multline}
\be
\begin{split}
C_6= & \ \mcN_\mcA\g_{34\wtmcA}/\g_{34\mcA}\lp\mcN_\mcO^{-1}\lp\lp M^{-1}\rp_\al^{\hph{\al}\al}-\lp M^{-1}\rp_\al^{\hph{\al}\beta}\rp\rp_{k+2\,00}\\
= & \frac{\lp D-\Delta_\mcA+k\rp\lp\Delta_3-\Delta_4+\Delta_\mcA+k+1\rp}{\lp D-\Delta_\mcA-2\rp\lp\Delta_\mcA+k+1\rp}\\
& \qquad\times\frac{\lp D-\Delta_\mcA-1\rp\lp D-\Delta_\mcA+k+1\rp-\lp\Delta_\mcA-1\rp\lp\Delta_3-\Delta_4\rp}{\lp D+\Delta_3-\Delta_4-\Delta_\mcA+k+1\rp\lp D-\Delta_3+\Delta_4-\Delta_\mcA+k+1\rp},
\end{split}
\ee
\begin{multline}
C_7=\mcN_\mcA\g_{34\wtmcA}/\g_{34\mcA}\lp\mcN_\mcO^{-1}\lp\lp M^{-1}\rp_\beta^{\hph{\beta}\beta}-\lp M^{-1}\rp_\beta^{\hph{\beta}\al}\rp\rp_{k+2\,00}=\frac{D-\Delta_\mcA+k}{D-\Delta_\mcA-2}\\
\times\frac{\lp D-\Delta_\mcA-1\rp\lp\Delta_\mcA+k+1\rp\lp D-\Delta_\mcA+k+1\rp-\lp\Delta_\mcA-1\rp\lp\Delta_3-\Delta_4\rp^2}{\lp\Delta_\mcA+k+1\rp\lp D+\Delta_3-\Delta_4-\Delta_\mcA+k+1\rp\lp D-\Delta_3+\Delta_4-\Delta_\mcA+k+1\rp},
\end{multline}
\begin{multline}
C_8=\mcN_\mcA\g_{34\wtmcA}/\g_{34\mcA}\lp\mcN_\mcO^{-1}\lp\lp M^{-1}\rp_\al^{\hph{\al}\al}-\lp M^{-1}\rp_\al^{\hph{\al}\beta}\rp\rp_{k\,00}\\
=\frac{\lp\Delta_\mcA+k\rp\lp\lp D-\Delta_\mcA-1\rp\lp D-\Delta_\mcA+k-1\rp-\lp\Delta_\mcA-1\rp\lp\Delta_3-\Delta_4\rp\rp}{\lp D-\Delta_\mcA-2\rp\lp D-\Delta_\mcA+k-1\rp\lp -\Delta_3+\Delta_4+\Delta_\mcA+k-1\rp},
\end{multline}
\begin{multline}
C_9=\mcN_\mcA\g_{34\wtmcA}/\g_{34\mcA}\lp\mcN_\mcO^{-1}\lp\lp M^{-1}\rp_\beta^{\hph{\beta}\beta}-\lp M^{-1}\rp_\beta^{\hph{\beta}\al}\rp\rp_{k\,00}\\
=\frac{\lp\Delta_\mcA+k\rp\lp\lp D-\Delta_\mcA-1\rp\lp\Delta_\mcA+k-1\rp\lp D-\Delta_\mcA+k-1\rp-\lp\Delta_\mcA-1\rp\lp\Delta_3-\Delta_4\rp^2\rp}{\lp D-\Delta_\mcA-2\rp\lp D-\Delta_\mcA+k-1\rp\lp\Delta_3-\Delta_4+\Delta_\mcA+k-1\rp\lp -\Delta_3+\Delta_4+\Delta_\mcA+k-1\rp}.
\end{multline}
In computing these constants we have used notation where a subscript on a quantity in parentheses, $(f)_{k'\,P\,Q}$ means that we should evaluate $f$ (which is given in terms of three-point function data) for external particles of weights $\Delta_3+P$ and $\Delta_4+Q$, and an exchange operator of spin $\ell=k'$ and dimension $\Delta_\mcO=\Delta_\mcA$.
\section{Operators appearing in symmetric exchange blocks}
\label{app:DpOps}

Define
\be
\d_1=\Delta_3-\Delta_4+2u\p_u+2v\p_v,\qquad\d_2=\Delta_1-\Delta_2-\Delta_3+\Delta_4-2v\p_v,\non
\ee
\be
\d_3=2v\p_v,\qquad\d_4=\Delta_1-\Delta_2-2u\p_u-2v\p_v.
\ee

Then the operators which appear in the expression (\ref{eq:SymmExchangeGeneralExpression}) are
\be
\mcD^{--}_0=\sqrt{u}\lp\d_1-1\rp,\quad\mcD^{-+}_0=\sqrt{u}\lp\d_2-2\rp,\quad\mcD^{+-}_0=-\frac{\sqrt{u}}{v}\d_3,\quad\mcD^{++}_0=-\sqrt{u}\lp\d_4+1\rp,
\ee
\begin{multline}
\mcD^{--}_{11}=-\sqrt{u}\lp\d_2+v\lp\d_1+1\rp\rp\lp\d_1-1\rp,\quad\mcD^{-+}_{11}=-\sqrt{u}\lp\d_2+v\lp\d_1+1\rp\rp\lp\d_2-2\rp,\\
\mcD^{+-}_{11}=\sqrt{u}\lp\d_1-1\rp\lp\d_3+\d_4+1\rp,\quad\mcD^{++}_{11}=\sqrt{u}\d_2\lp\d_3+\d_4+1\rp,
\end{multline}
\begin{multline}
\mcD^{--}_{12}=-\frac{1}{\sqrt{v}}\lp\d_3-v\lp\d_1-1\rp\rp\lp\d_2+v\lp\d_1-1\rp\rp,\\
\mcD^{-+}_{12}=\sqrt{v}\lp\d_2-2+v\lp\d_1+1\rp\rp\lp\d_2-\d_4-1\rp,\\
\mcD^{+-}_{12}=\frac{1}{\sqrt{v}}\lp\d_3-v\lp\d_1-1\rp\rp\lp\d_3+\d_4+1\rp,\quad\mcD^{++}_{12}=-\sqrt{v}\lp\d_2-\d_4-1\rp\lp\d_3+\d_4+1\rp,
\end{multline}
\begin{multline}
\mcD^{--}_{21}=u\sqrt{v}\lp\d_1-1\rp\lp\d_1+1\rp,\quad\mcD^{-+}_{21}=u\sqrt{v}\lp\d_1+1\rp\lp\d_2-2\rp,\\
\mcD^{+-}_{21}=-\frac{u}{\sqrt{v}}\lp\d_1-1\rp\d_3,\quad\mcD^{++}_{21}=-\frac{u}{\sqrt{v}}\d_2\d_3,
\end{multline}
\begin{multline}
\mcD^{--}_{22}=\sqrt{u}\lp\d_3-v\lp\d_1+1\rp\rp\lp\d_1-1\rp,\quad\mcD^{-+}_{22}=-\sqrt{u}v\lp\d_1+1\rp\lp\d_2-\d_4-1\rp,\\
\mcD^{+-}_{22}=-\frac{\sqrt{u}}{v}\lp\d_3-2-v\lp\d_1-1\rp\rp\d_3,\quad\mcD^{++}_{22}=\sqrt{u}\lp\d_2-\d_4-1\rp\d_3.
\end{multline}

\newpage


\begin{thebibliography}{10}

\bibitem{Rattazzi}
R.~Rattazzi, V.~S. Rychkov, E.~Tonni, and A.~Vichi, ``{Bounding scalar operator
  dimensions in 4D CFT}'',
  \href{http://dx.doi.org/10.1088/1126-6708/2008/12/031}{{\em JHEP} {\bfseries
  0812} (2008) 031}, \href{http://arxiv.org/abs/0807.0004}{{\ttfamily
  arXiv:0807.0004 [hep-th]}}.

\bibitem{Rychkov2009}
V.~S. Rychkov and A.~Vichi, ``Universal constraints on conformal operator
  dimensions'', \href{http://dx.doi.org/10.1103/PhysRevD.80.045006}{{\em
  Phys.Rev.} {\bfseries D80} (2009) 045006},
  \href{http://arxiv.org/abs/0905.2211}{{\ttfamily arXiv:0905.2211 [hep-th]}}.

\bibitem{Caracciolo2010}
F.~Caracciolo and V.~S. Rychkov, ``Rigorous limits on the interaction strength
  in quantum field theory'',
  \href{http://dx.doi.org/10.1103/PhysRevD.81.085037}{{\em Phys.Rev.}
  {\bfseries D81} (2010) 085037},
  \href{http://arxiv.org/abs/0912.2726}{{\ttfamily arXiv:0912.2726 [hep-th]}}.

\bibitem{Poland}
D.~Poland and D.~Simmons-Duffin, ``{Bounds on 4D conformal and superconformal
  field theories}'', \href{http://dx.doi.org/10.1007/JHEP05(2011)017}{{\em
  JHEP} {\bfseries 1105} (2011) 017},
  \href{http://arxiv.org/abs/1009.2087}{{\ttfamily arXiv:1009.2087 [hep-th]}}.

\bibitem{Rattazzi2011}
R.~Rattazzi, S.~Rychkov, and A.~Vichi, ``{Central charge bounds in 4D conformal
  field theory}'', \href{http://dx.doi.org/10.1103/PhysRevD.83.046011}{{\em
  Phys.Rev.} {\bfseries D83} (2011) 046011},
  \href{http://arxiv.org/abs/1009.2725}{{\ttfamily arXiv:1009.2725 [hep-th]}}.

\bibitem{Rattazzi2011a}
R.~Rattazzi, S.~Rychkov, and A.~Vichi, ``{Bounds in 4D conformal field theories
  with global symmetry}'',
  \href{http://dx.doi.org/10.1088/1751-8113/44/3/035402}{{\em J.Phys.}
  {\bfseries A44} (2011) 035402},
  \href{http://arxiv.org/abs/1009.5985}{{\ttfamily arXiv:1009.5985 [hep-th]}}.

\bibitem{Rychkov2011}
S.~Rychkov, ``{Conformal Bootstrap in Three Dimensions?}'',
\href{http://arxiv.org/abs/1111.2115}{{\ttfamily arXiv:1111.2115 [hep-th]}}.

\bibitem{Polanda}
D.~Poland, D.~Simmons-Duffin, and A.~Vichi, ``{Carving Out the Space of 4D
  CFTs}'', \href{http://dx.doi.org/10.1007/JHEP05(2012)110}{{\em JHEP}
  {\bfseries 1205} (2012) 110},
  \href{http://arxiv.org/abs/1109.5176}{{\ttfamily arXiv:1109.5176 [hep-th]}}.

\bibitem{El-Showk2012a}
S.~El-Showk, M.~F. Paulos, D.~Poland, S.~Rychkov, D.~Simmons-Duffin, and
  A.~Vichi, ``{Solving the 3D Ising Model with the Conformal Bootstrap}'',
  \href{http://dx.doi.org/10.1103/PhysRevD.86.025022}{{\em Phys.Rev.}
  {\bfseries D86} (2012) 025022},
  \href{http://arxiv.org/abs/1203.6064}{{\ttfamily arXiv:1203.6064 [hep-th]}}.

  \bibitem{El-Showk2014}
S.~El-Showk, M.~F. Paulos, D.~Poland, S.~Rychkov, D.~Simmons-Duffin, and
  A.~Vichi, ``{Solving the 3d Ising Model with the Conformal Bootstrap II. c-Minimization and Precise Critical Exponents}'',
  \href{http://dx.doi.org/10.1007/s10955-014-1042-7}{{\em J. Stat. Phys.}
  {\bfseries 157} (2014) 869-914},
  \href{http://arxiv.org/abs/1403.4545}{{\ttfamily arXiv:1403.4545 [hep-th]}}.
  
\bibitem{El-Showk2014a}
S.~El-Showk, M.~Paulos, D.~Poland, S.~Rychkov, D.~Simmons-Duffin, {\em et~al.},
  ``{Conformal Field Theories in Fractional Dimensions}'',
  \href{http://dx.doi.org/10.1103/PhysRevLett.112.141601}{{\em Phys.Rev.Lett.}
  {\bfseries 112} (2014) 141601},
\href{http://arxiv.org/abs/1309.5089}{{\ttfamily arXiv:1309.5089 [hep-th]}}.

\bibitem{Caracciolo2014}
F.~Caracciolo, A.~C. Echeverri, B.~von Harling, and M.~Serone, ``{Bounds on OPE
  Coefficients in 4D Conformal Field Theories}'',
\href{http://arxiv.org/abs/1406.7845}{{\ttfamily arXiv:1406.7845 [hep-th]}}.

\bibitem{Kos2014}
F.~Kos, D.~Poland, and D.~Simmons-Duffin, ``{Bootstrapping Mixed Correlators in
  the 3D Ising Model}'',
\href{http://arxiv.org/abs/1406.4858}{{\ttfamily arXiv:1406.4858 [hep-th]}}.

\bibitem{Fitzpatrick2012}
A.~L. Fitzpatrick, J.~Kaplan, D.~Poland, and D.~Simmons-Duffin, ``{The Analytic
  Bootstrap and AdS Superhorizon Locality}'',
  \href{http://dx.doi.org/10.1007/JHEP12(2013)004}{{\em JHEP} {\bfseries 1312}
  (2013) 004},
\href{http://arxiv.org/abs/1212.3616}{{\ttfamily arXiv:1212.3616 [hep-th]}}.

\bibitem{Komargodski2013}
Z.~Komargodski and A.~Zhiboedov, ``{Convexity and Liberation at Large Spin}'',
  \href{http://dx.doi.org/10.1007/JHEP11(2013)140}{{\em JHEP} {\bfseries 1311}
  (2013) 140},
\href{http://arxiv.org/abs/1212.4103}{{\ttfamily arXiv:1212.4103 [hep-th]}}.

\bibitem{Fitzpatrick2014a}
A.~L. Fitzpatrick, J.~Kaplan, and M.~T. Walters, ``{Universality of
  Long-Distance AdS Physics from the CFT Bootstrap}'',
  \href{http://dx.doi.org/10.1007/JHEP08(2014)145}{{\em JHEP} {\bfseries 1408}
  (2014) 145},
\href{http://arxiv.org/abs/1403.6829}{{\ttfamily arXiv:1403.6829 [hep-th]}}.

\bibitem{Kaviraj2015}
A.~Kaviraj, K.~Sen, and A.~Sinha, ``Analytic bootstrap at large spin'',
  \href{http://arxiv.org/abs/1502.01437}{{\ttfamily 1502.01437}}.

\bibitem{Alday2015}
L.~F. Alday, A.~Bissi, and T.~Lukowski, ``{Large spin systematics in CFT}'',
\href{http://arxiv.org/abs/1502.07707}{{\ttfamily arXiv:1502.07707 [hep-th]}}.

\bibitem{Kaviraj2015a}
A.~Kaviraj, K.~Sen, and A.~Sinha, ``{Universal anomalous dimensions at large
  spin and large twist}'',
\href{http://arxiv.org/abs/1504.00772}{{\ttfamily arXiv:1504.00772 [hep-th]}}.

\bibitem{Vichi}
A.~Vichi, ``{Improved bounds for CFT's with global symmetries}'',
  \href{http://dx.doi.org/10.1007/JHEP01(2012)162}{{\em JHEP} {\bfseries 1201}
  (2012) 162}, \href{http://arxiv.org/abs/1106.4037}{{\ttfamily arXiv:1106.4037
  [hep-th]}}.

 \bibitem{Hogervorst}
M.~Hogervorst, H.~Osborn, and S.~Rychkov, ``{Diagonal limit for conformal blocks in d dimensions}'', \href{http://dx.doi.org/10.1007/JHEP08(2013)014}{{\em JHEP}
  {\bfseries 1308} (2013) 014},
\href{http://arxiv.org/abs/1305.1321}{{\ttfamily arXiv:1305.1321}}.

\bibitem{Hogervorst2013}
M.~Hogervorst, and S.~Rychkov, ``{Radial coordinates for conformal blocks}'', \href{http://dx.doi.org/10.1103/PhysRevD.87.106004}{{\em Phys.Rev.}
  {\bfseries D87} (2013) 106004},
\href{http://arxiv.org/abs/1303.1111}{{\ttfamily arXiv:1303.1111}}.

\bibitem{Kos2014a}
F.~Kos, D.~Poland, and D.~Simmons-Duffin, ``{Bootstrapping the $O(N)$ vector
  models}'', \href{http://dx.doi.org/10.1007/JHEP06(2014)091}{{\em JHEP}
  {\bfseries 1406} (2014) 091},
\href{http://arxiv.org/abs/1307.6856}{{\ttfamily arXiv:1307.6856}}.

\bibitem{Bae2014}
J.-B. Bae and S.-J. Rey, ``{Conformal Bootstrap Approach to O(N) Fixed Points
  in Five Dimensions}'',
\href{http://arxiv.org/abs/1412.6549}{{\ttfamily arXiv:1412.6549 [hep-th]}}.

\bibitem{Nakayama2014}
Y.~Nakayama and T.~Ohtsuki, ``{Five dimensional $O(N)$-symmetric CFTs from
  conformal bootstrap}'',
\href{http://arxiv.org/abs/1404.5201}{{\ttfamily arXiv:1404.5201 [hep-th]}}.

\bibitem{Chester2014}
S.~M. Chester, S.~S. Pufu, and R.~Yacoby, ``{Bootstrapping O(N) Vector Models
  in 4 \&lt; d \&lt; 6}'',
\href{http://arxiv.org/abs/1412.7746}{{\ttfamily arXiv:1412.7746 [hep-th]}}.

\bibitem{Kos2015}
F.~Kos, D.~Poland, D.~Simmons-Duffin, and A.~Vichi, ``{Bootstrapping the O(N)
  Archipelago}'',
\href{http://arxiv.org/abs/1504.07997}{{\ttfamily arXiv:1504.07997 [hep-th]}}.

\bibitem{Beem2013}
C.~Beem, L.~Rastelli, and B.~C. van Rees, ``{$N=4$ Superconformal Bootstrap}'',
  \href{http://dx.doi.org/10.1103/PhysRevLett.111.071601}{{\em Phys. Rev.
  Lett.} {\bfseries 111} (2013) },
  \href{http://arxiv.org/abs/1304.1803}{{\ttfamily arXiv:1304.1803 [hep-th]}}.

\bibitem{Alday2014}
L.~F. Alday and A.~Bissi, ``{The superconformal bootstrap for structure
  constants}'', \href{http://dx.doi.org/10.1007/JHEP09(2014)144}{{\em JHEP}
  {\bfseries 1409} (2014) 144},
\href{http://arxiv.org/abs/1310.3757}{{\ttfamily arXiv:1310.3757 [hep-th]}}.

\bibitem{Bashkirov2013}
D.~Bashkirov, ``{Bootstrapping the $N=1$ SCFT in three dimensions}'',
\href{http://arxiv.org/abs/1310.8255}{{\ttfamily arXiv:1310.8255 [hep-th]}}.

\bibitem{Fitzpatrick:2014oza} 
  A.~L.~Fitzpatrick, J.~Kaplan, Z.~U.~Khandker, D.~Li, D.~Poland and D.~Simmons-Duffin,
  ``Covariant Approaches to Superconformal Blocks,''
  JHEP {\bf 1408}, 129 (2014),
  \href{http://arxiv.org/abs/1402.1167}{{\ttfamily arXiv:1402.1167 [hep-th]]}}.

\bibitem{Khandker:2014mpa} 
  Z.~U.~Khandker, D.~Li, D.~Poland and D.~Simmons-Duffin,
  JHEP {\bf 1408}, 049 (2014)
  [arXiv:1404.5300 [hep-th]].
  
\bibitem{Berkooz2014}
M.~Berkooz, R.~Yacoby, and A.~Zait, ``{Bounds on $ \mathcal{N} $ = 1
  superconformal theories with global symmetries}'',
  \href{http://dx.doi.org/10.1007/JHEP08(2014)008}{{\em JHEP} {\bfseries 1408}
  (2014) 008},
\href{http://arxiv.org/abs/1402.6068}{{\ttfamily arXiv:1402.6068 [hep-th]}}.

\bibitem{Alday2015a}
L.~F. Alday and A.~Bissi, ``{Generalized bootstrap equations for $
  \mathcal{N}=4 $ SCFT}'',
  \href{http://dx.doi.org/10.1007/JHEP02(2015)101}{{\em JHEP} {\bfseries 1502}
  (2015) 101},
\href{http://arxiv.org/abs/1404.5864}{{\ttfamily arXiv:1404.5864 [hep-th]}}.

\bibitem{Nakayama2014a}
Y.~Nakayama and T.~Ohtsuki, ``{Approaching the conformal window of $O(n)\times
  O(m)$ symmetric Landau-Ginzburg models using the conformal bootstrap}'',
  \href{http://dx.doi.org/10.1103/PhysRevD.89.126009}{{\em Phys.Rev.}
  {\bfseries D89} (2014) 126009},
\href{http://arxiv.org/abs/1404.0489}{{\ttfamily arXiv:1404.0489 [hep-th]}}.

\bibitem{Chester2014a}
S.~M. Chester, J.~Lee, S.~S. Pufu, and R.~Yacoby, ``{The $ \mathcal{N}=8 $
  superconformal bootstrap in three dimensions}'',
  \href{http://dx.doi.org/10.1007/JHEP09(2014)143}{{\em JHEP} {\bfseries 1409}
  (2014) 143},
\href{http://arxiv.org/abs/1406.4814}{{\ttfamily arXiv:1406.4814 [hep-th]}}.

\bibitem{Beem2014}
C.~Beem, M.~Lemos, P.~Liendo, L.~Rastelli, and B.~C. van Rees, ``{The
  ${\mathcal N}=2$ superconformal bootstrap}'',
\href{http://arxiv.org/abs/1412.7541}{{\ttfamily arXiv:1412.7541 [hep-th]}}.

\bibitem{Bobev2015a}
N.~Bobev, S.~El-Showk, D.~Mazac, and M.~F. Paulos, ``{Bootstrapping the
  Three-Dimensional Supersymmetric Ising Model}'',
\href{http://arxiv.org/abs/1502.04124}{{\ttfamily arXiv:1502.04124 [hep-th]}}.

\bibitem{Bobev2015}
N.~Bobev, S.~El-Showk, D.~Mazac, and M.~F. Paulos, ``{Bootstrapping SCFTs with
  Four Supercharges}'',
\href{http://arxiv.org/abs/1503.02081}{{\ttfamily arXiv:1503.02081 [hep-th]}}.

\bibitem{Iliesiu:2015qra} 
  L.~Iliesiu, F.~Kos, D.~Poland, S.~S.~Pufu, D.~Simmons-Duffin and R.~Yacoby,
  ``Bootstrapping 3D Fermions,''
  \href{http://arxiv.org/abs/1508.00012}{{\ttfamily arXiv:1508.00012 [hep-th]}}.
  
\bibitem{Dolan2001}
F.~Dolan and H.~Osborn, ``Conformal four point functions and the operator
  product expansion'',
  \href{http://dx.doi.org/10.1016/S0550-3213(01)00013-X}{{\em Nucl.Phys.}
  {\bfseries B599} (2001) 459--496},
  \href{http://arxiv.org/abs/hep-th/0011040}{{\ttfamily arXiv:hep-th/0011040
  [hep-th]}}.

\bibitem{Dolan2004}
F.~Dolan and H.~Osborn, ``Conformal partial waves and the operator product
  expansion'', \href{http://dx.doi.org/10.1016/j.nuclphysb.2003.11.016}{{\em
  Nucl.Phys.} {\bfseries B678} (2004) 491--507},
  \href{http://arxiv.org/abs/hep-th/0309180}{{\ttfamily arXiv:hep-th/0309180
  [hep-th]}}.

\bibitem{Dolan2011}
F.~Dolan and H.~Osborn, ``{Conformal Partial Waves: Further Mathematical
  Results}'', \href{http://arxiv.org/abs/1108.6194}{{\ttfamily arXiv:1108.6194
  [hep-th]}}.

\bibitem{Costa2011}
M.~S. Costa, J.~Penedones, D.~Poland, and S.~Rychkov, ``{Spinning Conformal
  Blocks}'', \href{http://dx.doi.org/10.1007/JHEP11(2011)154}{{\em JHEP}
  {\bfseries 1111} (2011) 154},
\href{http://arxiv.org/abs/1109.6321}{{\ttfamily arXiv:1109.6321 [hep-th]}}.

\bibitem{Simmons-Duffin2014}
D.~Simmons-Duffin, ``{Projectors, Shadows, and Conformal Blocks}'',
  \href{http://dx.doi.org/10.1007/JHEP04(2014)146}{{\em JHEP} {\bfseries 1404}
  (2014) 146},
\href{http://arxiv.org/abs/1204.3894}{{\ttfamily arXiv:1204.3894 [hep-th]}}.

\bibitem{Costa2014}
M.~S. Costa and T.~Hansen, ``{Conformal correlators of mixed-symmetry
  tensors}'',
\href{http://arxiv.org/abs/1411.7351}{{\ttfamily arXiv:1411.7351 [hep-th]}}.

\bibitem{Echeverri2015}
A.~C. Echeverri, E.~Elkhidir, D.~Karateev, and M.~Serone, ``{Deconstructing
  Conformal Blocks in 4D CFT}'',
\href{http://arxiv.org/abs/1505.03750}{{\ttfamily arXiv:1505.03750 [hep-th]}}.

\bibitem{Costa:2011mg} 
  M.~S.~Costa, J.~Penedones, D.~Poland and S.~Rychkov,
  ``Spinning Conformal Correlators,''
  JHEP {\bf 1111}, 071 (2011)
  \href{http://arxiv.org/abs/1107.3554}{{\ttfamily arXiv:1107.3554 [hep-th]}}.

\bibitem{Hijano:2015zsa} 
  E.~Hijano, P.~Kraus, E.~Perlmutter and R.~Snively,
  ``Witten Diagrams Revisited: The AdS Geometry of Conformal Blocks,''
  \href{http://arxiv.org/abs/1508.00501}{{\ttfamily arXiv:1508.00501 [hep-th]}}.
    
\bibitem{Dymarsky:2013wla} 
  A.~Dymarsky,
  ``On the four-point function of the stress-energy tensors in a CFT,''
  \href{http://arxiv.org/abs/1311.4546}{{\ttfamily arXiv:1311.4546 [hep-th]}}.
\bibitem{Geyer1999}
B.~Geyer, M.~Lazar and D.~Robaschik, ``Decomposition of nonlocal light cone operators into harmonic operators of definite twist'',
  \href{http://dx.doi.org/10.1016/S0550-3213(99)00334-X}{{\em Nucl.Phys.}
  {\bfseries B559} (1999) 339--377},
  \href{http://arxiv.org/abs/hep-th/9901090}{{\ttfamily arXiv:hep-th/9901090
  [hep-th]}}.
  
\bibitem{Geyer2000}
B.~Geyer and M.~Lazar, ``Twist decomposition of nonlocal light cone operators. 2. General tensors of 2nd rank'',
  \href{http://dx.doi.org/10.1016/S0550-3213(00)00227-3}{{\em Nucl.Phys.}
  {\bfseries B581} (2000) 341--390},
  \href{http://arxiv.org/abs/hep-th/0003080}{{\ttfamily arXiv:hep-th/0003080
  [hep-th]}}.
  
  
\bibitem{Eilers2006} 
  J.~Eilers,
  ``Geometric twist decomposition off the light-cone for nonlocal QCD operators,''
  \href{http://arxiv.org/abs/hep-th/0608173}{{\ttfamily arXiv:0608173 [hep-th]}}.
\end{thebibliography}

\providecommand{\href}[2]{#2}\begingroup\raggedright\endgroup

\end{document}